\documentclass[apj]{emulateapj}
\usepackage{amssymb,amsmath}
\usepackage{color,hyperref}
\definecolor{linkcolor}{rgb}{0,0,0.25}
\hypersetup{
  colorlinks=true,        % false: boxed links; true: colored links
  linkcolor=linkcolor,    % color of internal links
  citecolor=linkcolor,    % color of links to bibliography
  filecolor=linkcolor,    % color of file links
  urlcolor=linkcolor     % color of external links
}
\newcounter{address}
\newcommand{\ie}{i.e.}
\newcommand{\etal}{et al.}
\newcommand{\dd}{\mathrm{d}}
\newcommand{\eg}{e.g.}
\newcommand{\Eqnname}{Equation}
\newcommand{\eqnname}{Equation}
\newcommand{\equationname}{\eqnname}
\newcommand{\figurenames}{\figurename s}
\newcommand{\sectionname}{$\mathsection$}
\newcommand{\apogee}{APOGEE}

\newcommand{\sdssiii}{SDSS-III}
\newcommand{\twomass}{2MASS}
\newcommand{\wise}{WISE}

\newcommand{\feh}{\ensuremath{[\mathrm{Fe/H}]}}

\newcommand{\logg}{\ensuremath{\log g}}

\newcommand{\Ro}{\ensuremath{R_0}}
\newcommand{\ro}{\Ro}

\newcommand{\hR}{\ensuremath{h_R}}
\newcommand{\hr}{\hR}
\newcommand{\hs}{\ensuremath{h_\sigma}}

\newcommand{\vc}{\ensuremath{V_c}}

\newcommand{\va}{\ensuremath{V_a}}
\newcommand{\vlos}{\ensuremath{V_{\mathrm{los}}}}
\newcommand{\vsun}{\ensuremath{{\bf V}_\odot^{\mathrm{gal}}}}
\newcommand{\df}{\ensuremath{\mathrm{DF}}}

\newcommand{\dens}{\ensuremath{\nu_*}}
\newcommand{\iso}{\ensuremath{\mathrm{iso}}}

\newcommand{\ks}{\ensuremath{K_s}}
\newcommand{\kso}{\ensuremath{K_{s,0}}}
\newcommand{\dm}{\ensuremath{\mu}}
\newcommand{\platel}{\ensuremath{l_{\mathrm{field}}}}
\newcommand{\plateb}{\ensuremath{b_{\mathrm{field}}}}
\newcommand{\fdehnen}{\ensuremath{f_{\text{Dehnen}}}}
\newcommand{\sigmaR}{\ensuremath{\sigma_R}}
\newcommand{\rE}{\ensuremath{R_e}}
\newcommand{\Lc}{\ensuremath{L_c}}

\newcommand{\ndata}{3,365}
\newcommand{\nfields}{14}
\newcommand{\nfieldsexpand}{fourteen}

\newcommand{\kpc}{\,\mathrm{kpc}}
\newcommand{\kms}{\,\mathrm{km\ s}^{-1}}
\newcommand{\dex}{\ensuremath{\mathrm{dex}}}

\begin{document}

\submitted{}

\title{The Milky Way's circular velocity curve between $4$ and 
  $14\,\mathrm{KPC}$ from APOGEE data}
\author{Jo~Bovy\altaffilmark{1,2},
  Carlos~Allende~Prieto\altaffilmark{3,4},
  Timothy~C.~Beers\altaffilmark{5,6},
  Dmitry~Bizyaev\altaffilmark{7},
  Luiz~N.~da~Costa\altaffilmark{8,9},
  Katia~Cunha\altaffilmark{9,10},
  Garrett~L.~Ebelke\altaffilmark{7},
  Daniel~J.~Eisenstein\altaffilmark{11},
  Peter~M.~Frinchaboy\altaffilmark{12},
  Ana~Elia~Garc\'{\i}a~P\'{e}rez\altaffilmark{13},
  L\'{e}o~Girardi\altaffilmark{8,14},
  Fred~R.~Hearty\altaffilmark{13},
  David~W.~Hogg\altaffilmark{15,16},
  Jon~Holtzman\altaffilmark{17},
  Marcio~A.~G.~Maia\altaffilmark{8,9},
  Steven~R.~Majewski\altaffilmark{13},
  Elena~Malanushenko\altaffilmark{7},
  Viktor~Malanushenko\altaffilmark{7},
  Szabolcs~M{\'e}sz{\'a}ros\altaffilmark{3,4},
  David~L.~Nidever\altaffilmark{13},
  Robert~W.~O'Connell\altaffilmark{13},
  Christine~O'Donnell\altaffilmark{13},
  Audrey~Oravetz\altaffilmark{7},
  Kaike~Pan\altaffilmark{7},
  Helio~J.~Rocha-Pinto\altaffilmark{8,18},
  Ricardo~P.~Schiavon\altaffilmark{19},
  Donald~P.~Schneider\altaffilmark{20,21},
  Mathias~Schultheis\altaffilmark{22},
  Michael~Skrutskie\altaffilmark{13},
  Verne~V.~Smith\altaffilmark{5,9},
  David~H.~Weinberg\altaffilmark{23},
  John~C.~Wilson\altaffilmark{13},
  and Gail~Zasowski\altaffilmark{13,23}
}
\altaffiltext{\theaddress}{\label{IAS}\stepcounter{address} %1
  Institute for Advanced Study, Einstein Drive, Princeton, NJ 08540, USA; bovy@ias.edu~} 
\altaffiltext{\theaddress}{\label{Hubble}\stepcounter{address} %2
  Hubble fellow}
\altaffiltext{\theaddress}{\label{IAC}\stepcounter{address} %3
  Instituto de Astrof{\'{\i}}sica de Canarias (IAC), E-38200 La Laguna, Tenerife, Spain} %10
\altaffiltext{\theaddress}{\label{Tenerife}\stepcounter{address} %4
  Departamento de Astrof{\'{\i}}sica, Universidad de La Laguna (ULL), E-38206 La Laguna, Tenerife, Spain}
\altaffiltext{\theaddress}{\label{NOAO}\stepcounter{address} %5
  National Optical Astronomy Observatory, Tucson, AZ 85719, USA}
\altaffiltext{\theaddress}{\label{Michigan}\stepcounter{address} %6
  Department of Physics \& Astronomy and JINA (Joint Institute for
  Nuclear Astrophysics), Michigan State University, East Lansing, MI
  48824, USA}
\altaffiltext{\theaddress}{\label{APO}\stepcounter{address} %7
  Apache Point Observatory, P.O. Box 59, Sunspot, NM 88349, USA}
\altaffiltext{\theaddress}{\label{RioLab}\stepcounter{address} %8
  Laborat\'{o}rio Interinstitucional de e-Astronomia - LIneA, Rua Gal. Jos\'e Cristino 77, Rio de Janeiro, RJ - 20921-400, Brazil}
\altaffiltext{\theaddress}{\label{Rio}\stepcounter{address} %9
  Observat\'{o}rio Nacional, Rio de Janeiro, RJ 20921-400, Brazil}
\altaffiltext{\theaddress}{\label{Steward}\stepcounter{address} %10
  Steward Observatory, U. Arizona, Tucson, AZ 85719, USA}
\altaffiltext{\theaddress}{\label{Harvard}\stepcounter{address} %11
  Harvard-Smithsonian Center for Astrophysics, 60 Garden St., MS \#20, Cambridge, MA 02138, USA}
\altaffiltext{\theaddress}{\label{TCU}\stepcounter{address} %12
Department of Physics and Astronomy, Texas Christian University, Fort Worth, TX 76129, USA}
\altaffiltext{\theaddress}{\label{UVa}\stepcounter{address} %13
  Department of Astronomy, University of Virginia, Charlottesville, VA, 22904, USA}
\altaffiltext{\theaddress}{\label{Padova}\stepcounter{address} %14
  Osservatorio Astronomico di Padova - INAF, Vicolo dell'Osservatorio 5, I-35122 Padova, Italy}
\altaffiltext{\theaddress}{\label{NYU}\stepcounter{address} %15
  Center for Cosmology and Particle Physics, Department of Physics, New York
  University, 4 Washington Place, New York, NY 10003, USA}
\altaffiltext{\theaddress}{\label{MPIA}\stepcounter{address}%16
  Max-Planck-Institut f\"ur Astronomie, K\"onigstuhl 17, D-69117
  Heidelberg, Germany}
\altaffiltext{\theaddress}{\label{NMSU}\stepcounter{address}%17
  New Mexico State University, Las Cruces, NM 88003, USA}
\altaffiltext{\theaddress}{\label{Valongo}\stepcounter{address}%18
  Observatório do Valongo, Universidade Federal do Rio de Janeiro,
Rio de Janeiro RJ 20080-090, Brazil}
\altaffiltext{\theaddress}{\label{NMSU}\stepcounter{address}%19
  Gemini Observatory, 670 A'ohoku Place, Hilo, HI 96720, USA}
\altaffiltext{\theaddress}{\label{PennState}\stepcounter{address} %20
  Department of Astronomy and Astrophysics, The Pennsylvania State University, University Park, PA 16802, USA}
\altaffiltext{\theaddress}{\label{PennState2}\stepcounter{address} %21
  Institute for Gravitation and the Cosmos, The Pennsylvania State University, University Park, PA 16802, USA}
\altaffiltext{\theaddress}{\label{Besancon}\stepcounter{address} %22
  Institut Utinam, CNRS UMR6213, OSU THETA, Universit\'e de Franche-Comt\'e, 41bis, avenue de l'Observatoire , 25000 Besancon, France}
\altaffiltext{\theaddress}{\label{Besancon}\stepcounter{address} %23
  The Ohio State University, Department of Astronomy, Columbus OH 43210, USA}

\begin{abstract} 
  We measure the Milky Way's rotation curve over the Galactocentric
  range $4\kpc \lesssim R \lesssim 14\kpc$ from the first year of data
  from the Apache Point Observatory Galactic Evolution Experiment
  (\apogee). We model the line-of-sight velocities of \ndata\ stars
  in \nfieldsexpand\ fields with $b = 0^\circ$ between $30^\circ \leq
  l \leq 210^\circ$ out to distances of $10\,\kpc$ using an
  axisymmetric kinematical model that includes a correction for the
  asymmetric drift of the warm tracer population ($\sigma_R \approx
  35 \kms$). We determine the local value of the circular velocity to
  be $\vc(\ro) = 218 \pm 6\,\kms$ and find that the rotation curve is
  approximately flat with a local derivative between
  $-3.0\,\kms\,\kpc^{-1}$ and $0.4\,\kms\,\kpc^{-1}$. We also measure
  the Sun's position and velocity in the Galactocentric rest frame,
  finding the distance to the Galactic center to be $8\,\kpc < \ro <
  9\,\kpc$, radial velocity $V_{R,\odot} = -10 \pm 1\,\kms$, and
  rotational velocity $V_{\phi,\odot} = 242^{+10}_{-3}\,\kms$, in good
  agreement with local measurements of the Sun's radial velocity and
  with the observed proper motion of Sgr A$^*$. We investigate various
  systematic uncertainties and find that these are limited to offsets
  at the percent level, $\sim\!2\kms$ in \vc. Marginalizing over all
  the systematics that we consider, we find that $V_c(R_0) <
  235\,\kms$ at $>99\,\%$ confidence. We find an offset between the
  Sun's rotational velocity and the local circular velocity of $26 \pm
  3\,\kms$, which is larger than the locally-measured solar motion of
  $12\,\kms$. This larger offset reconciles our value for $V_c$ with
  recent claims that $V_c \gtrsim 240\,\kms$. Combining our results
  with other data, we find that the Milky Way's dark-halo mass within
  the virial radius is $\sim\!8\times10^{11}\,M_\odot$.
\end{abstract}

\section{Introduction}\label{sec:intro}

The Milky Way's inner rotation curve $V_c(R)$, and in particular its
value $V_c \equiv V_c(R_0)$ at the Sun's Galactocentric radius $R_0$,
is crucial for our understanding of many Galactic and extra-galactic
observations. It provides an important constraint on mass models for
the Milky Way and the question of whether the Galactic disk is
maximal. The shape of the rotation curve is an important ingredient
for realistic models of the disk's formation and evolution. The
circular velocity at the Sun, which is located approximately $2.5$
disk scale lengths from the Galactic center
\citep{Bovy12c}, is also important for placing the Milky Way in a
cosmological context, for example, when asking whether the Milky Way
follows the Tully-Fisher relation
\citep[\eg,][]{Klypin02a,Flynn06a,Hammer07a}. $V_c$ is often
considered to be an important parameter for dark-matter
direct-detection experiments and for correcting the motion of
extra-galactic objects for the motion of the Sun, although we seek to
dispel these notions below (\sectionname~\ref{sec:discuss-solar}).

Traditionally, the local circular velocity has been obtained by
measuring the Sun's motion with respect to an object that is assumed
to be at rest with respect to the Galaxy. The most robust of those
measurements is that derived from the observed proper motion of Sgr
A$^*$ \citep{Reid04a}, even though this requires an estimate of
$R_0$. An alternative method that does not require knowledge of $R_0$
consists of measuring the Sun's reflex motion using a stellar tidal
stream with orbital pole near $l =
270^\circ$ \citep{Majewski06a}. Similar measurements using samples of
halo stars
\citep{Sirko04a}, or the globular cluster system \citep{Woltjer75a},
may be contaminated by the residual, presumably prograde, motion of
these populations and, thus, only provide a lower limit. As discussed
in detail in \sectionname~\ref{sec:discuss-solar}, such 
measurements intrinsically measure the \emph{Sun's} rotational
velocity, $V_{\phi,\odot}$, rather than the circular velocity. To
arrive at $V_c$, these measurements depend on a highly uncertain
correction for the Sun's motion with respect to $V_c$. The measurement
of $V_c$ by observing the extreme line-of-sight velocity of HI
emission toward Galactic longitudes $l\!\sim\!90^\circ$ in principle
also directly measures $V_{\phi, \odot}$ \citep{KTG79}, although the
HI density drops too steeply with radius for $V_{\phi, \odot}$ to be
directly observed, and more intricate modeling is necessary to turn
the HI emission toward $l\!\sim\!90^\circ$ into a constraint on
$V_c$.

Other measurements of $V_c$ have in the past been limited to
measurements of the Oort constants, due to a lack of data at large
distances from the Sun \citep[\eg,][]{Feast97a}. Recently, new
estimates have been obtained from the kinematics of masers in the
Galactic disk \citep{Reid09a,Bovy09a,Mcmillan10a}, and from fitting an
orbit to the cold stellar stream GD-1 \citep{Koposov10a}. However,
until now, no consensus has been reached as to whether $V_c$ lies near
the IAU-recommended value of $V_c = 220\,\kms$ \citep{Kerr86a}, or
whether it needs to be revised upward to around $V_c =
250\,\kms$ \citep[\eg,][]{Ghez08a,Reid09a,Bovy09a,Schoenrich12a}.

Most of the information relating to the shape of the rotation curve is
based upon the observed kinematics of the HI emission, either through
the tangent-point method at $l < 90^\circ$ and $l > 270^\circ$
\citep{vdH54a,GKT79,Levine08a}, or through the observed thickness of
the HI layer at $90^\circ < l < 270^\circ$ \citep{Merrifield92a}. Such
measurements are purely geometrical and, in particular, cannot
constrain the component of Galactic rotation that has a uniform
angular speed. This also holds for measurements of $V_c$ based on the
line-of-sight velocities of open clusters
\citep[\eg,][]{Frinchaboy08a}. The best constraints on the local
slope of the circular-velocity curve therefore also come from
measurements of the Oort constants \citep[\eg,][]{Feast97a}, although
the HI observations are crucial to measuring the rotation curve over
the full radial range of the disk.

In this paper we present the first measurement of the Milky Way's
circular velocity curve using kinematically-warm stellar tracers out
to heliocentric distances of $10\,\kpc$, from the Sloan Digital Sky
Survey III's Apache Point Observatory Galactic Evolution Experiment
(\sdssiii/\apogee; \citealt{Eisenstein11a}; S.~R.~Majewski, \etal\
2012, in preparation). \apogee\ is a high-resolution spectroscopic
survey covering all of the major components of the Galaxy
that---crucially---operates in the near-infrared, which allows stars
to be observed to large distances in the dust-obscured regions of the
inner Milky Way disk. While warm stellar-disk tracers do not on
average rotate at the circular velocity like the HI emission discussed
above, their offset from $V_c(R)$---the so-called asymmetric drift
\citep{Stromberg46a}---is a dynamical effect that can be calculated
from their observed velocity dispersion. Because our measurement
relies on a dynamical effect, we are sensitive to $V_c$ in the sense
of the radial-force component at $R$, rather than to $V_{\phi,\odot}$,
although the large range in $l$ of our sample allows us to
independently measure $V_{\phi,\odot}$. Additional benefits of using
the intermediate-age to old stellar population to trace the dynamics
of the disk are that these populations are much less sensitive to
non-axisymmetric streaming motions than cold gas, and that the large
asymmetric drift can be used to constrain the component of Galactic
rotation that possesses a uniform angular speed.

Using a data set of \ndata\ stars at $b = 0^\circ$ along
\nfields\ lines of sight covering $30^\circ \leq l \leq 210^\circ$, we find
that the Milky Way's rotation curve is approximately flat over
$4\,\kpc < R < 14\,\kpc$, with $V_c(R_0) = 218 \pm 6\,\kms$. A value
of $V_c > 235\,\kms$ is ruled out at $>99\%$ confidence by our
data. Our measurement of $V_{\phi,\odot} = 242^{+10}_{-3}\,\kms$
agrees with the observed proper motion of Sgr A$^*$, but the Sun's
offset from $V_c$ is larger than the locally-measured value
\citep{Schoenrich10a} by $14 \pm 3\,\kms$.

\begin{figure}[htbp]
\includegraphics[width=0.5\textwidth,clip=]{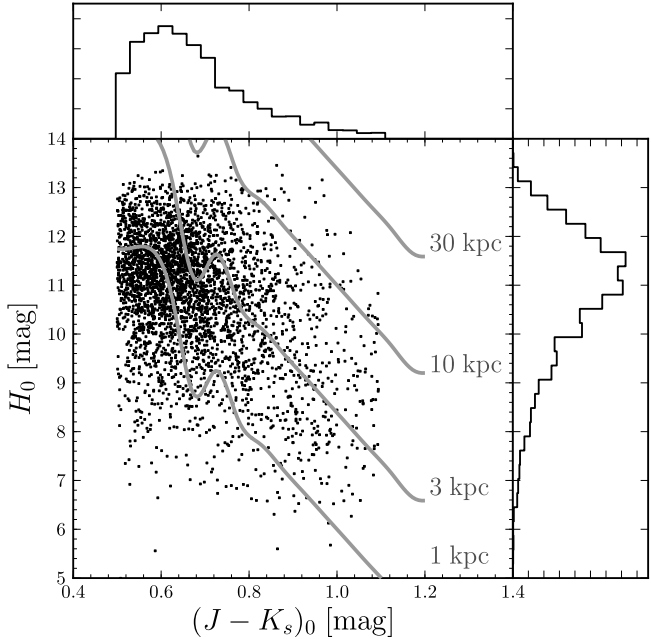}
\caption{Distribution of the \ndata\ mid-plane data points in extinction-corrected color and magnitude. The gray curves give approximate distances based on
  the peak of the distance likelihood
  in \figurename~\ref{fig:imf_h_jk}.}\label{fig:data_h_jk}
\end{figure}

The outline of this paper is as follows. In
\sectionname~\ref{sec:data}, we describe the \apogee\ data set. The methodology employed to model the observed kinematics of
the kinematically-warm tracer population is presented
in \sectionname~\ref{sec:method}. We discuss our results
in \sectionname~\ref{sec:results}, including a detailed discussion of
potential systematics in
\sectionname~\ref{sec:systematics}. The influence of
non-axisymmetric streaming motions on our data is assessed in
\sectionname~\ref{sec:discuss-nonaxi}. We compare our new measurement
of $V_c(R)$ with previous determinations in
\sectionname~\ref{sec:discuss-compare}. In
\sectionname~\ref{sec:discuss-solar}, we discuss the implications of
our measurement of the Sun's offset from circular motion and, in
\sectionname~\ref{sec:discuss-mass}, we estimate the mass of the Milky
Way implied by our data, as well as other recent data. We conclude in
\sectionname~\ref{sec:conclusion}. \appendixname~\ref{sec:appdist}
describes how photometric distances for the stars in our sample are
estimated. In \appendixname~\ref{sec:appmock} we discuss extensive
mock-data tests used to test our methodology and to determine the
data's sensitivity to $V_c(R)$. In what follows we sometimes refer to
the circular velocity at the Sun's location $V_c(R_0)$ by the
shorthand notation $V_c$.

\section{\apogee\ Data}\label{sec:data}

The \sdssiii/\apogee\ is a near-infrared (NIR; $H$-band; 1.51 to 1.70
$\mu$m), high-resolution (R $\approx$ 22,500), spectroscopic survey,
targeting primarily red giants in all Galactic environments, with
emphasis on the disk and the bulge. The \apogee\ instrument
(\citealt{Wilson10a}, J.~Wilson \etal\ 2012, in preparation) consists
of a spectrograph with 300 $2\arcsec$ fibers that reaches a
signal-to-noise ratio of $100$ per pixel (at about Nyquist sampling)
at $H \leq 12.2$ in three one-hour visits during bright time on the
2.5-meter Sloan telescope, at the Apache Point Observatory in Sunspot,
NM \citep{Gunn06a}. A detailed account of the target selection and
data reduction pipeline is given in G.~Zasowski \etal\ 2012 (in
preparation) and D.~Nidever
\etal\ 2012 (in preparation), respectively. Here we summarize the, for
our purposes, most important aspects of the target selection and data
reduction.

\begin{figure*}
\includegraphics[width=\textwidth,clip=]{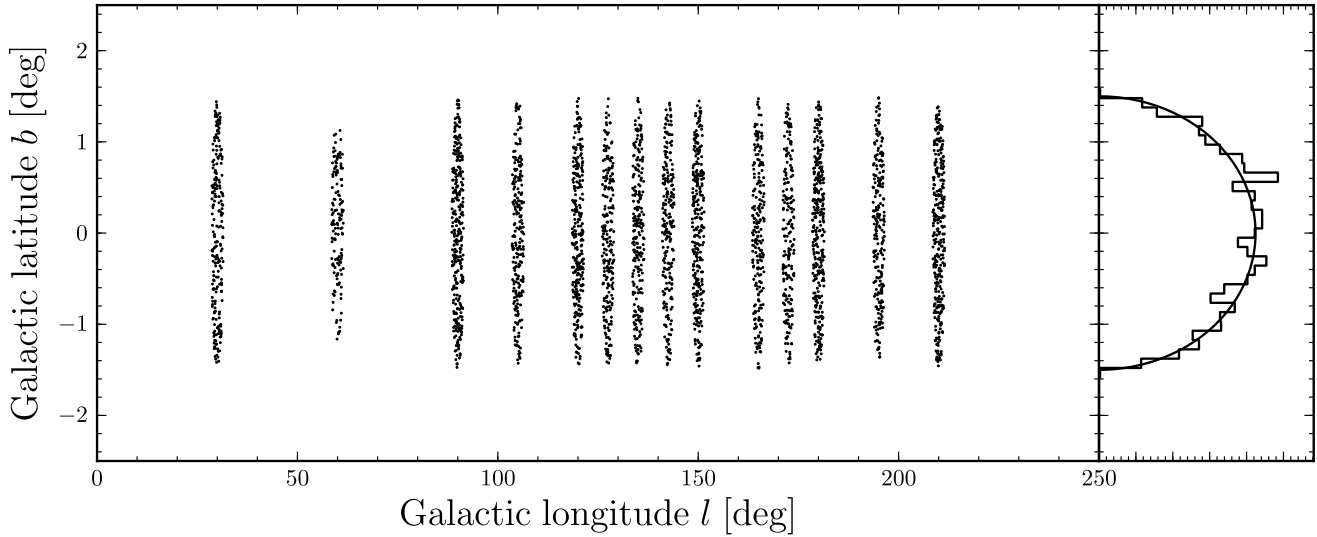}
\caption{Distribution of the data in $(l,b)$. The smooth curve in the
  right panel is a uniform distribution in ($l,b$) over each
  field.}\label{fig:data_lb}
\end{figure*}

Our analysis is based on data from \apogee's first year of regular
survey operations (09-11-2011 to 05-07-2012). We use data from fields
centered on $b = 0^\circ$; \apogee\ fields have a radius of
1.5$^\circ$. By only selecting fields with $30^\circ \leq l \leq
330^\circ$ we avoid the bulge region. We resolve multiple visits to
the same field by choosing the highest signal-to-noise ratio
measurement of the Doppler shift, rather than combining the multiple
epochs, because the typical uncertainty in the line-of-sight velocity
is well below 1 km s$^{-1}$. Only primary survey targets are selected,
excluding targets flagged as possible cluster members and stars
observed as part of any of various special programs. We exclude stars
with $(J-\ks)_0 > 1.1$, as these are problematic for the isochrone
model we adopt to marginalize over their distances
(see \appendixname~\ref{sec:appdist}). The resulting sample
has \ndata\ stars in \nfields\ different fields with $30^\circ
\leq l \leq 210^\circ$. The distribution of the data in
extinction-corrected color and magnitude is shown in
\figurename~\ref{fig:data_h_jk} and the distribution of the stars on the
sky is shown in \figurename~\ref{fig:data_lb}. The properties of the
sample in the various fields are given
in \tablename~\ref{table:fields}.

Spectroscopic targets are selected from the \twomass\ point-source
catalog \citep{Skrutskie06a}, with the following quality restrictions
applied: photometric uncertainties less than 0.1 mag in $J$, $H$, and
\ks; quality flag `A' or `B' in $JH\ks$; nearest neighbor more than $6\arcsec$ away; confusion flag `0' for $JH\ks$; galaxy contamination flag
`0'; read flag `1' or `2' for $JH\ks$; extkey equal to $-1$. The
extinction corrections for targets in the Galactic mid-plane use
mid-IR photometry from either \wise\ \citep{Wright10a} or Spitzer-IRAC
GLIMPSE-I
\citep{Churchwell09a}; therefore, both mid-IR detections and mid-IR
photometric uncertainties less than 0.1 mag are required\footnote{The
  first version of the \apogee\ target selection did not consistently
  insist on the availability of extinction corrections, such that a
  small number of mid-plane targets do not have extinction
  estimates. We have removed 17 stars without extinction estimates
  from the sample.}. \apogee's magnitude range---$7 \leq H <
  13.8$---is within the completeness limits for both the IRAC
  and \wise\ surveys. Photometry for all targets is
  extinction-corrected using the Rayleigh Jeans Color Excess method
\citep[RJCE;][]{Majewski11a}, which provides extinction values $A_K$ with
typical random uncertainties $\lesssim 0.05$ mag for individual stars
using a combination of near- and mid-IR photometry. Variations in the
adopted extinction law among different lines of sight can lead to
differences of up to 7\% \citep{Zasowski09a}.

\begin{deluxetable*}{rrrrrr}[bt]
\tablecaption{}
\tablecolumns{6}
\tablewidth{0pt}
\tabletypesize{\footnotesize}
\tablecaption{Properties of the Sample}
\tablehead{\colhead{Field location} & \colhead{Stars} & \colhead{$H < 12.2$} & \colhead{$12.2 \leq H < 12.8$} & \colhead{$12.8 \leq H < 13.8$} & \colhead{median $A_K$}\\
\colhead{$l$ [degrees]} & \colhead{} & \colhead{} & \colhead{} & \colhead{} & \colhead{}}
\startdata

$30^\circ$ & 230& 108& 33& 89& 0.8 \\
$60^\circ$ & 141& 50& 33& 58& 0.6 \\
$90^\circ$ & 310& 178& 46& 86& 0.4 \\
$105^\circ$ & 222& 222& 0& 0 & 0.2 \\
$120^\circ$ & 289& 157& 84& 48& 0.3 \\
$127^\circ$ & 228& 228& 0& 0 & 0.3 \\
$135^\circ$ & 229& 229& 0& 0 & 0.3 \\
$142^\circ$ & 227& 227& 0& 0 & 0.6 \\
$150^\circ$ & 279& 150& 87& 42& 0.3 \\
$165^\circ$ & 225& 225& 0& 0 & 0.1 \\
$172^\circ$ & 199& 199& 0& 0 & 0.3 \\
$180^\circ$ & 314& 174& 90& 50& 0.2 \\
$195^\circ$ & 227& 227& 0& 0 & 0.2 \\
$210^\circ$ & 315& 173& 91& 51& 0.2

\enddata
\label{table:fields}
\end{deluxetable*}

The color range that we select, $0.5 \leq (J-\ks)_0 \leq 1.1$,
includes the red clump (($J-\ks)_0 = 0.5$ to 0.7) and the red giant
branch, for all metallicities. Dwarf contamination is most severe at
($J-\ks)_0 < 0.8$, with most redder dwarfs being faint M and brown
dwarfs. Most \apogee\ observations consist of three or more individual
``visits'', with only three visits for stars with 7 $\leq H < 12.2$,
and six visits for $12.2 \leq H < 12.8$. About 10\% of our sample
consists of even more visits for stars with $12.8 \leq H <
13.8$. The \apogee\ sampling is random in color, and a combination of
random and systematic in apparent magnitude (selecting every $N$-th
star in a magnitude-ordered list for each field); details of the
selection function are unimportant for our purposes and it suffices to
note that the spectroscopic sample is a representative sample of the
underlying (non-extinction-corrected) magnitude distribution. At $H <
11$, dwarf contamination in the \apogee\ fields is expected to be less
than 10\% based on population-synthesis models
\citep{Girardi05a} in the \apogee\ fields (L.~Girardi, 2012, in
preparation). Models predict that this contamination increases to
about 20\% at $11 < H < 12$. We explicitly include dwarf contamination
as a free parameter in the analysis below.

Line-of-sight velocities are determined for each individual visit by
cross-correlating against a set of $\approx$ 100 synthetic template
spectra that sparsely cover the stellar-parameter range $3,500$ K $<
T_{\mathrm{eff}} < 25,000$ K in effective temperature,
$T_{\mathrm{eff}}$, $-2 < \feh < 0.3$ in metallicity, \feh, and $2 <
\logg < 5$ in surface gravity, \logg\ (see D.~Nidever \etal\ 2012, in
preparation, for details on the exact procedure). The distribution of
rms scatter in the measured line-of-sight velocity for stars with
multiple observations peaks around $0.1\kms$. Tests of field-to-field
variations indicate that the zero point of the velocity scale is
stable at the $0.05\kms$ level. A preliminary comparison between the
\apogee-measured and literature line-of-sight velocity of 53 stars in M3, M13, and
M15 shows that the \apogee\ zeropoint accuracy is
$\approx0.26\pm0.20\kms$ (see D.~Nidever \etal, 2012, in preparation
for full details). In what follows we ignore the uncertainties on the
line-of-sight velocities entirely (that is, we assume that they are
zero), because the uncertainties are much smaller than any velocity
difference that we could hope to measure with our sample of
$\approx3,000$ stars having a typical dispersion of $\approx35\kms$.

Repeated visits to the same star will eventually allow binary stars to
be flagged as such based on the variability of their line-of-sight
velocities. However, currently multiple visits are not available for
all \nfields\ \apogee\ fields that we employ here, and removing fields
without multiple visits would significantly reduce the longitudinal
coverage of our sample. Using the stars in our sample that have
multiple visits, we find that $\sim\!4\,\%$ of stars show velocity
variability at the $10$ to $70\,\kms$ level---the level at which they
could be confused with the dispersion of a disk population. Binary
contamination would spuriously increase the inferred velocity
dispersion and therefore likely decrease the inferred circular
velocity, due to the dispersion-dependence of the asymmetric-drift
correction used below. However, as discussed below, the asymmetric
drift for our sample is $\lesssim 6\,\kms$ at $R < 10\,\kpc$, such
that the effect of binary contamination on the velocity dispersion is
$\ll 1\,\kms$. Therefore, we do not remove binaries from our sample, but
we discuss the results obtained when removing them
in \sectionname~\ref{sec:systematics}.

\begin{figure}[htbp]
\includegraphics[width=0.5\textwidth]{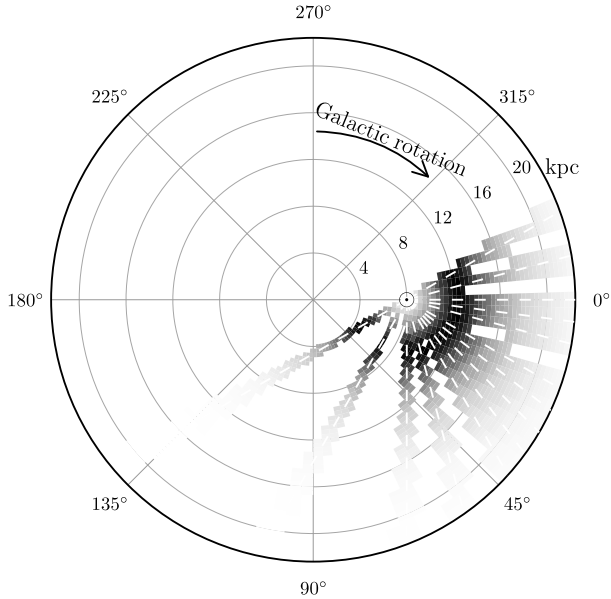}
\caption{Sum of all photometric distance distributions
  $p(d|l,b,(J-\ks)_0,H_0,\feh,\df,\iso)$ for stars in our sample,
  viewed from the north Galactic pole, with the Sun at azimuth
  $0^\circ$. These photometric-distance distributions are used
  in \eqnname~(\ref{eq:pvlos}) to marginalize over the distance to
  each individual star. We emphasize that this is \emph{not} the
  underlying distance distribution of our sample, as individual
  distances are poorly constrained for the giants in our
  sample. The \nfields\ fields that we use are indicated by dashed
  white lines.}\label{fig:distanceprior}
\end{figure}

Stellar parameters and elemental abundances are determined by the
\apogee\ Stellar Parameters and Chemical Abundances Pipeline
(ASPCAP; A.~E.~Garc\'{\i}a~P\'{e}rez \etal, 2013, in
preparation). After determining the spectral type, each star's
microturbulence, $T_{\mathrm{eff}}$, \feh, \logg,
$[\alpha/\mathrm{Fe}]$, $[\mathrm{C/Fe}]$, and $[\mathrm{N/Fe}]$ are
determined through $\chi^2$ minimization of the difference between the
observed and synthetic spectra derived from ATLAS9 model-atmosphere
grids (\citealt{Kurucz79a} and more recent updates). We only use \feh\
in our analysis below, and in particular, we do \emph{not} make use
of \logg\ for the purpose of selecting giants, because the measurement
of \logg\ has not been finalized yet within ASPCAP. Preliminary
comparisons with standard stars and globular clusters indicate that
the metallicities are currently accurate, to a precision of
$\sim\!0.1\,\dex$ at $\feh > -1$. All but $21$ of the disk stars in
our sample described below have higher metallicities than this, and
all but $160$ have $\feh > -0.5$.

Because many stars in our sample are on the red giant branch, and
hence do not follow narrow magnitude-color relations like dwarfs and
red clump stars, estimating precise distances to these stars is
difficult. In what follows, we avoid estimating distances to the
individual stars in our sample by marginalizing a kinematical model
over the distance to the star. This marginalization is performed by
integrating over the full photometric-distance probability
distribution function (PDF) for each individual star, obtained from
models for the stellar isochrones and initial mass function (IMF) of
the giant stars in the sample, as well as a prior that the stars are
in an exponential disk with a scale length of 3
kpc \citep{Bovy12a}. This technique is discussed in detail
in \appendixname~\ref{sec:appdist}. For dwarf stars, we assume that
they are close enough that their distances are irrelevant for their
kinematics. As such, we cannot show the spatial distribution of the
stars in our sample. As an alternative, we show in
\figurename~\ref{fig:distanceprior} the sum of all of the photometric
distance distributions for stars in our sample (assuming that they are
all giants).

\section{Methodology}\label{sec:method}

\subsection{General Considerations}\label{sec:general}

Our approach to determining the Milky Way's rotation curve from the
\apogee\ data is to fit a kinematical, axisymmetric model to the
observed line-of-sight velocities. In its most basic form, such a
kinematical model consists of (a) a model for the rotation curve, (b) a
model for the distribution of peculiar velocities with respect to
circular motion as a function of Galactocentric radius, and (c) a set
of parameters describing the transformation of positions and
velocities from the heliocentric to the Galactocentric frame. These
latter parameters are the distance \Ro\ from the Sun to the Galactic
center, and the Sun's velocity with respect to the (dynamical) center
of the Galaxy. We will ignore the small projection effects that arise
at non-zero Galactic latitude and the vertical dependence of $V_c$,
since these are all at the sub-km s$^{-1}$ level for $|b| < 1.5^\circ$
and distances $\lesssim 10\,\kpc$ \citep[\eg,][]{Bovy12d}. Therefore,
we are only concerned with motions in the plane of the Galactic disk,
and we ignore vertical motions.

The kinematical model provides the probability distribution of
line-of-sight velocities as a function of position (Galactocentric
radius $R$ and azimuth $\phi$, or distance $d$ and Galactic longitude
$l$), after marginalizing over the component of the velocity
tangential to the line of sight. This requires knowing the distance,
which is only weakly constrained for giants (see
\appendixname~\ref{sec:appdist}). Therefore, we additionally marginalize the
kinematical model over the distance to each star, to obtain the
probability of the observed line-of-sight velocity of a star, given
its position $(l,b)$, photometry ($J_0,H_0,\kso$), the iron
abundance \feh\ (taken as representative of the overall metallicity
of the star), the kinematical model (represented by the
circular-velocity curve $\vc(R)$ and the distribution of peculiar
velocities, \df), and the coordinate-transformation parameters \Ro\
and $\vsun = (V_{R,\odot},V_{\phi,\odot})$ (where we label this
parameter as ``gal'' to emphasize that this is the \emph{full}
velocity with respect to the center of the Galaxy, not just the Sun's
motion with respect to the local circular velocity):
\begin{equation}\label{eq:pvlos}
\begin{split}
& p(\vlos  |l,b, (J-\ks)_0,H_0,\feh,\vc(R),\Ro,V_{R,\odot},V_{\phi,\odot},\df,\iso)  \\
&= \sum_{\mathrm{dwarf/giant}} P(\mathrm{dwarf/giant})\\
& \ \int \dd d\, p(\vlos | d,l,b,\vc(R),\Ro,V_{R,\odot},V_{\phi,\odot},\df)\\
& \ \  \times
 p(d|l,b,(J-\ks)_0,H_0,\feh,\df,\iso,\mathrm{dwarf/giant}).
\end{split}
\end{equation}
The first factor within the integral is produced by marginalizing the
kinematical model over the velocity component tangential to the line of
sight at a given position. The second factor is given by the
photometric-distance PDF, discussed in
\appendixname~\ref{sec:appdist}; we include `iso' in the prior
information to emphasize that we use an isochrone and IMF model to
obtain the photometric-distance PDFs. We also marginalize over whether
a star is a giant or a dwarf star, which changes the
photometric-distance PDF; the dwarf contamination probability
$P(\mathrm{dwarf})$ is a single free parameter in our
model. \Eqnname~(\ref{eq:pvlos}) is the likelihood for a single star,
and the total likelihood for the sample is calculated by multiplying
together the individual likelihoods for all the \ndata\ data
points. This total likelihood is what we optimize to fit models to the
data.

We can also calculate the distribution of line-of-sight velocities for
individual fields, which we will use later to show comparisons between
the best-fit model and the data. For the distribution of \vlos\ of a
field centered at (\platel,\plateb) we obtain
\begin{equation}\label{eq:pvlosplate}
\begin{split}
& p(\vlos |\platel,\plateb,\vc(R),\Ro,V_{R,\odot},V_{\phi,\odot},\df,\iso)  \\
& = 
\int \dd l\,\dd b \int \dd (J-\ks)_0\,\dd H_0\, \dd \feh\\
& \qquad  p((J-\ks)_0,H_0,\feh,l,b|\platel,\plateb,\df,\iso)\\
& \ \times 
\int \dd d\,p(\vlos|l,b,(J-\ks)_0,H_0,\feh,\vc(R),\Ro,\\
& \qquad \qquad \qquad \ \ \ V_{R,\odot},V_{\phi,\odot},\df,\iso)\,,
\end{split}
\end{equation}
where $p(J-\ks,H,\feh,l,b|\platel,\plateb,\df,\iso)$ is taken directly
from the field's data (that is, we sum over the data
($l$,$b$,$(J-\ks)_0$,$H_0$,\feh)).

\subsection{Kinematical Model}\label{sec:kinematicmodel}

We now describe in detail the full model that is fit to the data. Our
fiducial model for the distribution of velocities in the
Galactocentric rest frame is that it is a single biaxial Gaussian,
with a mean radial velocity of zero because of the assumption of
axisymmetry, and a mean rotational velocity given by the local
circular velocity $\vc(R)$, adjusted for the local asymmetric drift
$\va(R;\sigma_R,\hR,\hs)$, which is a function of the velocity
dispersion, $\sigma_R$, and the radial and radial-velocity dispersion
scale lengths of the population (\hR\ and \hs, respectively, see
below). The only free parameters of the distribution of peculiar
velocities are therefore the radial and rotational velocity
dispersions, as the mixed radial--azimuthal moment vanishes, again
because of the assumption of axisymmetry. The mixed moment, or
equivalently the vertex deviation, in the solar neighborhood does not
vanish for the warm disk population \citep{Dehnen98a}, but this seems
likely to result from the presence of moving groups, and not from the
smooth underlying distribution having a strong non-zero vertex
deviation \citep{Bovy09b}.

Assuming that the distribution of velocities in the Galactocentric
cylindrical frame is Gaussian allows us to analytically marginalize
the velocity distribution over the velocity component tangential to
the line of sight. The resulting one-dimensional distribution is
itself Gaussian, with mean
$(\vc(R)-\va(R;\sigma,\hR,\hs))\sin(\phi+l)$ and variance
$\sigma_R^2\left(1+\sin^2(\phi+l)\,(X^2-1)\right)$, where $X^2$ is the
ratio of the rotational and radial velocity variances
$X^2\equiv \sigma_\phi^2/\sigma_R^2$. The Galactocentric line-of-sight
velocity for a star is calculated from its heliocentric line-of-sight
velocity
\begin{equation}\label{eq:meanvlos}
  V_{\mathrm{los}}^{\mathrm{gal}} = V_{\mathrm{los}}^{\mathrm{helio}}  - V_{R,\odot}\,\cos l
  + \Omega_{\odot} \,\Ro\,\sin l\,,
\end{equation}
where we have written the Sun's velocity in terms of its angular
velocity $\Omega_\odot$. This angular velocity is expected to be close
to the observed proper motion of Sgr A$^*$ in the plane of the Galaxy
(30.24 km s$^{-1}$ kpc$^{-1}$; \citealt{Reid04a}), but we leave it as
a free parameter, in addition to $V_{R,\odot}$.

\begin{figure}[htbp]
\includegraphics[width=0.5\textwidth]{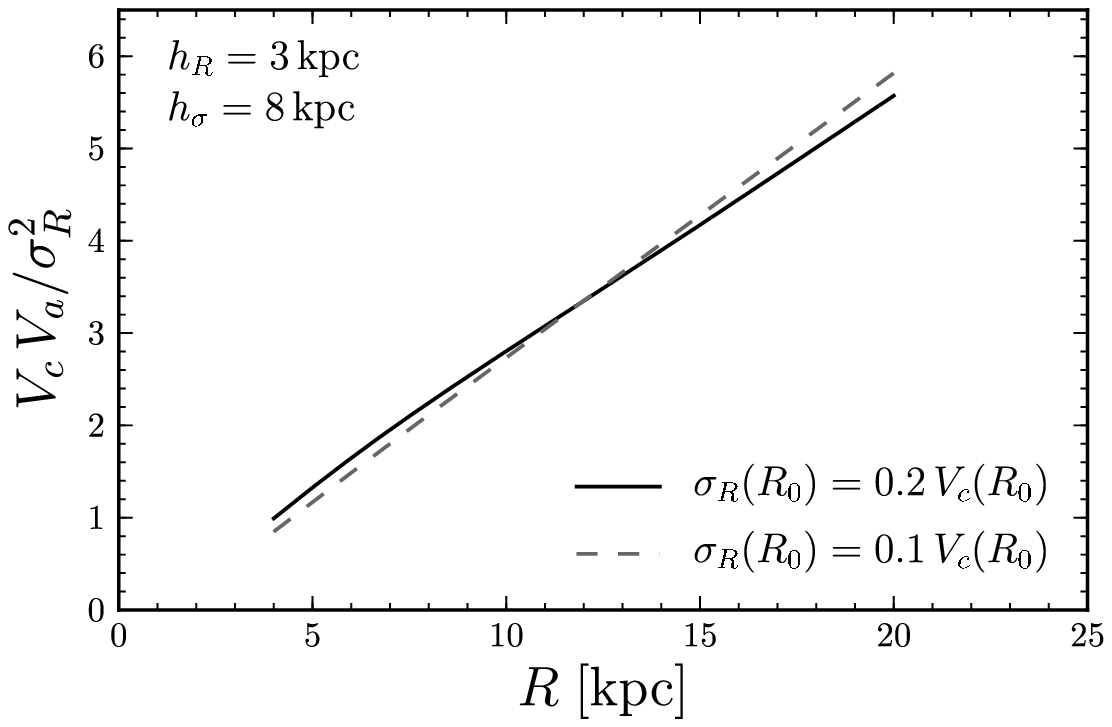}\\
\includegraphics[width=0.5\textwidth]{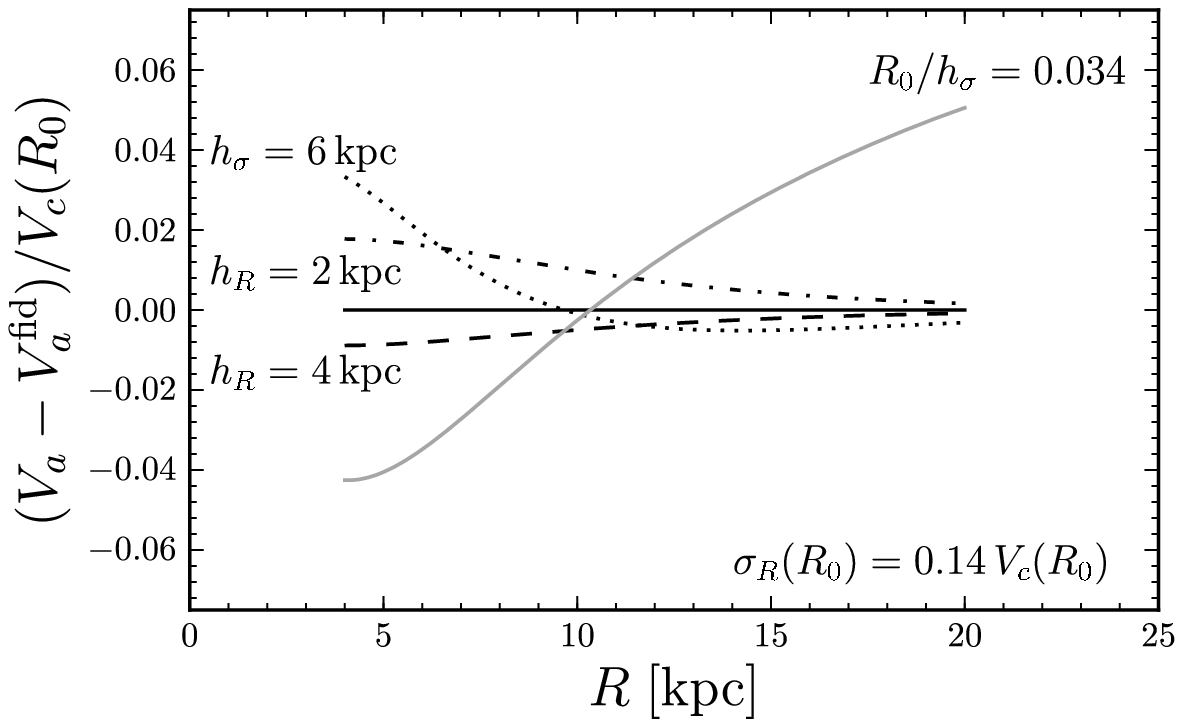}\\
\includegraphics[width=0.5\textwidth]{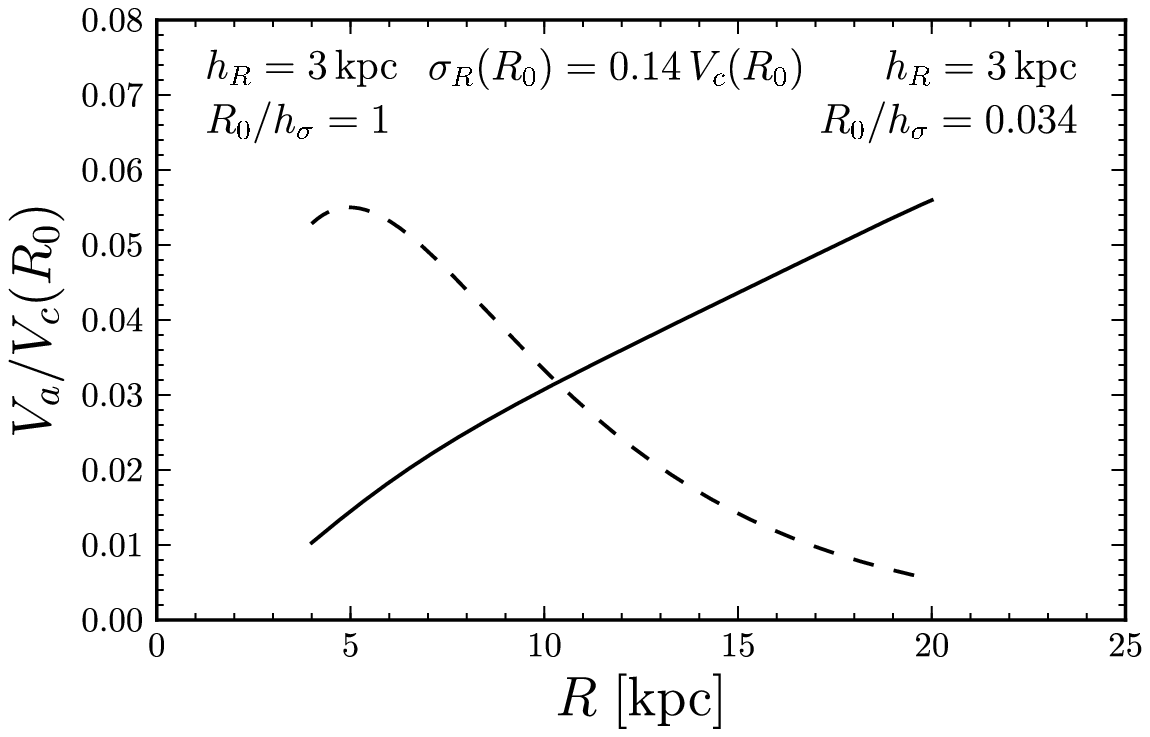}
\caption{Asymmetric-drift model used in this paper. The top panel
  shows the asymmetric drift $V_a$ in a Dehnen DF
  (\eqnname~[\ref{eq:fdehnen}]), expressed dimensionlessly as
  $V_c\,V_a/\sigma_R^2$. We use the $\sigma_R(\Ro) = 0.2 \vc(\Ro)$
  curve and show the $\sigma_R(\Ro) = 0.1 \vc(\Ro)$ for
  comparison. The middle panel shows the difference in asymmetric
  drift for different values of \hr\ and \hs, using the solid curve
  from the top panel as the fiducial model.  The bottom panel shows
  the actual asymmetric-drift correction as a fraction of the circular
  velocity used for the best-fit model below (\emph{solid line}), and
  also shows the correction for $R_0/h_\sigma = 1$ (\emph{dashed
  line}). All curves in the middle and bottom panels have
  $\sigma_R(\Ro) = 0.14 \vc(\Ro)$ (the best-fit value for the \apogee\
  sample, see below).}\label{fig:va}
\end{figure}

The asymmetric drift in the plane of the Galaxy of a population of
stars is given by \citep[\eg,][]{binneytremaine}
\begin{equation}\label{eq:va1}
\frac{\vc(R)\va(R)}{\sigma^2_R(R)} = \frac{1}{2}\left[X^2-1-\frac{\partial \ln\left(\dens\sigma^2_R\right)}{\partial \ln R}\right]\,,
\end{equation}
assuming that the radial--vertical velocity moment $\sigma_{RZ} = 0$
in the Galactic plane, for symmetry reasons. For an exponential-disk
tracer population (stellar tracer density $\dens \propto e^{-R/h_R}$)
with an exponentially declining radial-velocity dispersion
($\sigma_R(R) \propto e^{-R/h_\sigma}$), this formula becomes
\begin{equation}\label{eq:va}
\frac{\vc(R)\va(R)}{\sigma^2_R(R)} = \frac{1}{2}\left[X^2-1+R\left(\frac{1}{\hR}+\frac{2}{\hs}\right)\right]\,.
\end{equation}
The main unknown in this relation is $X^2$ and its dependence on
$R$. In the fit below, we assume a constant $X^2$ with $R$ for
calculating the model's Gaussian line-of-sight velocity dispersion
(see above), but to model the asymmetric drift more realistically, we
use an $X^2(R)$ coming from an axisymmetric equilibrium model for the
distribution function $f(E,L)$ in a disk with a constant circular
velocity with radius. For this distribution function, we use a Dehnen
distribution function \citep{dehnen99b} given by
\begin{equation}\label{eq:fdehnen}
\fdehnen(E,L) \propto \frac{\dens(\rE)}{\sigmaR^2(\rE)} \, \exp\left[ \frac{\Omega(\rE)\left[L-\Lc(E)\right]}{\sigmaR^2(\rE)}\right]\,,
\end{equation}
where \rE, \Lc, and $\Omega(\rE)$ are the radius, angular momentum,
and angular velocity, respectively, of the circular orbit with energy
$E$. Using the procedure given in \sectionname~3.2 of
\citet{dehnen99b}, we choose the $\dens(R)$ and $\sigmaR(R)$ functions
such that they reproduce a disk with exponential surface density and
velocity-dispersion profiles to an accuracy of a fraction of a percent
at all radii. 

The asymmetric drift for the Dehnen distribution function (DF) for a
population with $\sigma_R(\Ro) = 0.2\,\vc(\Ro) \approx 44$ km s$^{-1}$
is shown in \figurename~\ref{fig:va} for $\hR= 3\,\kpc$, $\hs =
8\,\kpc$. The same is shown for a population with $\sigma_R(\Ro) =
0.1\,\vc(\Ro)
\approx 22$ km s$^{-1}$, and the difference, normalized for the
difference in $\sigma_R$ between them, is small. For different values
of \hR\ and \hs\ we correct this function using
\eqnname~(\ref{eq:va}). As different values for \hR\ and \hs\ lead to
a different $X^2(R)$, this approach is slightly incorrect, but tests
show that the difference in \va\ is $\lesssim 5\%$ for reasonable
values of
\hR\ and \hs. As the asymmetric drift is typically $\lesssim 20$ km
s$^{-1}$ for our sample, this leads to changes $\lesssim 1$ km
s$^{-1}$, which we can ignore. Similarly, differences in $X^2$ for our
sample from this model are typically $\lesssim 0.2$, also leading to
changes $\lesssim 1$ km s$^{-1}$, such that our assumption of a
constant $X^2(R)$ does not strongly bias the value of the inferred
circular velocity. In the exponential-disk model, the asymmetric drift
for our sample is $\lesssim 25$ km s$^{-1}$ at all radii, even in the
innermost disk \citep{Freeman87a}.

The middle panel of \figurename~\ref{fig:va}---now for $\sigma_R(R_0)
= 0.14\,V_c$, the best-fit value below---shows the difference in
asymmetric drift for different values of \hR\ and \hs; the main
difference is at small $R$. This figure shows that the dependence of
the asymmetric drift correction on \hR\ and \hs\ at $R_0$ is mostly
$\lesssim 0.02\,\vc$---as it is unlikely that the scale length is as
small as 2 kpc for this sample of stars \citep{Bovy12a}, or that the
radial-velocity dispersion scale length is much shorter than 8
kpc---which is the same as the statistical uncertainty on the circular
velocity that we infer below. Therefore, the systematic uncertainty
due to the scale length of the population of stars is $\lesssim 4$ km
s$^{-1}$ at $R_0$, and likely $\lesssim 2$ km s$^{-1}$. We could
directly measure the scale length of our sample, but as this requires
a detailed understanding of the dust distribution in the plane of the
Galaxy, and as it does not contribute greatly to the error budget of
our measurement, we do not attempt it here.

Our best-fit model below has a radial velocity dispersion that is
approximately flat over the range in $R$ considered here. In this
case, the asymmetric-drift correction \emph{increases} with $R$, as is
clear from \eqnname~(\ref{eq:va}), and the difference between this
dispersion profile and that considered above is shown in the middle
panel of \figurename~\ref{fig:va}. The bottom panel of
\figurename~\ref{fig:va} shows the actual asymmetric-drift correction
applied in our best-fit model below. It is clear that this correction
is small ($\lesssim 6\,\kms$ at $R < 10\,\kpc$); hence it does not
drive our inferred value for $V_c$ below. The bottom panel of
\figurename~\ref{fig:va} also illustrates that, if the dispersion scale
length were smaller than that in our best-fit model below, then the
circular velocity curve would drop more steeply than what we infer
below.

In addition to the kinematical model discussed so far, we include an
outlier model that consists of a Gaussian with a width of $\sigma =
100\,\kms$, centered on a mean value that is a free parameter in the
model. This outlier model exists primarily to deal with
data-processing outliers, stars with a color and magnitude that places
them beyond the disk, and close binaries with large velocity
amplitudes (as we do not currently have a sufficient number
of \apogee\ epochs for each star to confidently remove these from the
sample). In all of the fits below, the outlier fraction is $\lesssim
1\,\%$, thus it does not influence our fits.

\subsection{Dwarf Contamination}

The dwarf contribution to the likelihood in \eqnname~(\ref{eq:pvlos})
could in principle be treated in the same way as the giant
contribution, by integrating over the
$p(d|l,b,(J-\ks)_0,H_0,\feh,\df,\iso,\mathrm{dwarf})$ obtained from a
similar isochrone and IMF model as for the giants. However, this is
problematic, because the main sequence is much narrower than the giant
branch, such that this integral is dominated by a narrow range of
distances. As all of the dwarfs in our sample are faint main-sequence
stars, they are expected to all be within a few 100 pc from the Sun,
and thus be relatively unaffected by the radial and azimuthal
gradients in velocity distribution that we model for the giant
stars. Our approach is therefore to replace the integration over
distance with a single evaluation at zero distance from the Sun for
the dwarf part of the likelihood. The dwarf stars are otherwise
modeled using the same distribution of peculiar velocities as the
giant stars, \ie, we assume that they are drawn from a similar
population. The dwarf contamination in our best-fit model below is
$7.5\%$, and is similar for other fits considered below. This is close
to the value expected from stellar-population-synthesis models
(see \sectionname~\ref{sec:data}).

\subsection{Mock Data Tests}

The procedure described above makes non-trivial approximations to an
ideal axisymmetric fit to the data. The most important of these is
that, although we include the asymmetric drift as part of our model,
we do not include the deviations from a Gaussian rotational-velocity
distribution that are expected for a warm population of stars (with
the same physical origin as the asymmetric drift). In addition to this
assumption, we make small approximations to the full asymmetric drift
expression in \eqnname~(\ref{eq:va1}). These simplifications are
expected to have small effects on our analysis: The skew of the
rotational-velocity distribution is only fully visible near the
tangent point where the line-of-sight velocity is equal to the
rotational velocity, but our sample covers a large range of distances
for each line of sight, and 85\% of our sample lies at $90^\circ < l <
270^\circ$, where there is no tangent point. As such, most stars are
drawn from a distribution that is close to Gaussian.

To investigate these simplifications, we conduct a series of mock data
tests. In these tests, we sample line-of-sight velocities from the
best-fit flat-rotation-curve model below, except that we use the full
Dehnen DF of \eqnname~(\ref{eq:fdehnen})---uncorrected, such that it
does not quite reproduce the assumed exponential disk---rather than
the Gaussian approximation used in the fit (we ignore the best-fit
$X^2$ as the Dehnen DF has $X^2$ built-in). We then fit these mock
data using the approximate methodology.

These mock data tests are described in detail in
\appendixname~\ref{sec:appmock}. The results from these tests show
that the approximations made in the analysis do not produce any
significant bias in the fitted parameters. Fits with non-flat rotation
curves to the mock data samples indicate that we can reliably
determine the shape of the rotation curve between $4\,\kpc < R <
14\,\kpc$. The uncertainties in the fitted parameters for the mock
data are approximately the same as those for the real data (see
below), which indicates that the uncertainties in the Galactic
parameters derived below for the real data are correct.

\subsection{Parameter Sensitivity}\label{sec:sensitivity}

We briefly discuss here how our data are sensitive to the parameters
of the model. The mean heliocentric line-of-sight velocity at a
position $(R,\phi) = (d,l)$ is given by (see \eqnname~(\ref{eq:meanvlos}))
\begin{equation}
  \bar{V}_{\mathrm{los}} =
  \bar{V}_\phi\,\sin\left(\phi+l\right)+V_{R,\odot}\,\cos
  l-V_{\phi,\odot}\,\sin l\,,
\end{equation}
where $\bar{V}_\phi = V_c - V_a$. We have treated the position and
velocity of the Sun as free parameters, without imposing any prior on
$R_0$, and without assuming that the solar rotational velocity is
given by the local circular velocity corrected for the (previously
measured) velocity of the Sun with respect to the local standard of
rest (LSR). We can determine the full space motion of the Sun from our
sample, because the longitudinal dependence of the correction for the
solar motion of the line-of-sight velocity of a star at
$(R,\phi)\equiv(d,l)$ is sinusoidal ($\propto V_{\phi,\odot}\,\sin
l$), while the dependence on the rotational velocity at the position
of the star varies differently for the $d \approx 1$ to $10\,\kpc$
sample of
\apogee\ stars ($\propto (V_c-V_a)\,\sin(\phi+l)$). Mock data tests in
\appendixname~\ref{sec:appmock} show that our sample spans a
sufficiently wide range in Galactic longitude and distance to
disentangle these two effects, and measure the solar Galactocentric
motion. Because we know the Sun's rotational frequency from assuming
that the radio source Sgr A$^*$ at the Galactic center is at rest with
respect to the disk, we have an independent check on our inferred
value for the solar motion (but \emph{we do not use the measured
proper motion of Sgr A$^*$ as a prior on the solar motion}).

We can then turn a measurement of $\bar{V}_{\phi}(R)$ into a
measurement of $V_c(R)$ by correcting for the asymmetric drift,
$V_a(R)$, using \eqnname~(\ref{eq:va}). This transformation requires a
measurement, foremost, of the radial-velocity dispersion,
$\sigma_R(R)$. We can measure $\sigma_R$ almost directly using the $l
= 180^\circ$ line of sight. By modeling $\sigma_R(R)$ as an
exponential and assuming a constant $\sigma_\phi^2 / \sigma_R^2$, we
can measure $\sigma_R(R)$ from the line-of-sight-velocity dispersion
at each $l$. The only subtlety here is that, in the first quadrant ($R
< R_0, l < 90^\circ$), we typically sample positions where
$V_{\mathrm{los}}$ is close to $V_{\phi}$, while in the second and
third quadrants ($R > R_0, 90^\circ < l < 270^\circ$),
$V_{\mathrm{los}}$ is mainly composed of $V_{R}$. We can then use the
measurement of $\sigma_R(R)$ to correct for $V_a$ and derive $V_c$.

The fact that our data sample is composed of the intermediate-age to
old disk population is crucial for our ability to measure the full
shape of the rotation curve. Traditional tangent-point analyses of the
HI emission at $l < 90^\circ$, or the measurement of $V_c(R)$ at
$90^\circ < l < 270^\circ$ using the thickness of the HI layer, are
insensitive to solid-body (uniform angular speed) contributions to the
rotation curve, because these do not give rise to line-of-sight
velocities for circular orbits. Formally, the HI velocities are
invariant under the transformation $V_c(R) \rightarrow V_c(R)
+ \Omega\,R$. However, this is not the case for a warm tracer
population, as, for example, the mean rotational velocity transforms
as
\begin{equation}\label{eq:vlostrans}
  \bar{V}_{\mathrm{los}} \rightarrow \bar{V}_{\mathrm{los}} +
  V_a(\Omega_0)\left(1-\frac{1}{1+\Omega/\Omega_0(R)}\right)\,\sin\left(\phi+l\right)\,.
\end{equation}
where $\Omega_0(R) = V_c(R)/R$. Thus, the addition of $\Omega\,R$ to
$V_c(R)$ changes $\bar{V}_{\mathrm{los}}$ by
$\sim\!V_a\,\Omega/\Omega_0$, and $\Omega/\Omega_0\!\sim\!10\%$ gives
rise to a few $\kms$ changes in $\bar{V}_{\mathrm{los}}$, which we can
detect with our data. Equation~(\ref{eq:vlostrans}) clearly shows that
a warmer population, \ie, a population with a larger asymmetric drift,
is \emph{more sensitive} to the local slope of the rotation curve than
a colder population.

\section{Results}\label{sec:results}

\subsection{Basic Models}\label{sec:basic}

\begin{figure*}[htbp]
\includegraphics[width=0.23\textwidth,clip=]{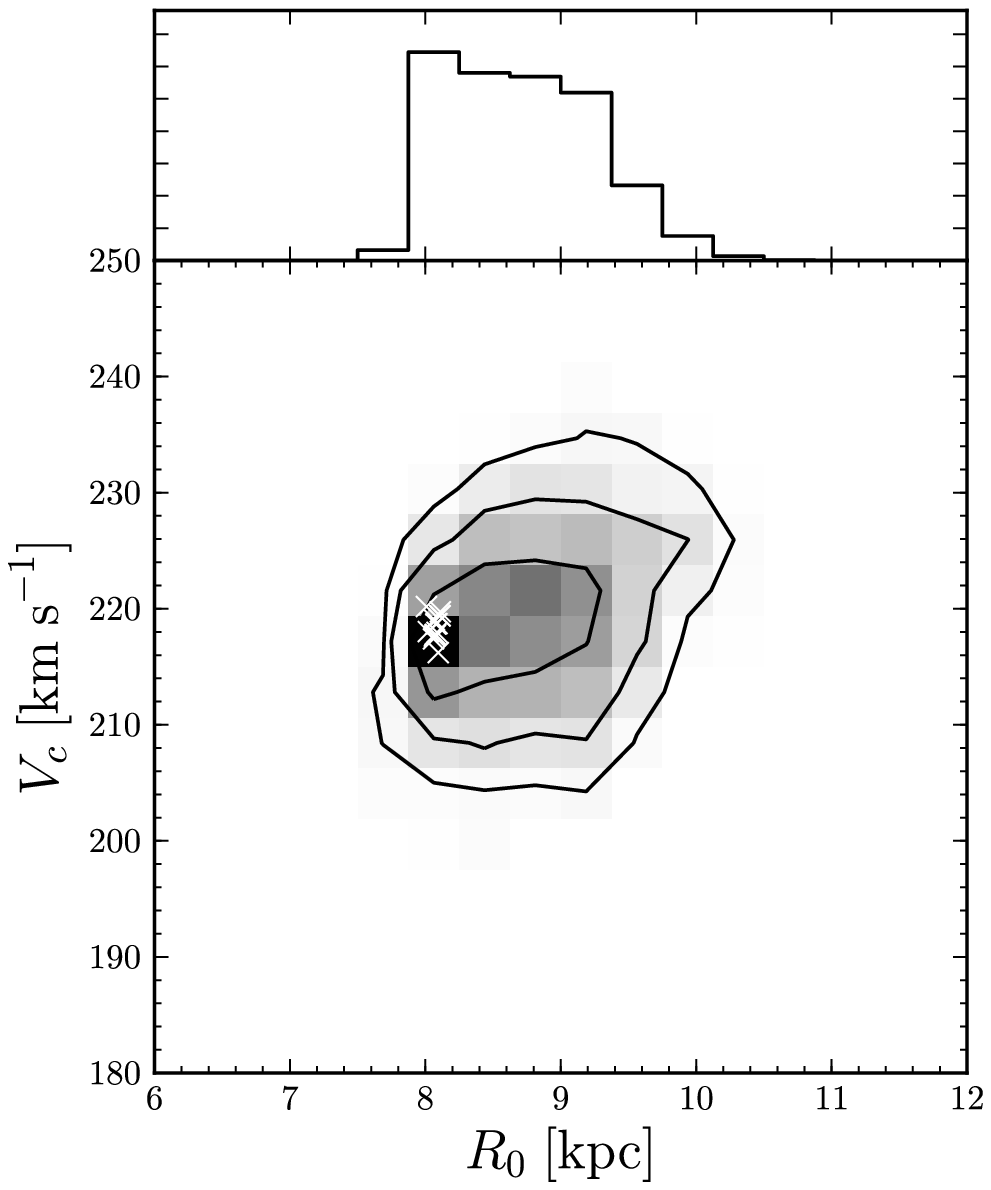}
\includegraphics[width=0.23\textwidth,clip=]{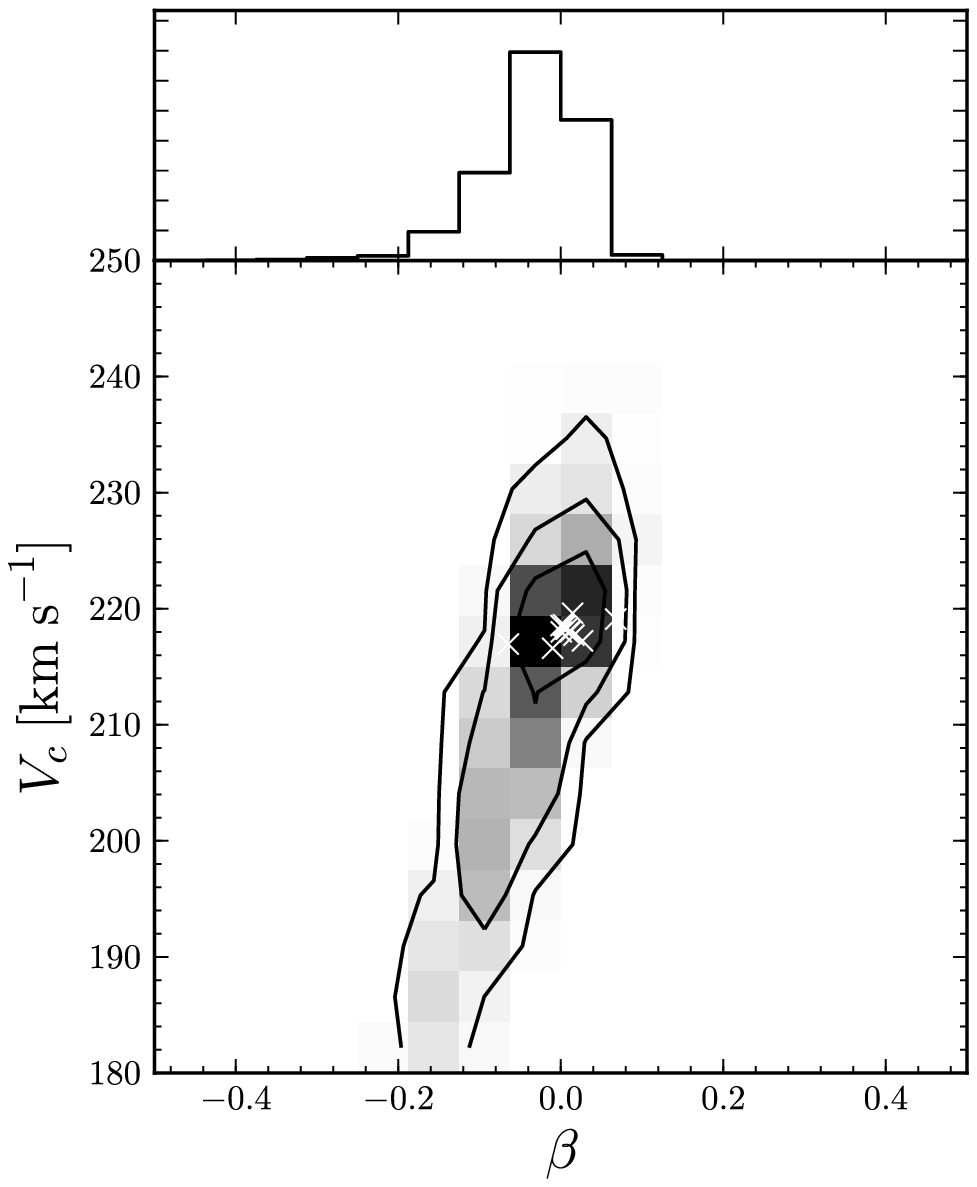}
\includegraphics[width=0.23\textwidth,clip=]{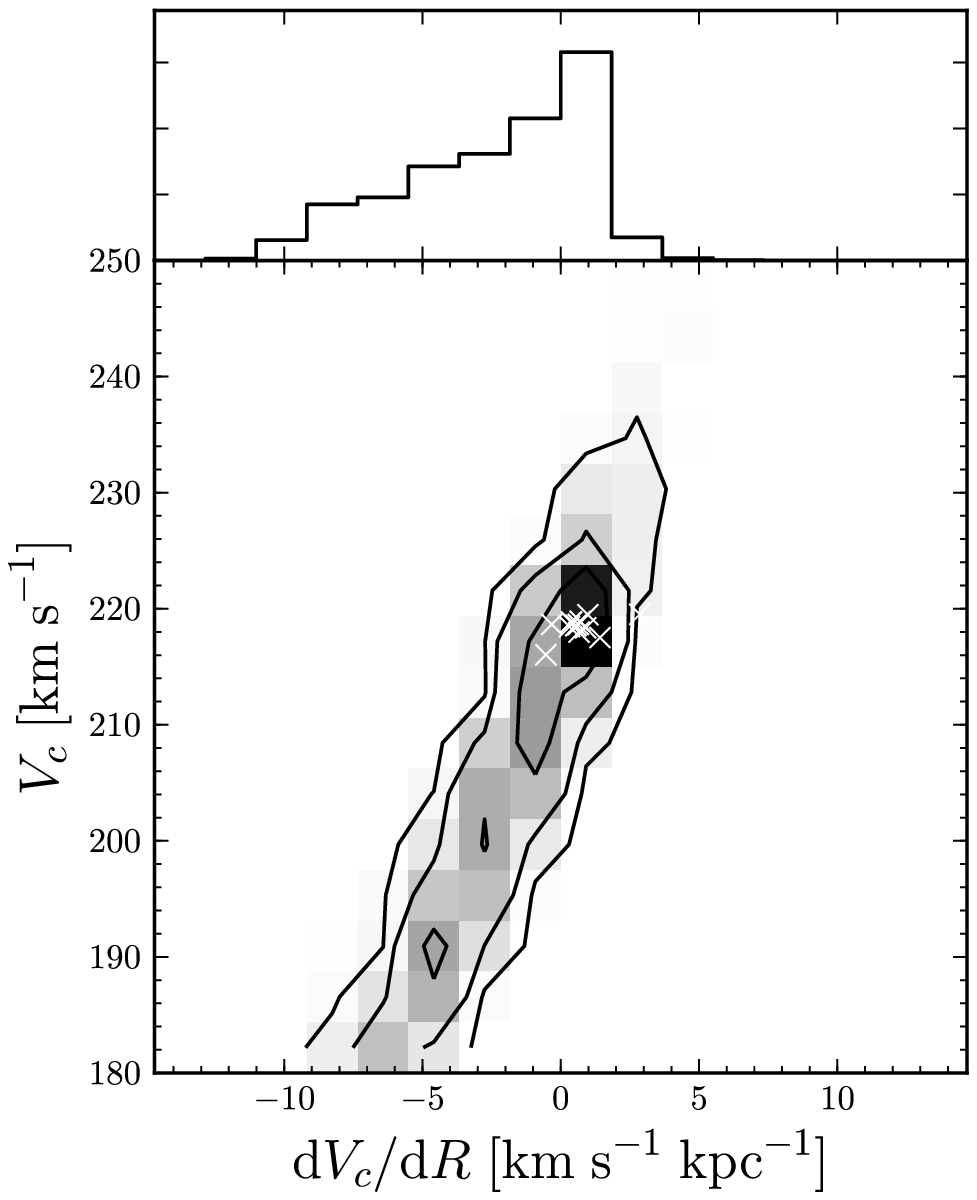}
\includegraphics[width=0.23\textwidth]{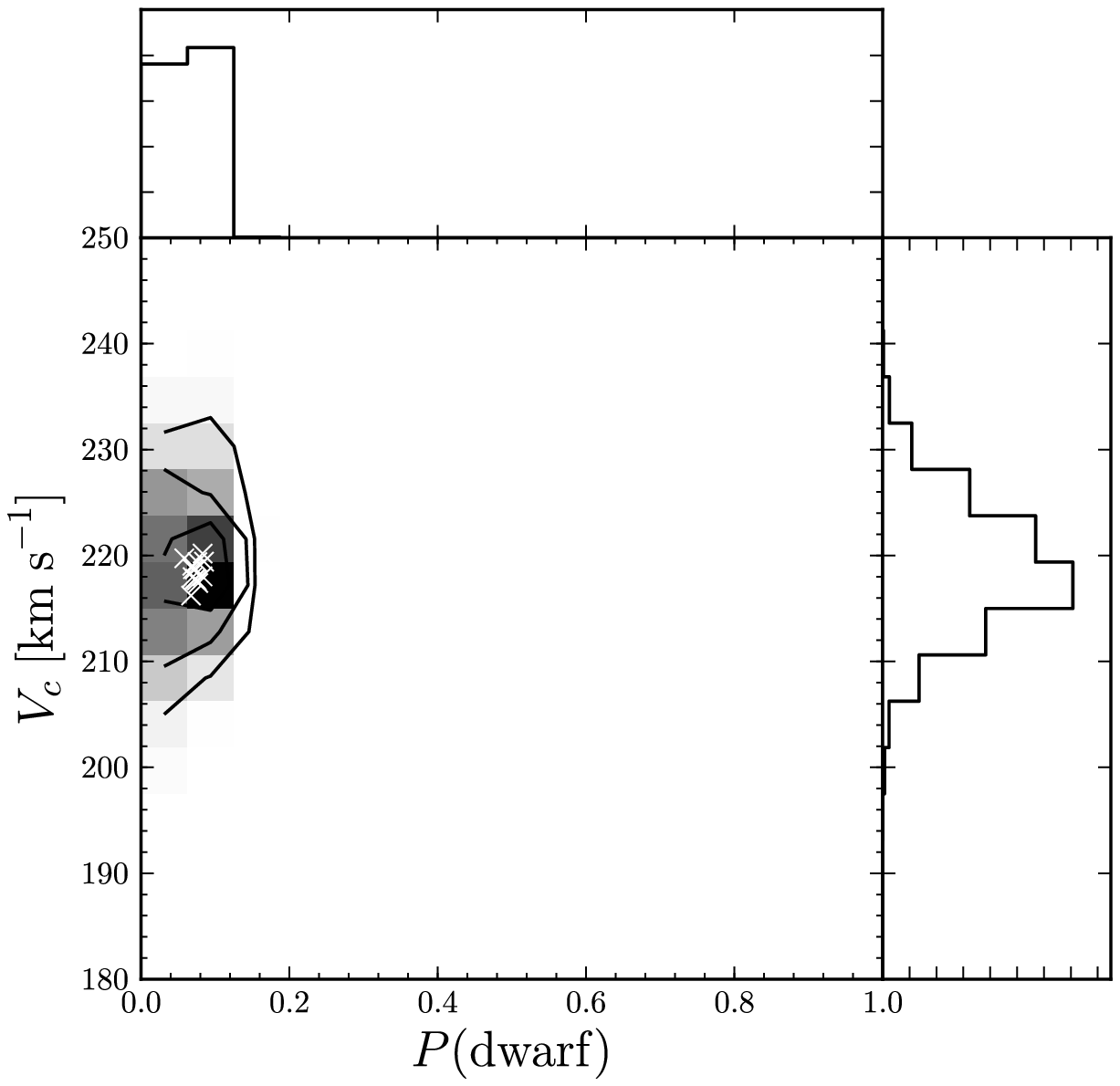}\\
\includegraphics[width=0.23\textwidth,clip=]{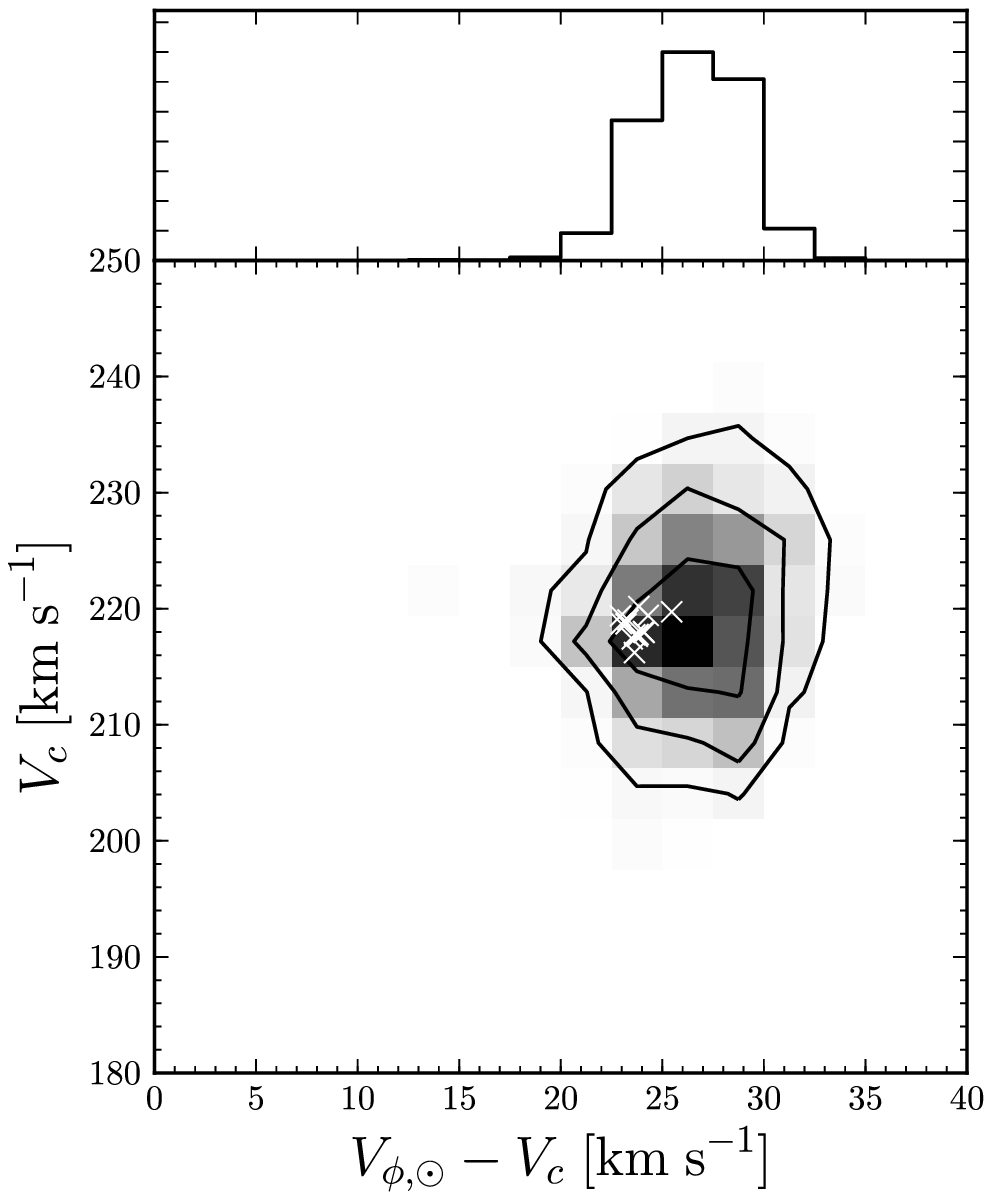}
\includegraphics[width=0.23\textwidth,clip=]{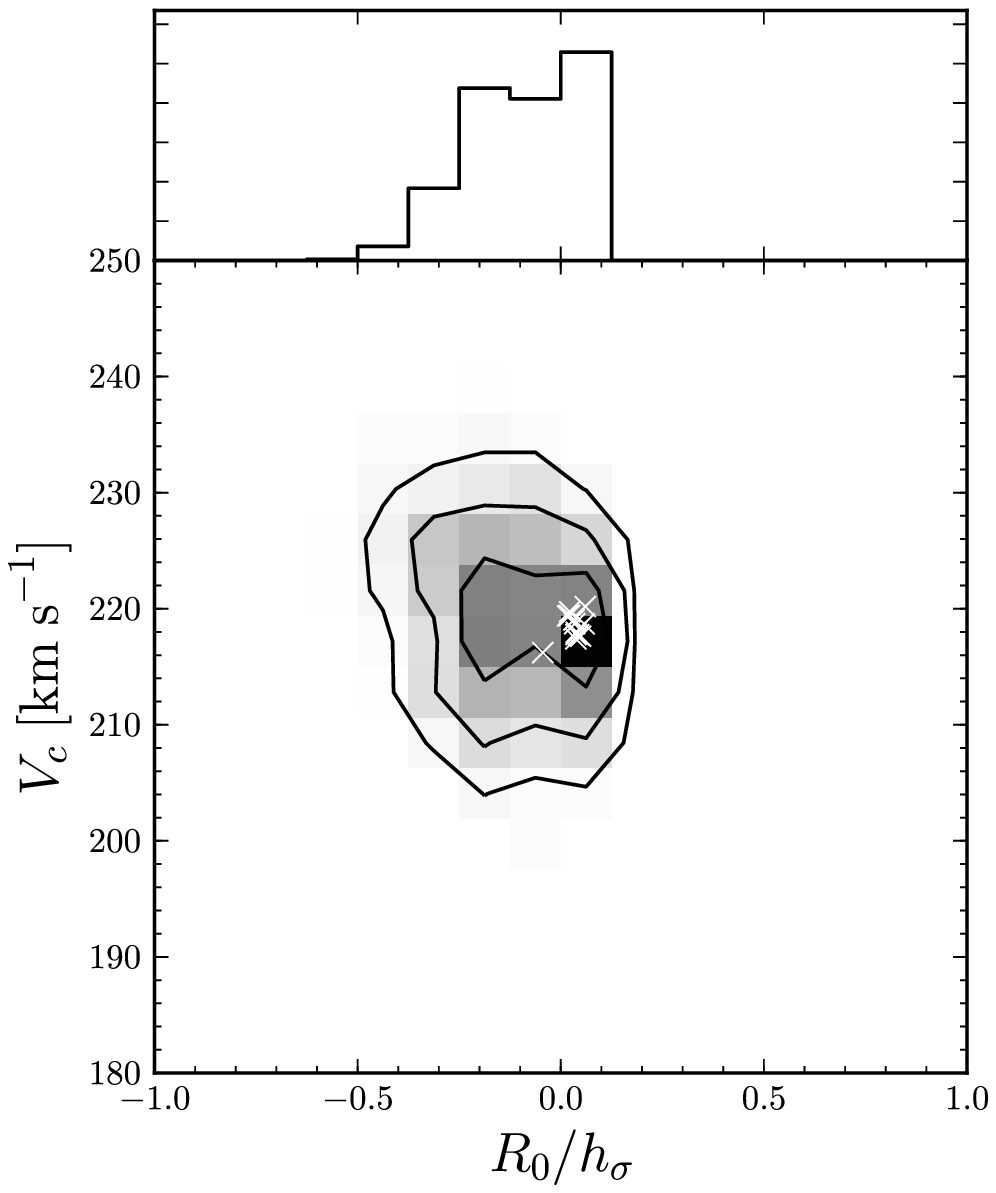}
\includegraphics[width=0.23\textwidth,clip=]{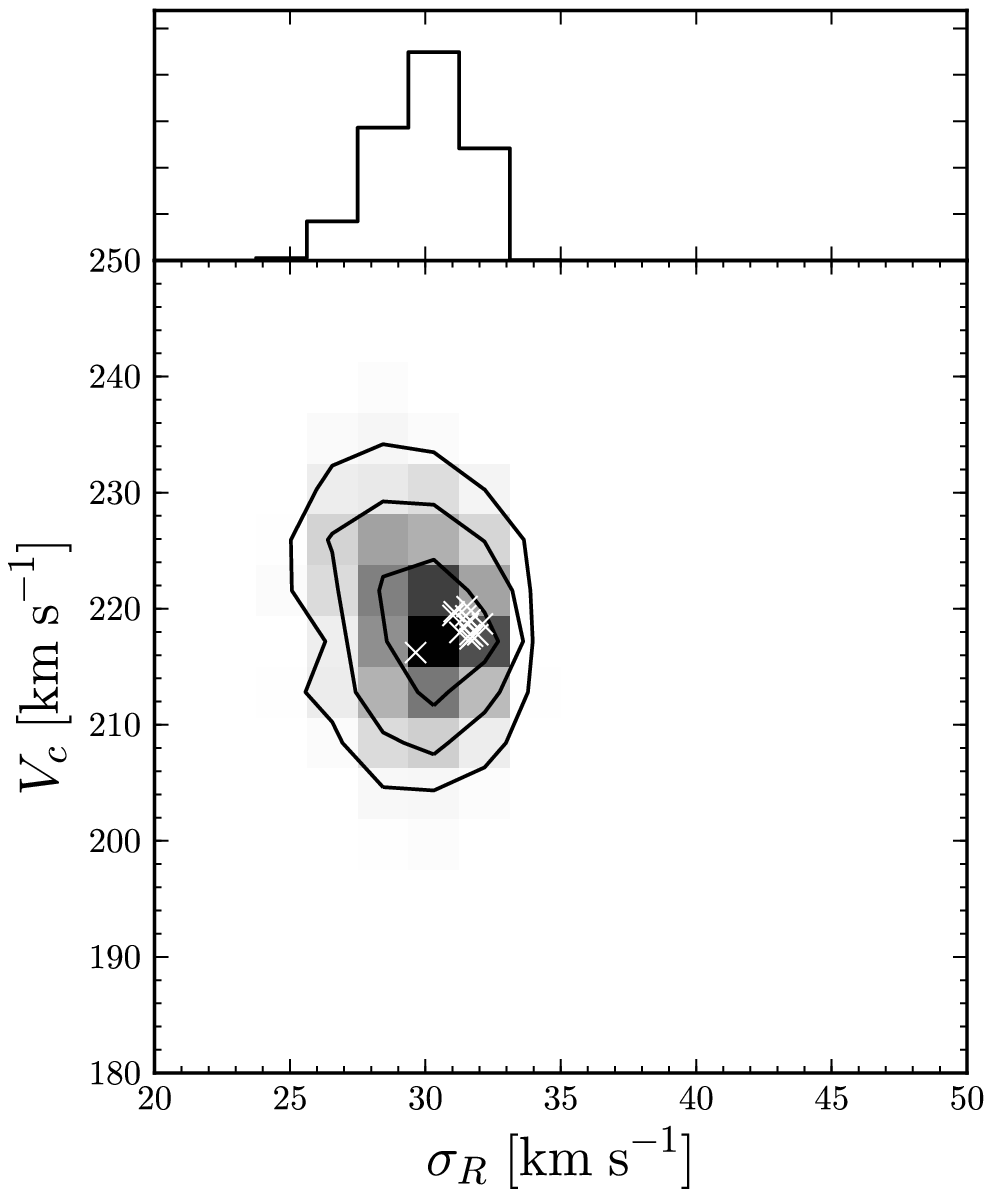}
\includegraphics[width=0.23\textwidth]{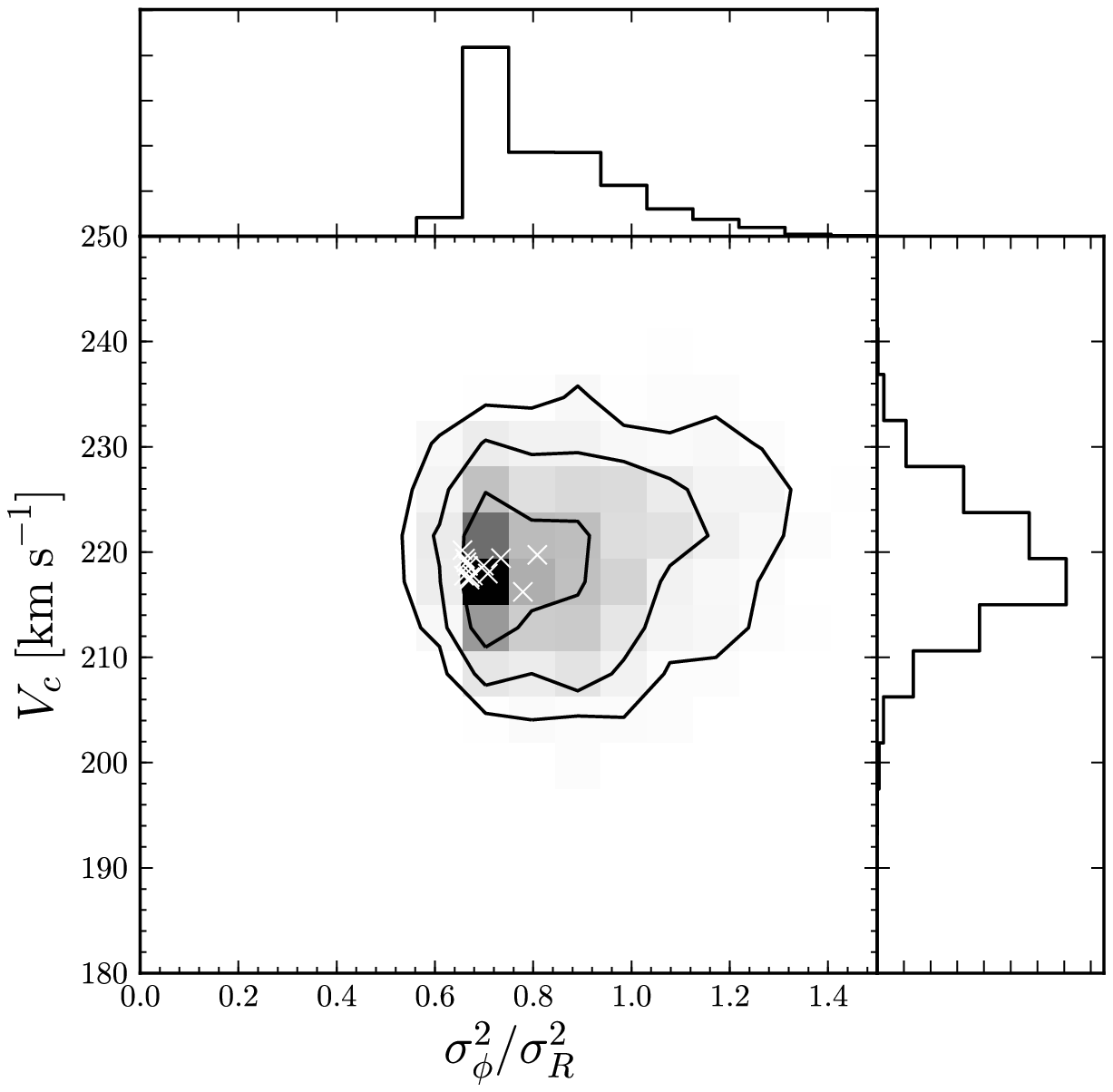}\\
\includegraphics[width=0.23\textwidth,clip=]{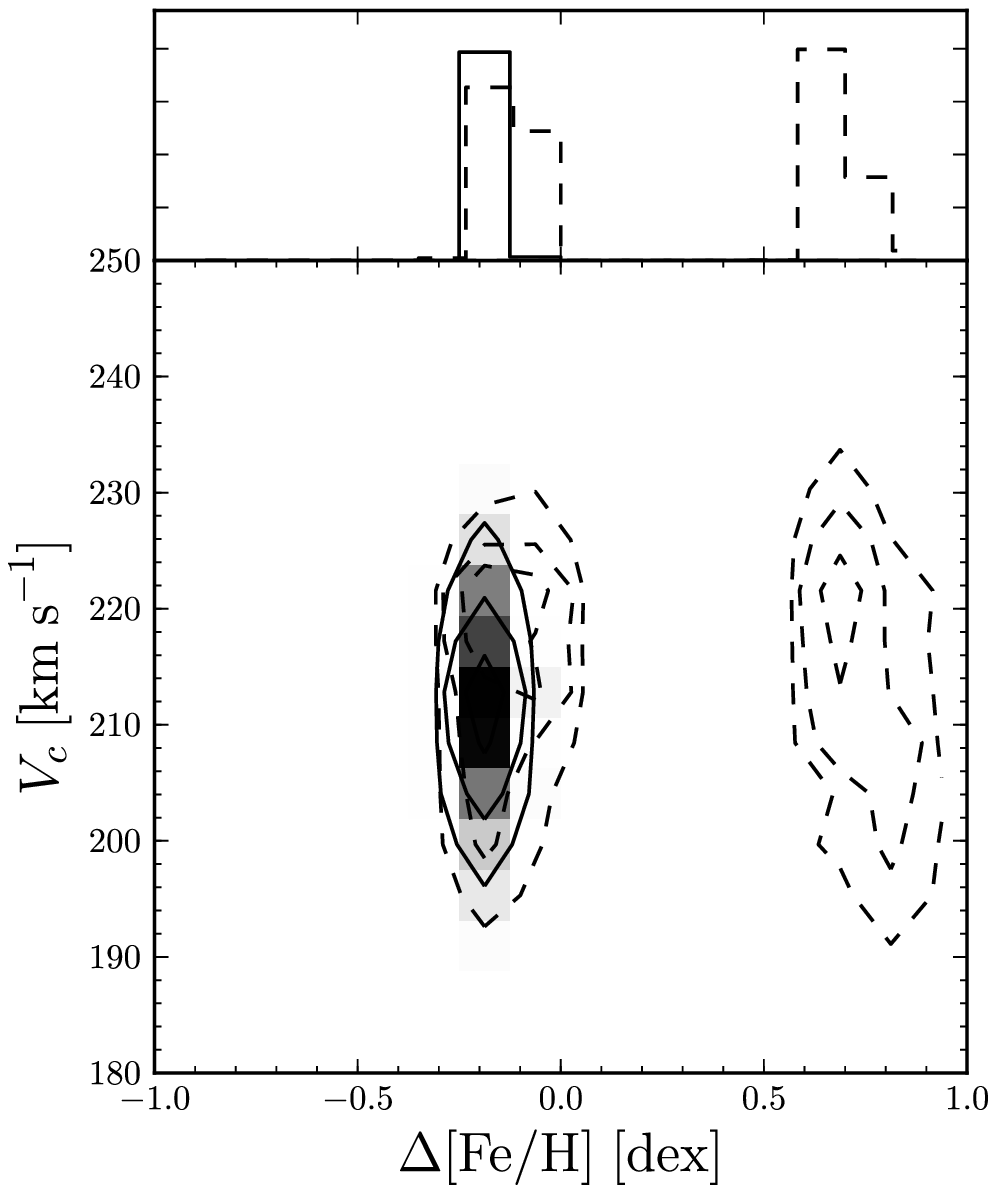}
\includegraphics[width=0.23\textwidth,clip=]{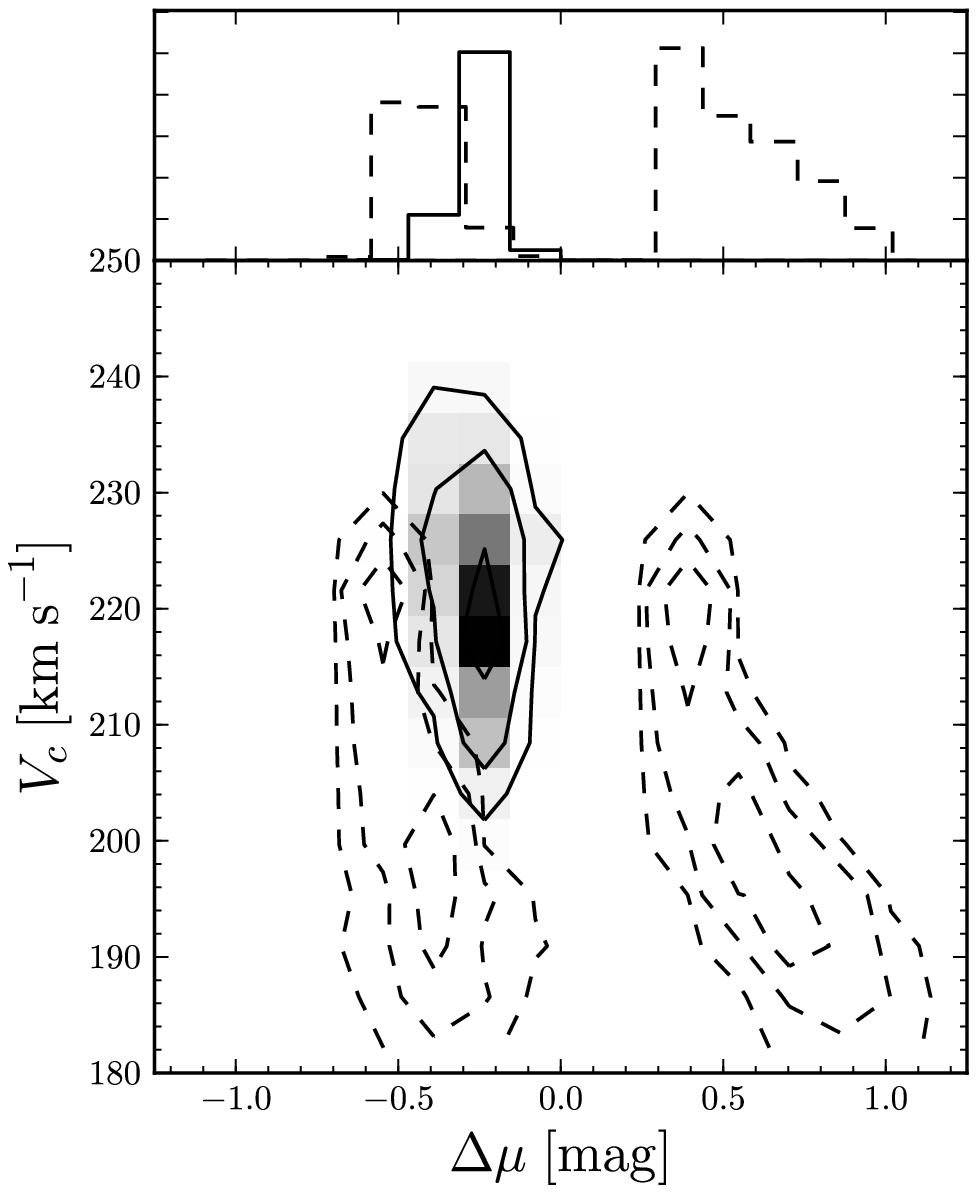}
\includegraphics[width=0.23\textwidth,clip=]{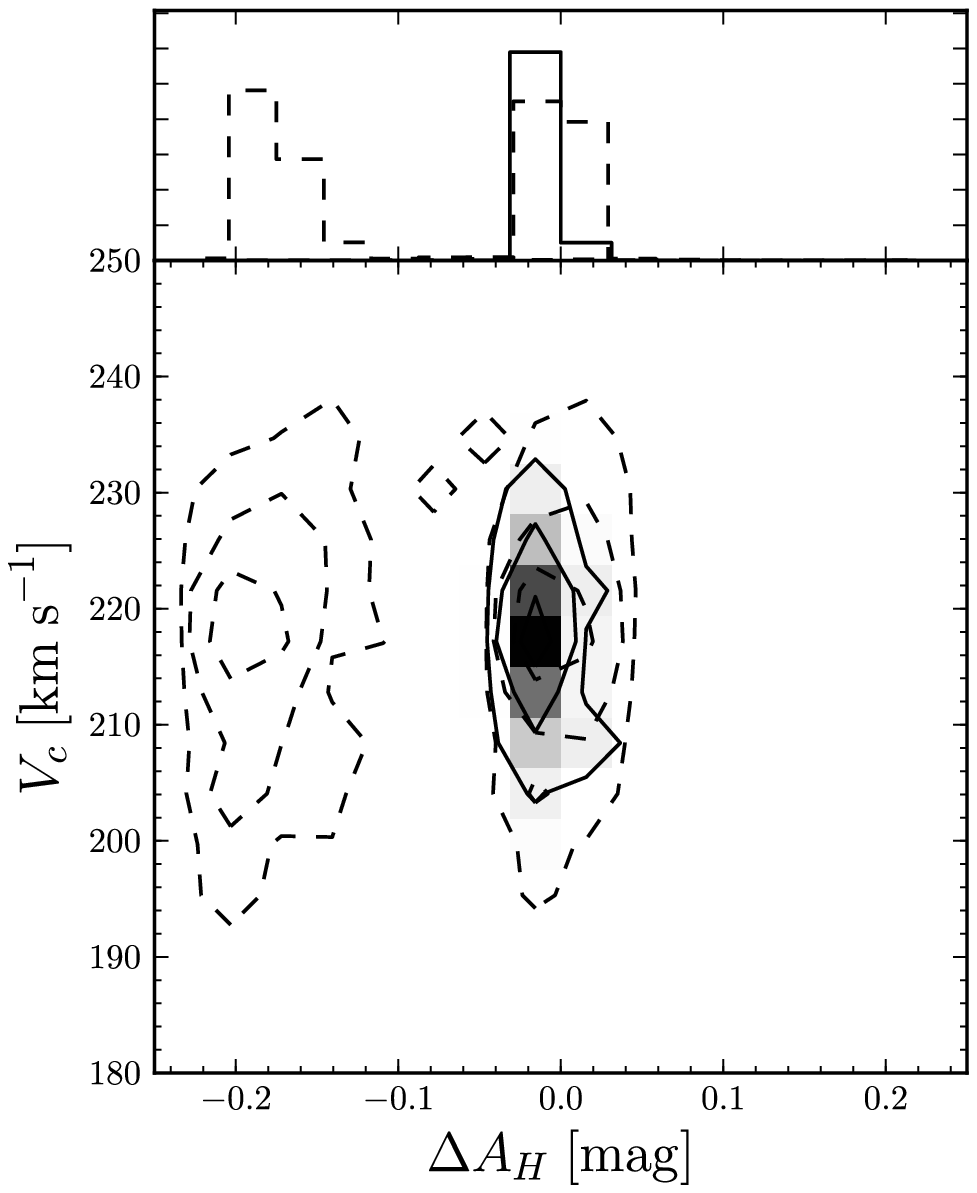}
\includegraphics[width=0.23\textwidth]{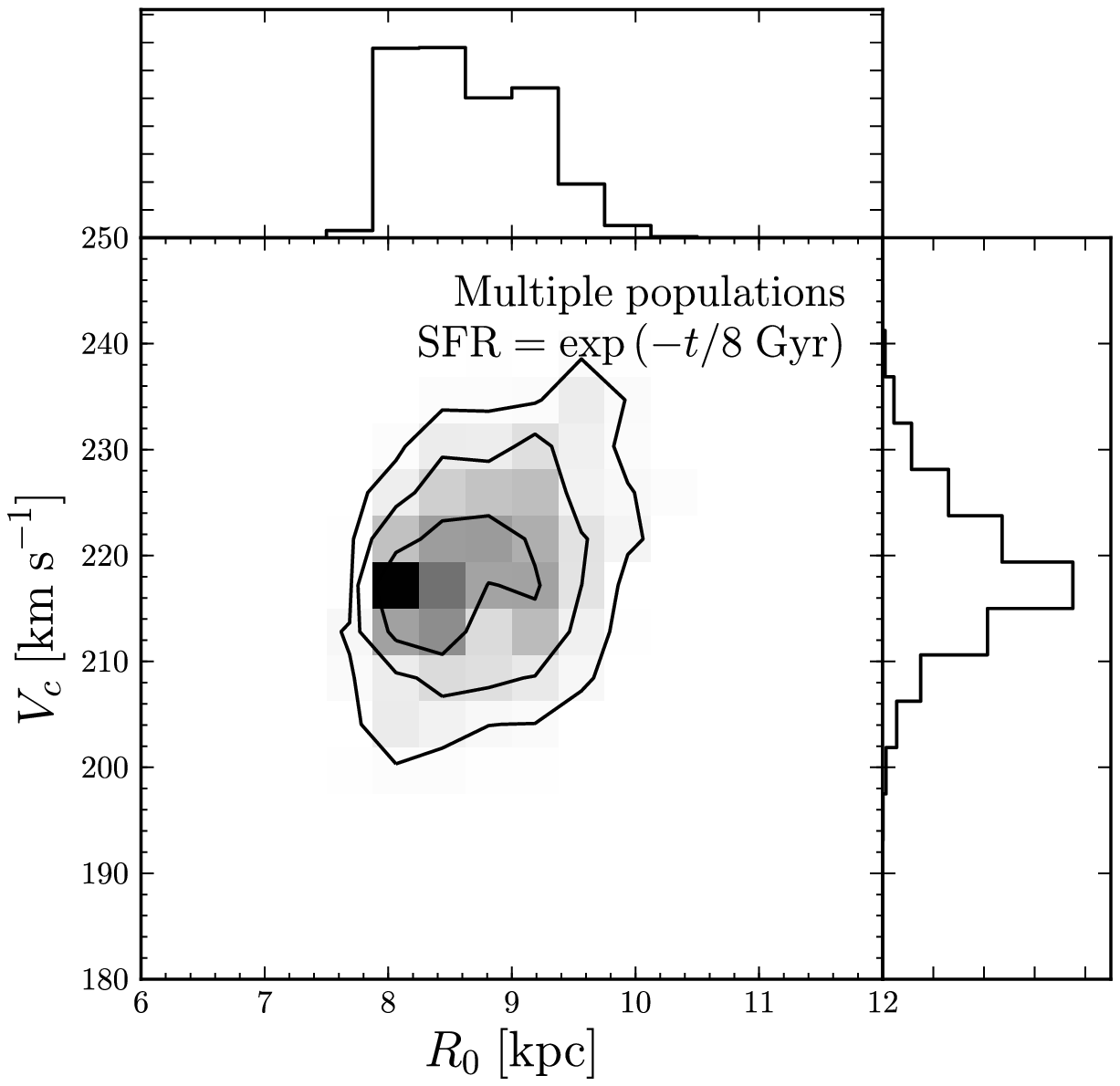}
\caption{PDFs for the parameters of the model. All PDFs in the top two
  rows are for the model with a flat rotation curve, except for the
  middle two panels in the top row, which are for the power-law fit
  and the linear-polynomial fit. The top row has the (marginalized)
  joint PDF for $V_c$ and (from left to right) $R_0$, power-law index
  $\beta$, linear derivative $\dd V_c / \dd R$, and the dwarf
  contamination fraction $P(\mathrm{dwarf})$. The middle row has (from
  left to right): the Sun's peculiar rotational velocity
  $V_{\phi,\odot}-V_c$, the ratio of $R_0$ to the radial-dispersion
  scale length, the radial-velocity dispersion, and the ratio of
  tangential to radial-velocity dispersion squared. The bottom row
  explores various systematics: A systematic offset in \feh, in
  distance modulus $\mu$, in extinction $A_H$, and using 20
  populations between $1$ and $10\,\mathrm{Gyr}$ with an
  exponentially-declining star-formation history and an increasing
  velocity dispersion as in \eqnname~(\ref{eq:multivdisp}). In the
  leftmost three panels in the bottom row, the PDF for a single
  systematic offset is shown as the linear density and the solid
  contours, while the contours of the PDF when allowing for a
  different systematic offset in inner ($l \leq 90^\circ$) and outer
  ($l > 90^\circ$) fields are represented as dashed lines. The figures in
  the top two rows also include the best-fit result from
  leave-one-longitudinal-field-out fits to the data as white crosses
  (that is, leaving one of the \nfields\ fields
  in \tablename~\ref{table:fields} out of the data
  set).}\label{fig:pdfs}
\end{figure*}

In this section, we discuss the results from fitting the basic
kinematical model of a single population with exponential $\dens(R)$
and $\sigma_R(R)$ profiles to the \apogee\ disk mid-plane data
from \sectionname~\ref{sec:data}. These are the main results of this
paper. In a subsequent section, we discuss the effects of various
systematics on these basic model fits, but we find in the end that
they do not significantly bias the results for the Galactic parameters
in this section.

\begin{figure}[htbp]
\includegraphics[width=0.5\textwidth]{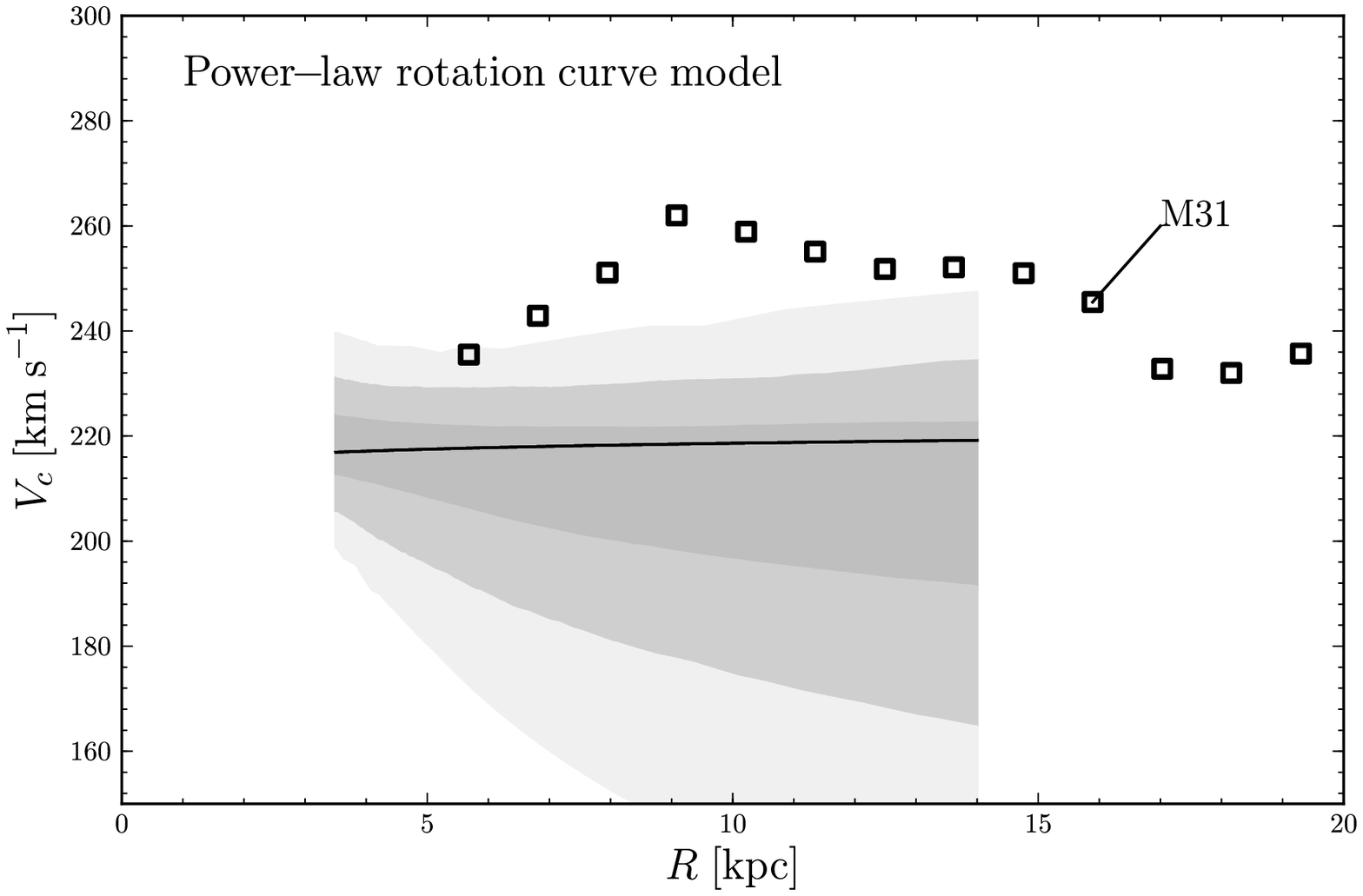}\\
\includegraphics[width=0.5\textwidth]{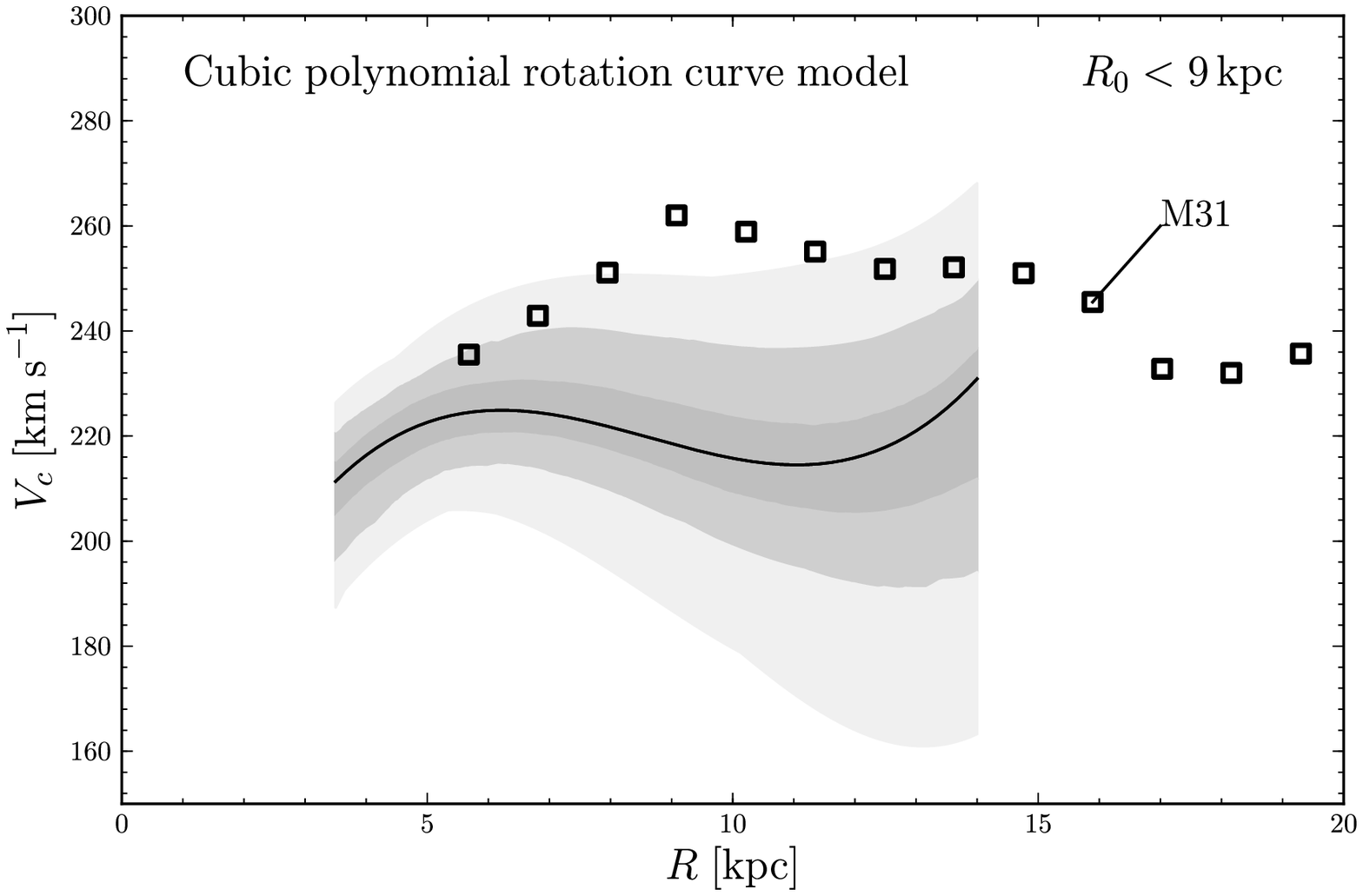}
\caption{The Milky Way's rotation curve in the range $4 < R < 14$ kpc
  as inferred from our data for different forms of the shape of the
  rotation curve. At each $R$, the range in $V_c(R)$ from 10,000
  samples of the PDF---\emph{assuming a power-law or cubic-polynomial
    model for the shape of the rotation curve}---is determined and the
  68\%, 95\%, and 99\% intervals are shown in varying shades of
  gray. For the cubic-polynomial model, we impose a prior that $R_0 <
  9\,\kpc$ (see text). For comparison, the squares show the rotation
  curve of M31 from the compilation of
  \citet{Carignon06a}.}\label{fig:rotcurve}
\end{figure}

\begin{deluxetable*}{lr@{}lr@{}l}
\tablecaption{}
\tablecolumns{5}
\tablewidth{0pt}
\tabletypesize{\footnotesize}
\tablecaption{Results for Galactic Parameters and Tracer Properties}
\tablehead{\colhead{Parameter} & \multicolumn{2}{c}{Flat rotation curve} & \multicolumn{2}{c}{Power-law $V_c(R) = V_c(R_0)\,\left(R/R_0\right)^\beta$}}

\startdata
$V_c(R_0)\ [\mathrm{km\ s}^{-1}]$ & $218$&$\pm 6$ & $218$&$^{+4}_{-19}$\\
$\beta$ & \ldots &  & $0.01$&$^{+0.01}_{-0.10}$\\
$\mathrm{d} V_c / \mathrm{d} R \left(R_0 \right)\ [\mathrm{km\ s}^{-1}\ \mathrm{kpc}^{-1}]$ & \ldots &  & $0.2$&$^{+0.2}_{-2.8}$\\
$A\ [\mathrm{km\ s}^{-1}\ \mathrm{kpc}^{-1}]$ & $13.5$&$^{+0.2}_{-1.7}$ & $13.5$&$^{+0.2}_{-1.0}$\\
$B\ [\mathrm{km\ s}^{-1}\ \mathrm{kpc}^{-1}]$ & $-13.5$&$^{+1.7}_{-0.2}$ & $-13.7$&$^{+3.3}_{-0.1}$\\
$(B^2-A^2)/(2\pi G)\ [\mathrm{M_{\odot}\ pc}^{-3}]$ & \ldots &  & $0.0002$&$^{+0.0002}_{-0.0025}$\\
$\Omega_0\ [\mathrm{km\ s}^{-1}\ \mathrm{kpc}^{-1}]$ & $27.0$&$^{+0.3}_{-3.5}$ & $27.3$&$^{+0.4}_{-4.2}$\\\\
$R_0\ [\mathrm{kpc}]$ & $8.1$&$^{+1.2}_{-0.1}$ & $8.0$&$^{+0.8}_{-0.1}$\\
$V_{R,\odot}\ [\mathrm{km\ s}^{-1}]$ & $-10.5$&$^{+0.5}_{-0.8}$ & $-10.3$&$^{+1.1}_{-0.1}$\\
$V_{\phi,\odot}\ [\mathrm{km\ s}^{-1}]$ & $242$&$^{+10}_{-3}$ & $241$&$^{+5}_{-17}$\\
$V_{\phi,\odot}-V_c\ [\mathrm{km\ s}^{-1}]$ & $23.9$&$^{+5.1}_{-0.5}$ & $23.1$&$^{+3.6}_{-0.5}$\\
$\mu_{\mathrm{Sgr\ A}^{^*}}\ [\mathrm{mas\ yr}^{-1}]$ & $6.32$&$^{+0.07}_{-0.70}$ & $6.36$&$^{+0.09}_{-0.86}$\\\\
$\sigma_R(R_0)\ [\mathrm{km\ s}^{-1}]$ & $31.4$&$^{+0.1}_{-3.2}$ & $32.2$&$^{+0.2}_{-2.6}$\\
$R_0/h_\sigma$ & $0.03$&$^{+0.01}_{-0.27}$ & $0.06$&$^{+0.01}_{-0.17}$\\
$X^2 \equiv \sigma_\phi^2 / \sigma_R^2$ & $0.70$&$^{+0.30}_{-0.01}$ & $0.64$&$^{+0.18}_{-0.02}$
\enddata

\tablecomments{\protect{M}aximum-likelihood best-fit results;
  uncertainties are given as 68\% intervals of the marginalized PDF
  for each parameter. The first block of parameters are the basic
  Galactic parameters: circular velocity $V_c(R_0)$ at the Sun's
  location, the derivative of the circular-velocity curve with $R$,
  and various other representations of these two basic parameters: the
  Oort parameters $A$ and $B$, the rotational frequency $\Omega_0$ at
  the Sun's location, and the combination $(B^2-A^2)/(2\pi G)$ that is
  relevant in the determination of the local dark matter density using
  the cylindrical Poisson equation. The second block of parameters
  describe the Sun's phase-space position in the Galaxy: distance
  $R_0$ to the Galactic center, the Sun's radial velocity
  $V_{R,\odot}$ (away from the Galactic center), the Sun's
  Galactocentric rotational velocity $V_{\phi,\odot}$, and the Sun's
  peculiar velocity with respect to local Galactic rotation
  $V_{\phi,\odot}-V_c$. We also give the proper motion of Sgr A$^*$
  derived from the previous parameters assuming that Sgr A$^*$ is at
  rest with respect to the disk. The final block of parameters
  describe the tracer population sampled by \apogee: radial velocity
  dispersion $\sigma_R(R_0)$ at $R_0$, the radial scale length
  $h_\sigma$ of the radial velocity dispersion (given as
  $R_0/h_\sigma$), and the ratio of the tangential to the radial
  velocity variance $\sigma_\phi^2/\sigma_R^2$.}\label{table:results}
\end{deluxetable*}

The results from fitting a flat rotation curve to the \apogee\ data
are given in the left column of \tablename~\ref{table:results}. The
right column of this table gives the results when fitting a power-law
model for the rotation curve $V_c(R) =
V_c(R_0)\,\left(R/R_0\right)^\beta$. The best-fit power-law index in
this case is approximately zero, such that both models for the
rotation curve yield similar best-fit values for all parameters. The
only significant difference between the two models for the rotation
curve is the uncertainty interval for $V_c(R_0)$, which extends to
lower values of $V_c$ in the case of the power-law fit, although the
upper limit on $V_c$ stays approximately the same. We find that $V_c$
is tightly constrained in the flat rotation-curve model: $V_c =
218\pm6\,\kms$ (68\% confidence). When fitting a power law we find
$V_c = 218^{+4}_{-19}\,\kms$, with a strong correlation between the
power-law index $\beta$ (or, equivalently, the local derivative $\dd
V_c/\dd R$ of the circular velocity with respect to $R$) and
$V_c$. Other parameters do not exhibit any strong correlation with
$V_c$ (see discussion below).

When fitting a power-law model for the rotation curve, the local slope
of the rotation curve is constrained to be $-3\,\kms\kpc^{-1} < \dd
V_c / \dd R < 0.4\,\kms\kpc^{-1}$ (68\% confidence), with a best-fit
value near the upper end of that range, $0.2\kms\kpc^{-1}$. When
fitting a linear model for the circular velocity the constraints are
similar, and we do not discuss them further; the top, middle panels of
\figurename~\ref{fig:pdfs} show the joint PDF for $V_c$ and $\beta$
(for the power-law fit) and $\dd V_c / \dd R$ (for the linear fit). In
\tablename~\ref{table:results}, we also give the Oort constants $A$
and $B$ corresponding to the $V_c$ and $\dd V_c / \dd R$ inferred
directly for our sample (for the case of a flat rotation curve, $A$
and $B$ are equal in magnitude and opposite in sign, by
definition). We also present the combination $(B^2-A^2)/(2\pi G)$,
which is the correction term that must be added to the local
dark-matter density to account for a non-flat rotation curve, when the
local dark-matter density is inferred from local vertical kinematics
using the cylindrical Poisson equation \citep[\eg,][]{Bovy12d}. It is
clear that we constrain this correction term to be smaller than a
fraction of the local dark-matter density:
$-0.0025\,M_\odot\,\mathrm{pc}^{-3} < (B^2-A^2)/(2\pi G) <
0.0004\,M_\odot\,\mathrm{pc}^{-3}$, compared to a local dark-matter
density of $0.008\pm0.003\,M_\odot\,\mathrm{pc}^{-3}$
\citep{Bovy12d}. Thus, the correction for the non-flatness of the
rotation curve for our best-fit parameters is approximately zero, and
at 68\% confidence it lies between $-0.1\,\mbox{GeV\ cm}^{-3}$ and
$0.0\,\mbox{GeV\ cm}^{-3}$. We also give the rotational frequency
$\Omega_0$ at the Sun's location.

Using a power-law rotation-curve model, we estimate the uncertainty in
the rotation curve by evaluating the range of $V_c(R)$ at each $R$ for
10,000 samples from the PDF shown in \figurename~\ref{fig:pdfs}. Thus,
every sample has a power-law rotation curve, but the range at each $R$
does not have to follow a power law itself. This exercise shows
whether, \emph{in the power-law model}, the value of $V_c(R)$ is
well-constrained or not for each $R$; The result is shown in the left
panel of \figurename~\ref{fig:rotcurve}. The same is shown in the
right panel of that figure for a fit assuming a cubic-polynomial model
for the rotation curve. However, in the latter model we have imposed a
prior that $R_0 < 9\,\kpc$, as a large number of samples from the PDF
have $R_0 > 9\,\kpc$, which we assume to be implausible from prior
data \citep[\eg,][]{Ghez08a,Gillessen09a}. We emphasize that we do not
include such a prior in any other part of the analysis presented
here. We see that the rotation curve is best constrained at small
radii ($4\,\kpc < R < 8\,\kpc$), and that it is largely consistent
with being close to flat over the full range of Galactocentric radii
considered here.

We also determine the full planar phase-space position of the Sun in
the Galactocentric reference frame. The current \apogee\ data are
relatively insensitive to the value of $R_0$. The 68\% interval for
$R_0$ for both the flat-rotation-curve and power-law fits is about
$8\,\kpc < R_0 < 9\,\kpc$, which is similar to that found for the mock
data in \appendixname~\ref{sec:appmock}. The best-fit value is at the
lower end of this range: $8.1\,\kpc$ when assuming a flat rotation
curve, $8\,\kpc$ for the power-law rotation-curve fit. The fully
marginalized PDF for $R_0$ in the top, left panel
of \figurename~\ref{fig:pdfs} is almost entirely flat between
$8.0\,\kpc < R_0 < 9.3\,\kpc$.

The Galactocentric motion of the Sun is approximately $V_{R,\odot} =
-10.5\pm1.0\,\kms$, and $V_{\phi,\odot} = 242^{+10}_{-3}\,\kms$ (in
the case of a flat rotation curve) or $V_{\phi,\odot} =
242^{+5}_{-17}\,\kms$ (for a power-law fit to the rotation curve). In
the latter fit, there is a strong correlation between $V_c$ and
$V_{\phi,\odot}$; the difference between the two is much better
constrained: $V_{\phi,\odot}-V_c = 23.1^{+3.6}_{-0.5}\,\kms$. The
marginalized PDF for $V_{\phi,\odot}-V_c$ in
\figurename~\ref{fig:pdfs} is well-described by $V_{\phi,\odot}-V_c =
26\pm3\,\kms$. We discuss the consequences of this solar motion in
detail in
\sectionname~\ref{sec:discuss-solar}, but we note here that the
estimate of the angular motion of the Galactic center that we obtain
from combining our fits for $V_{\phi,\odot}$ and $R_0$ is consistent
with the proper motion of Sgr A$^*$ as measured by \citet{Reid04a}:
our estimate is $\mu_{\mathrm{Sgr\ A}^*} =
6.3^{+0.1}_{-0.7}\,\mathrm{mas\ yr}^{-1}$, compared to the direct
measurement of $6.379\pm0.024\,\mathrm{mas\ yr}^{-1}$. We discuss the
apparent discrepancy between our agreement with the proper motion of
Sgr A$^*$ and our low value for $V_c$ in
\sectionname~\ref{sec:discuss-solar}. 

The final block of parameters in \tablename~\ref{table:results}
describe the tracer population. The velocity dispersion that we infer
for the tracer stars is close to that expected for an old disk
population: $\sigma_R(R_0) \approx 32.0^{+0.5}_{-3}\,\kms$ for the
flat-rotation-curve and power-law fits. The ratio of the
tangential-to-radial velocity dispersions squared is $0.69 < X^2 <
1.0$, with the best-fit value at the lower end of this range. This
value is higher than expected from the epicycle approximation for a
flat or falling rotation curve, which is $X^2 \leq 0.5$. However, this
expectation holds only for a cold disk, and corrections due to the
temperature of the old disk population always increase $X^2$ near
$R_0$ \citep{Kuijken91a}: the Dehnen disk distribution functions of
\eqnname~(\ref{eq:fdehnen}) have $X^2$ that varies spatially, and
reaches approximately 0.65 near $R_0$ \citep{dehnen99b}. The best-fit
value for $R_0/h_\sigma$ is approximately zero, with non-zero positive
values ruled out by the data: the 68\% interval is $-0.24 <
R_0/h_\sigma < 0.03$. Thus, the radial-velocity dispersion does not
drop exponentially with radius with a scale length between $2\,R_0/3$
and $R_0$; such a drop would be expected from previous measurements of
the radial dispersion as a function of $R$ \citep{Lewis89a}, or from
the observed exponential decline of the vertical velocity dispersion
\citep{Bovy12c} combined with the assumption of constant
$\sigma_z/\sigma_R$. We have attempted fits with two populations of
stars with different radial scale lengths ($3\,\kpc$ and $5$ or
$6\,\kpc$) and radial-velocity dispersions, but the same
radial-dispersion scale length. The best-fit $R_0/h_\sigma$ remains
zero, such that it does not seem that we are seeing a mix of multiple
populations that conspire to form a flat $\sigma_R$ profile.

Even with the best-fit flat radial-dispersion profile, the disk is
stable over most of the range in $R$ considered here. The Toomre Q
parameter---$Q=\sigma_R\,\kappa /(3.36 G \Sigma)$
\citep{Toomre64a}---for a flat rotation curve is
\begin{equation}
\begin{split}
Q = 1.72\,& \left(\frac{\sigma_R}{32\,\kms}\right)\,
\left(\frac{V_c}{220\,\kms}\right)\\
&  \left(\frac{R}{8\,\kpc}\right)^{-1}\,
\left(\frac{\Sigma}{50\,M_\odot\,\mathrm{pc}^{-2}}\right)^{-1}\,.
\end{split}
\end{equation}
This expression has $Q > 1$ down to $4.9\,\kpc$ and $Q=0.91$ at
$R=4\,\kpc$ for a constant $\sigma_R(R)$ and a surface density
$\Sigma \propto e^{-R/(3\,\kpc)}$. Although the disk is marginally
unstable in our best-fit model, this conclusion depends strongly on
the assumed radial scale length: for $h_R=3.25\,\kpc$, $Q > 1$
everywhere at $R > 4\,\kpc$. The flatness of the inferred $\sigma_R$
profile also depends on the assumed constancy of $X^2$. Actual
equilibrium axisymmetric disks, such as those having a Dehnen
distribution function (\eqnname~(\ref{eq:fdehnen})), have a
radially-dependent $X^2$, with $X^2$ at $R=4\,\kpc$ typically smaller
than at $R=8$ to $16\,\kpc$ \citep[][Figure 4]{dehnen99b}. At
$R=4\,\kpc$, which for the present data sample is only reached for the
$l=30^\circ$ line of sight, the line-of-sight velocity is entirely
composed of the tangential-velocity component, such that any decrease
in $X^2$ leads to an \emph{increase} in $\sigma_R$ to sustain the same
$\sigma_\phi$.  Therefore, the true $\sigma_R(R)$ profile
presumably \emph{is} falling with $R$, and the entire disk at $R >
4\,\kpc$ should be stable in our model.

Full PDFs for all of the parameters of the basic models discussed in
this section are given in \figurename~\ref{fig:pdfs}. It is clear that
with the exception of $V_c$ and the derivative of the rotation
curve---$\beta$ in the power-law model and $\mathrm{d} V_c /
\mathrm{d} R$ in the linear model---there are no strong degeneracies
among the parameters. Also included in this figure are the results
from fitting all but one of the \nfields\ \apogee\ field for each
field: these leave-one-out results show that no single field drives
the analysis for any of the parameters.

\begin{figure*}[hbtp]
\begin{center}
\includegraphics[width=0.75\textwidth]{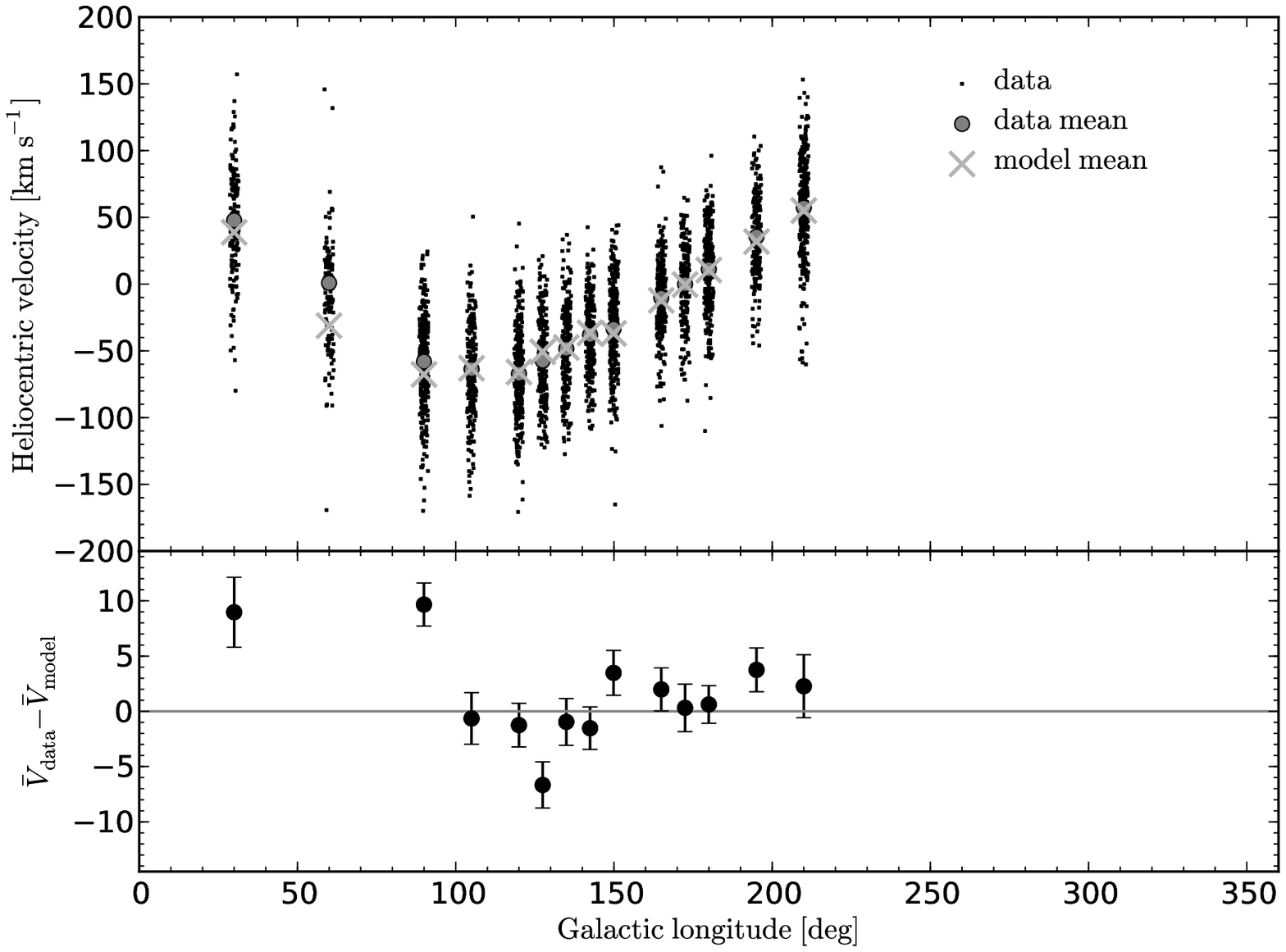}
\caption{Comparison between our best-fit flat-rotation-curve model and the data. The top
  panel shows the individual heliocentric velocities of stars in our
  sample as a function of Galactic longitude, with the mean for
  each \apogee\ field indicated by a gray dot. The prediction for this
  mean from the best-fit model is shown as the gray cross. The bottom
  panel shows the difference between the data mean
  $\bar{V}_{\mathrm{data}}$ and the model mean
  $\bar{V}_{\mathrm{model}}$. Note that the residual for the
  $l=60^\circ$ field in the bottom panel is missing as it is larger
  than $15\,\kms$.}\label{fig:bestfit}
\end{center}
\end{figure*}

\begin{figure*}[htbp]
\includegraphics[width=0.24\textwidth,clip=]{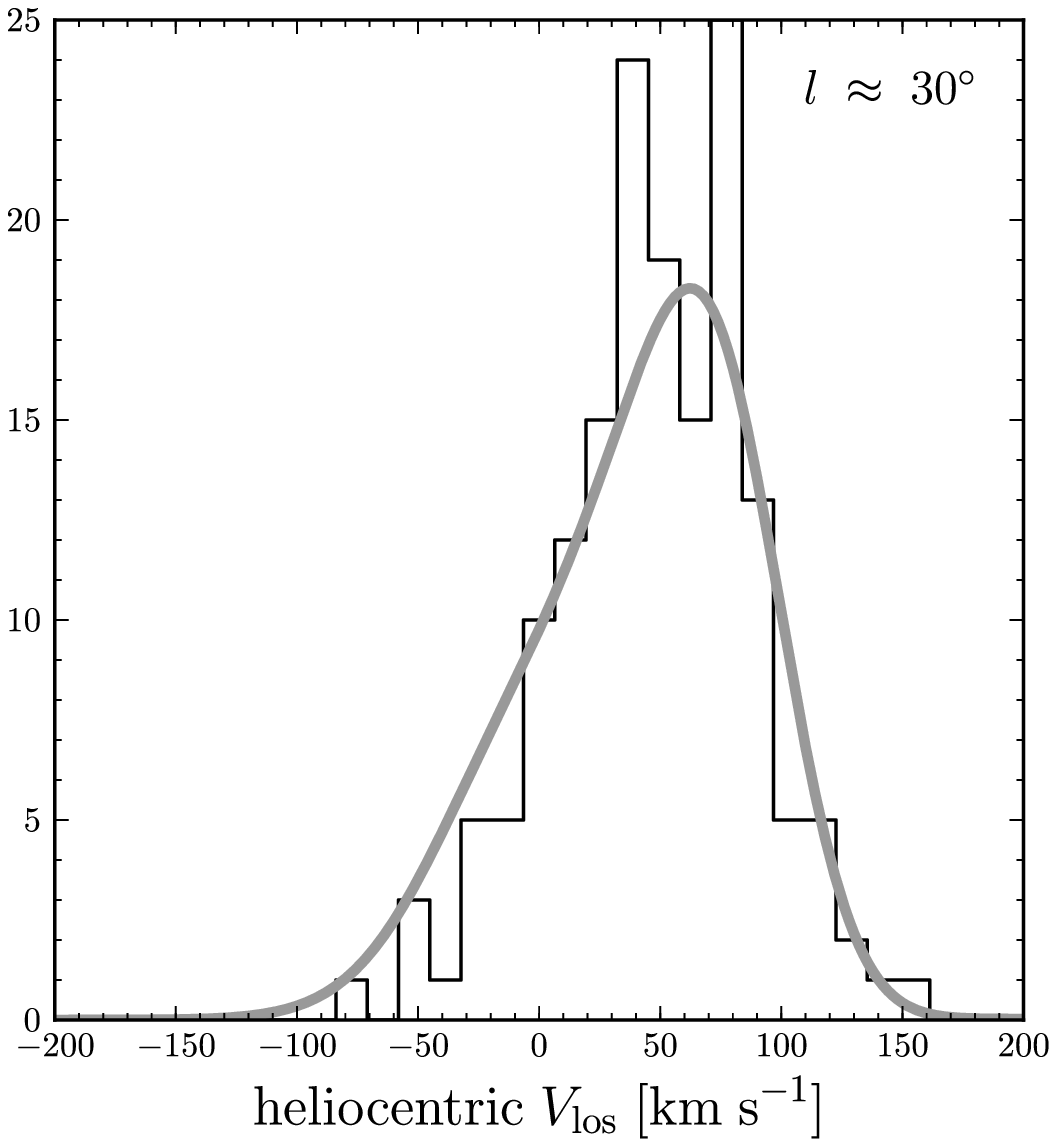}
\includegraphics[width=0.24\textwidth,clip=]{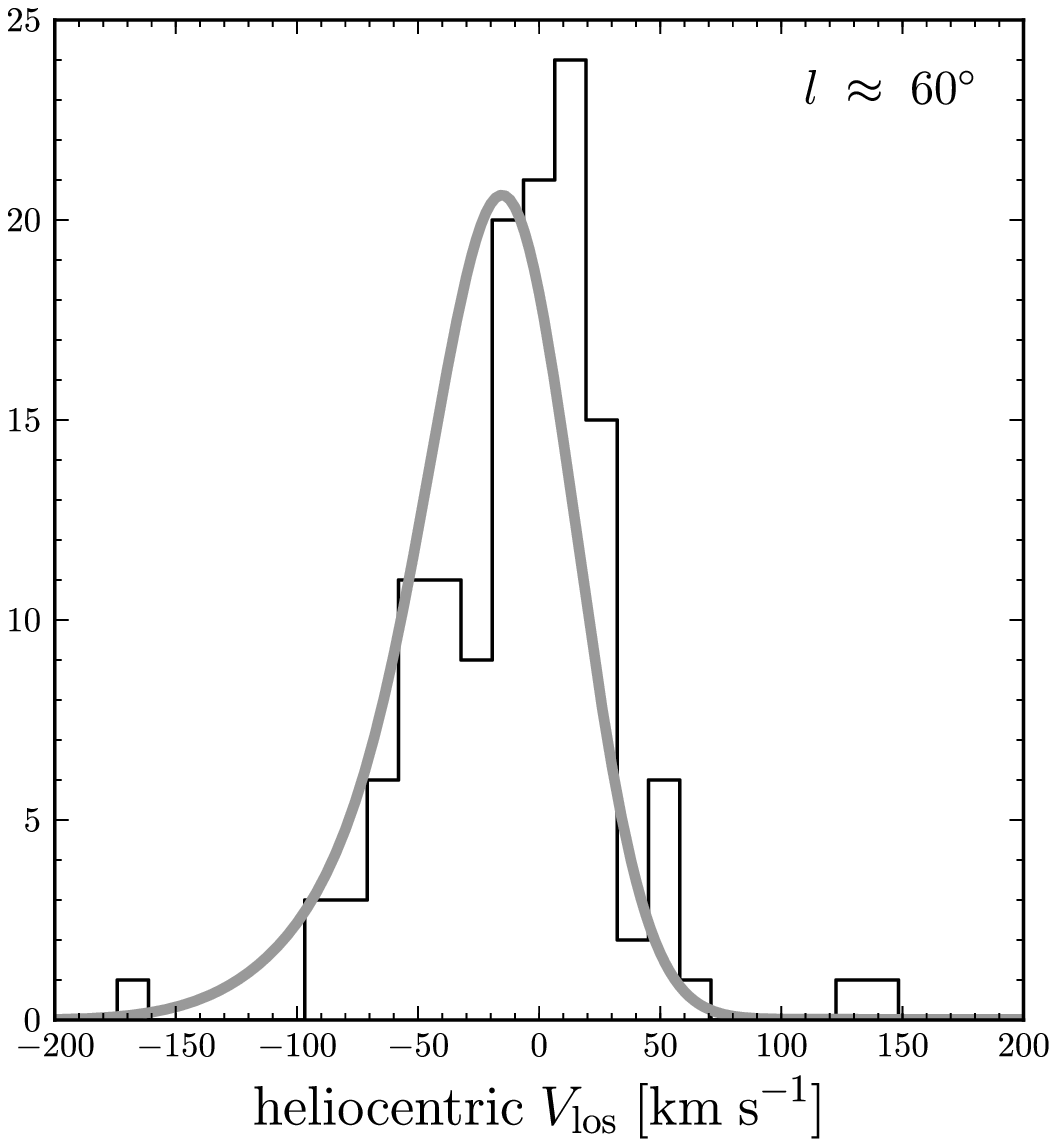}
\includegraphics[width=0.24\textwidth,clip=]{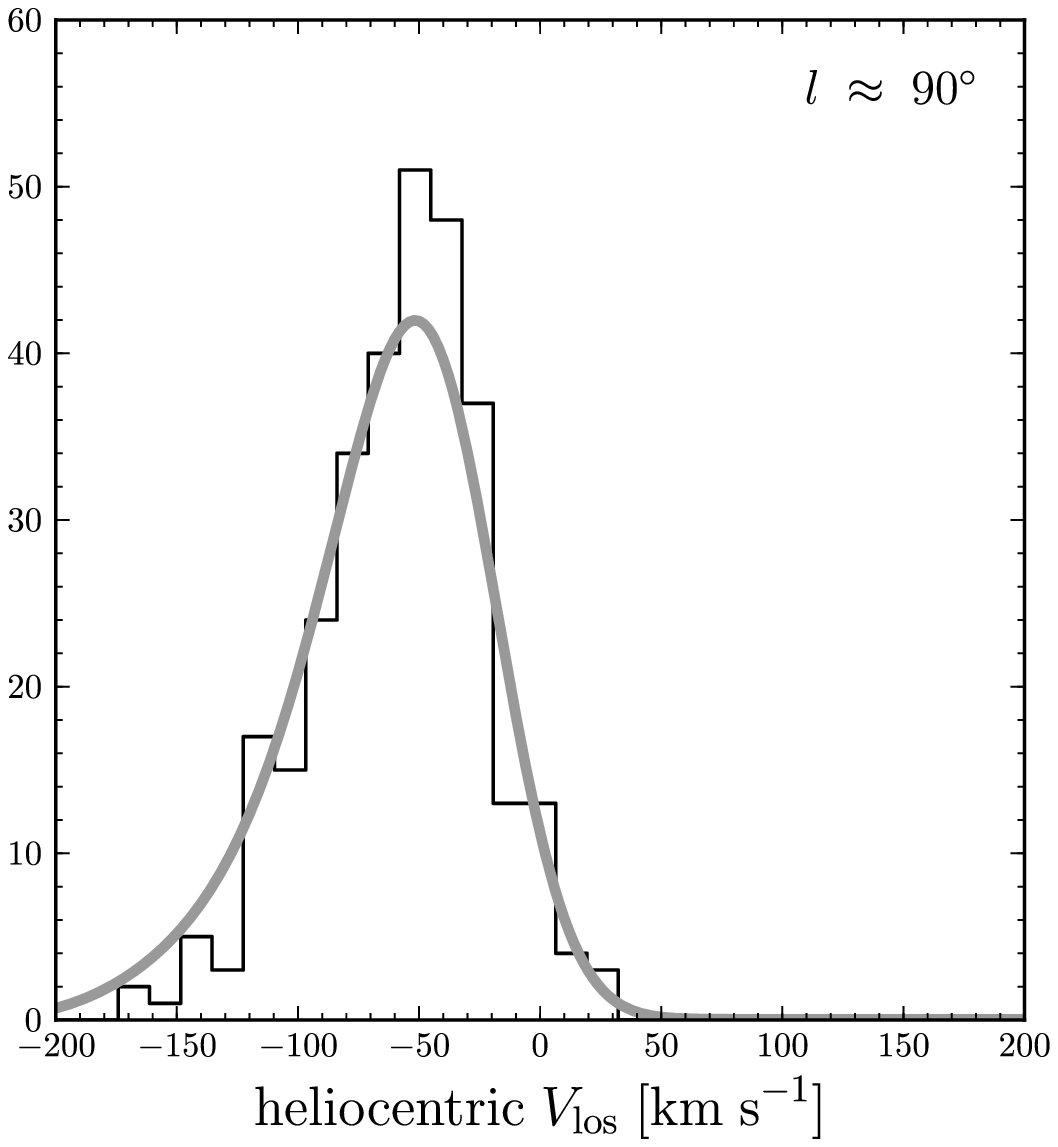}
\includegraphics[width=0.24\textwidth,clip=]{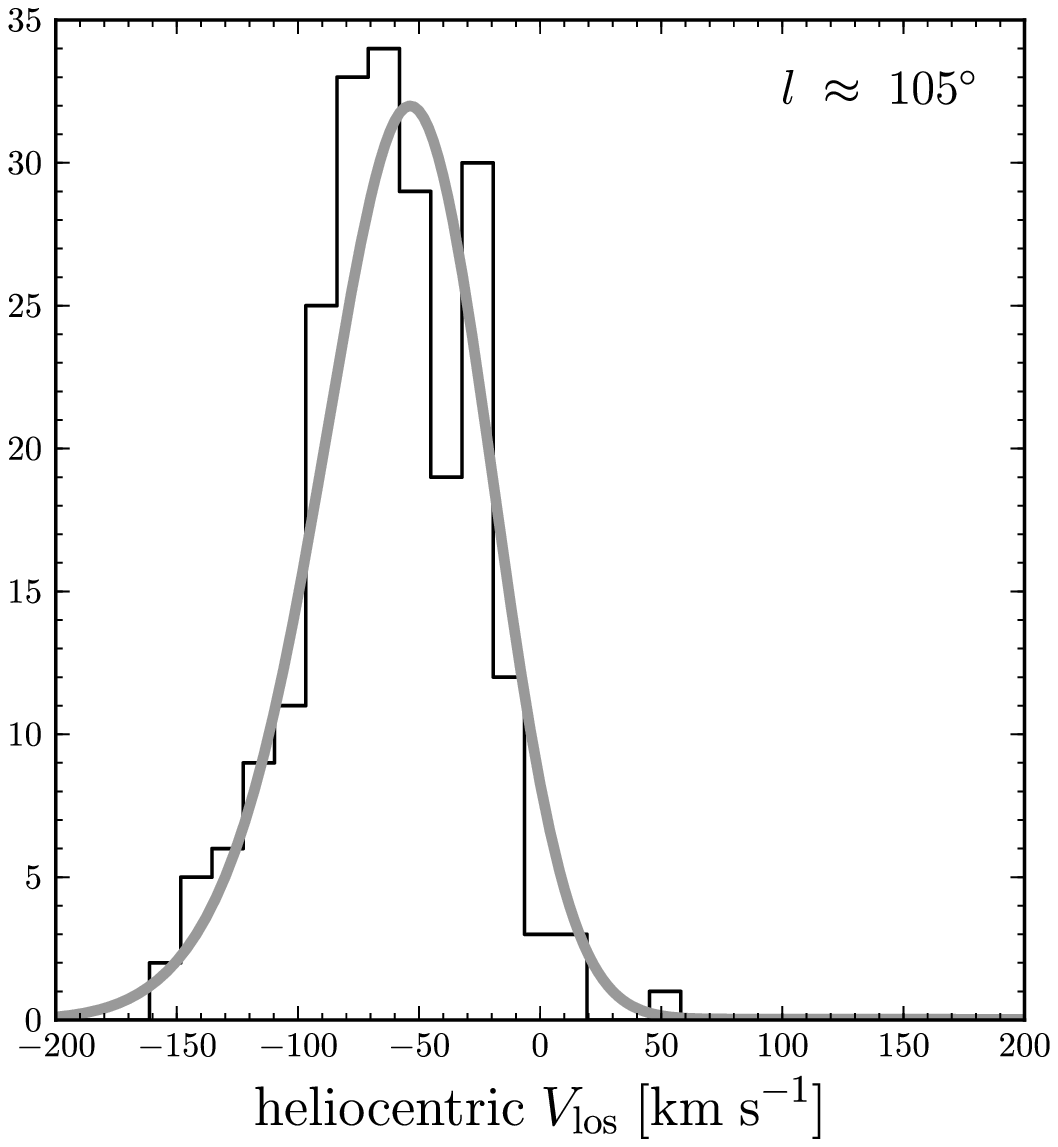}\\
\includegraphics[width=0.24\textwidth,clip=]{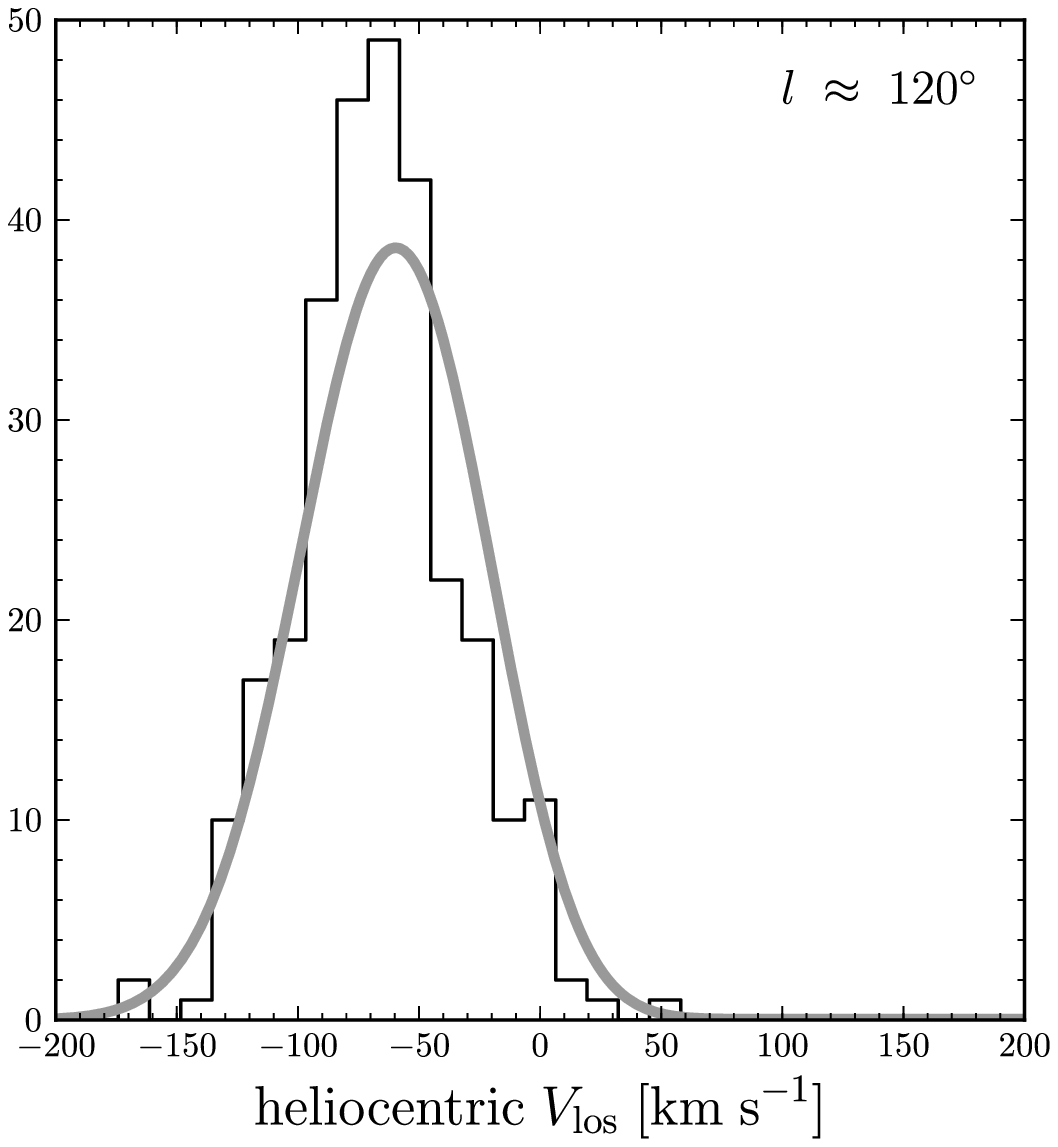}
\includegraphics[width=0.24\textwidth,clip=]{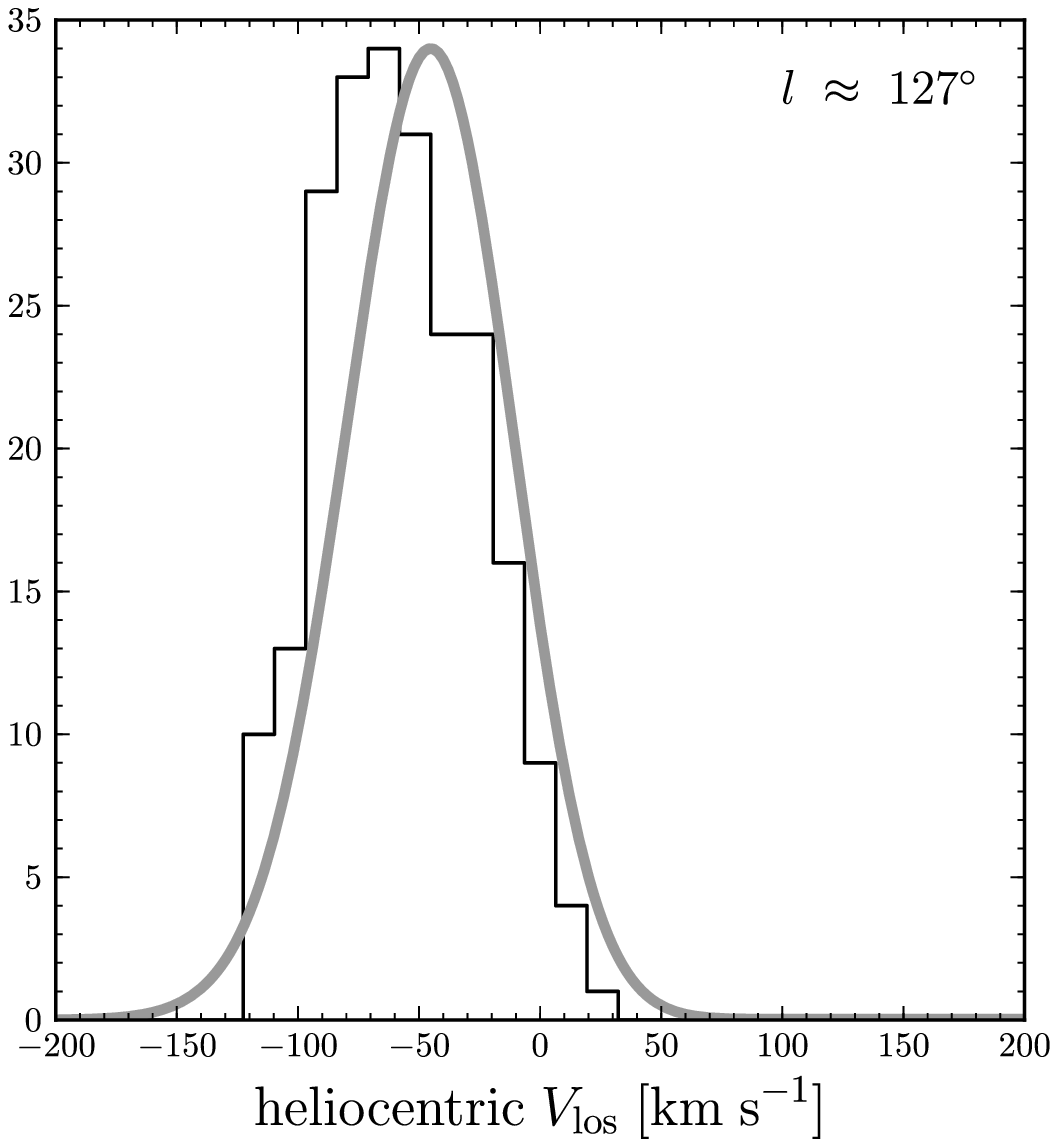}
\includegraphics[width=0.24\textwidth,clip=]{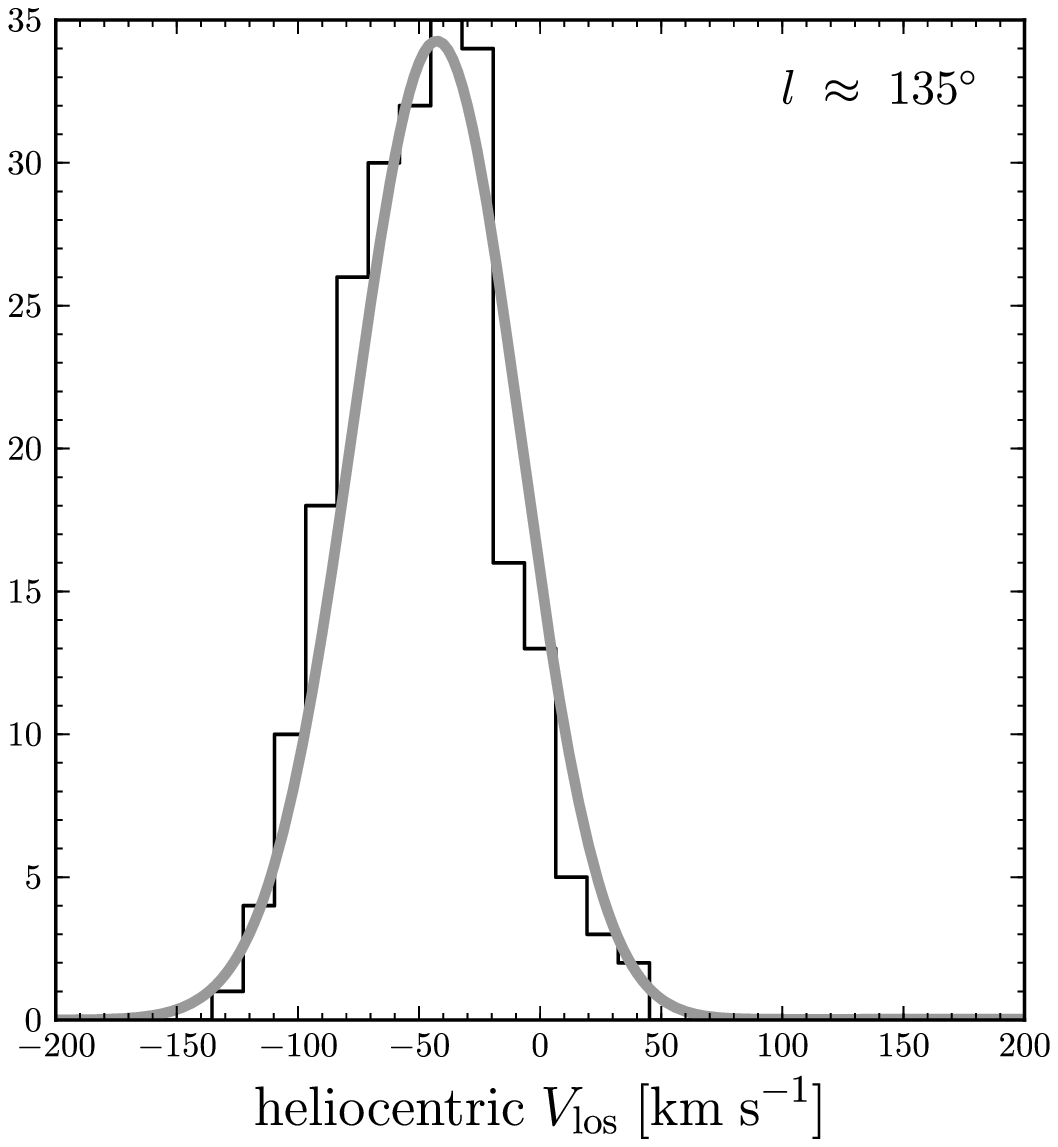}
\includegraphics[width=0.24\textwidth,clip=]{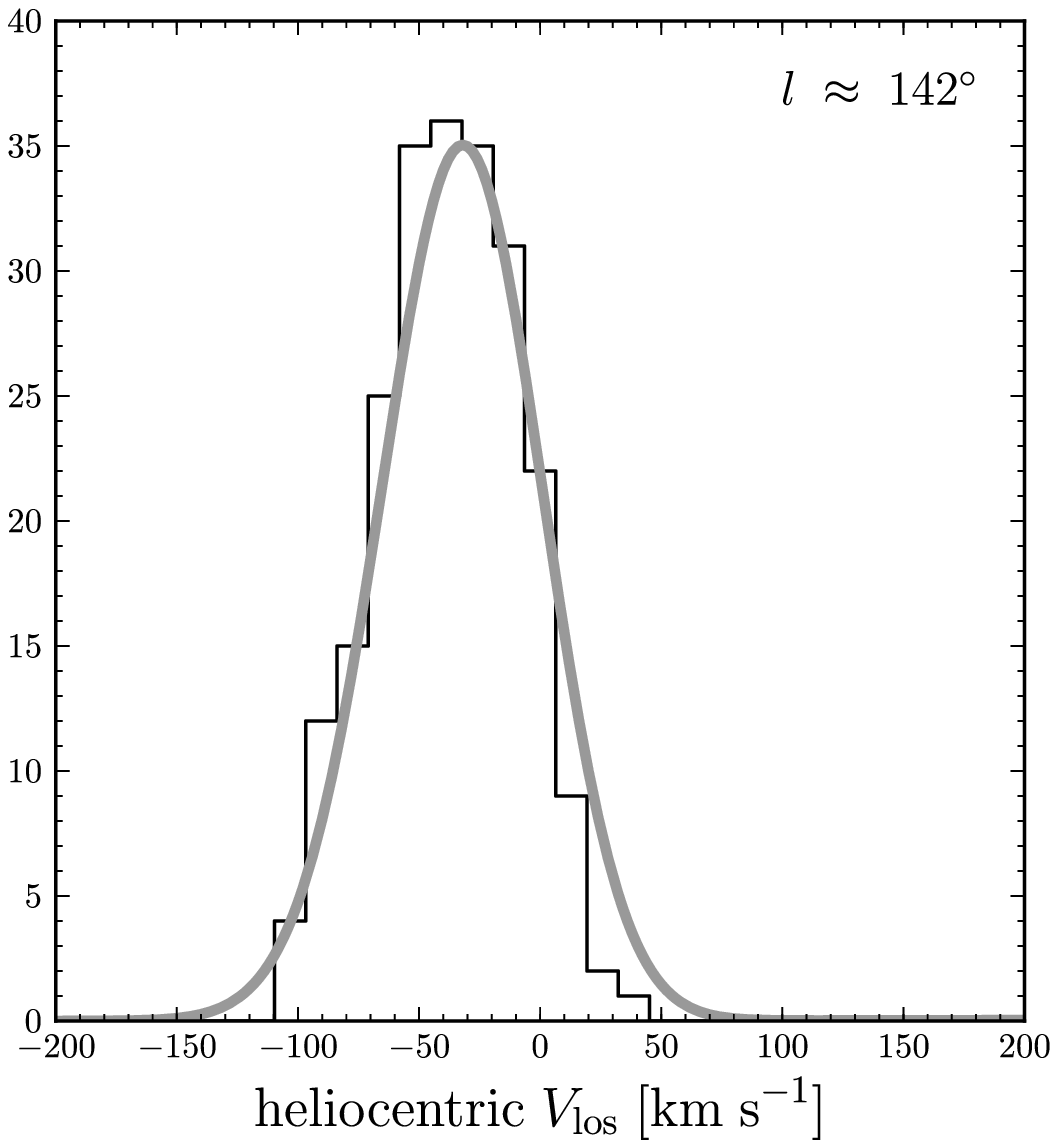}\\
\includegraphics[width=0.24\textwidth,clip=]{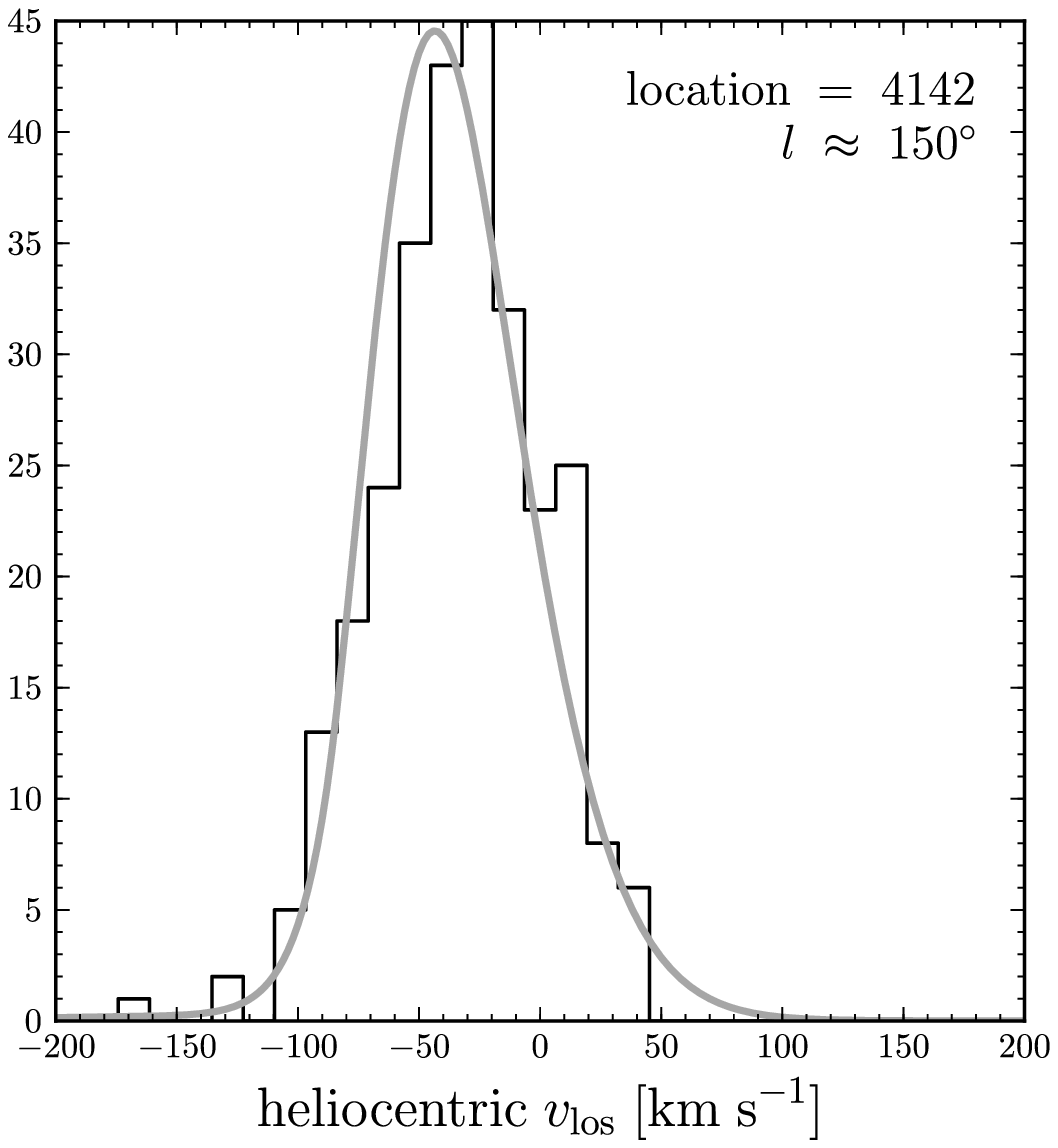}
\includegraphics[width=0.24\textwidth,clip=]{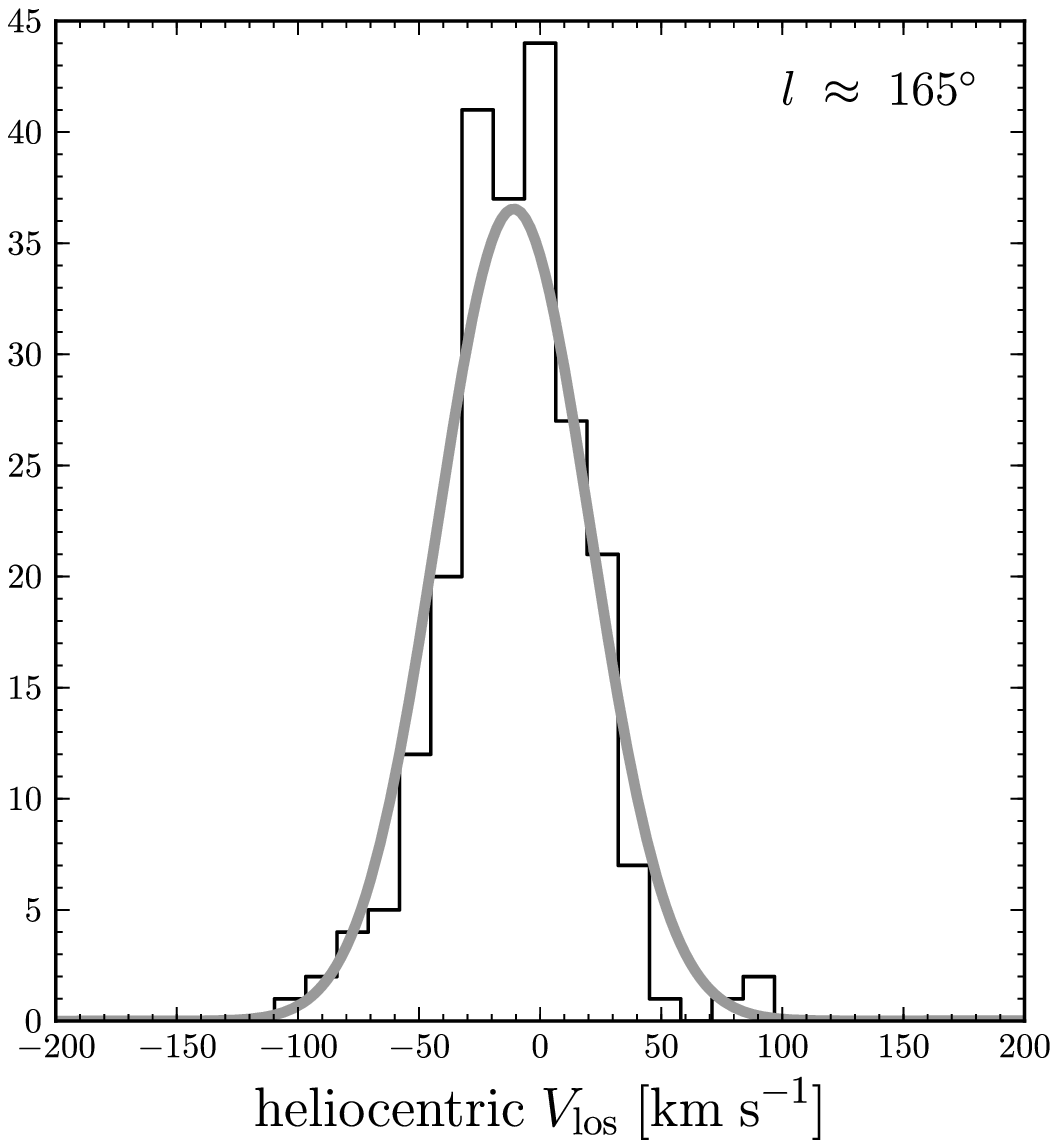}
\includegraphics[width=0.24\textwidth,clip=]{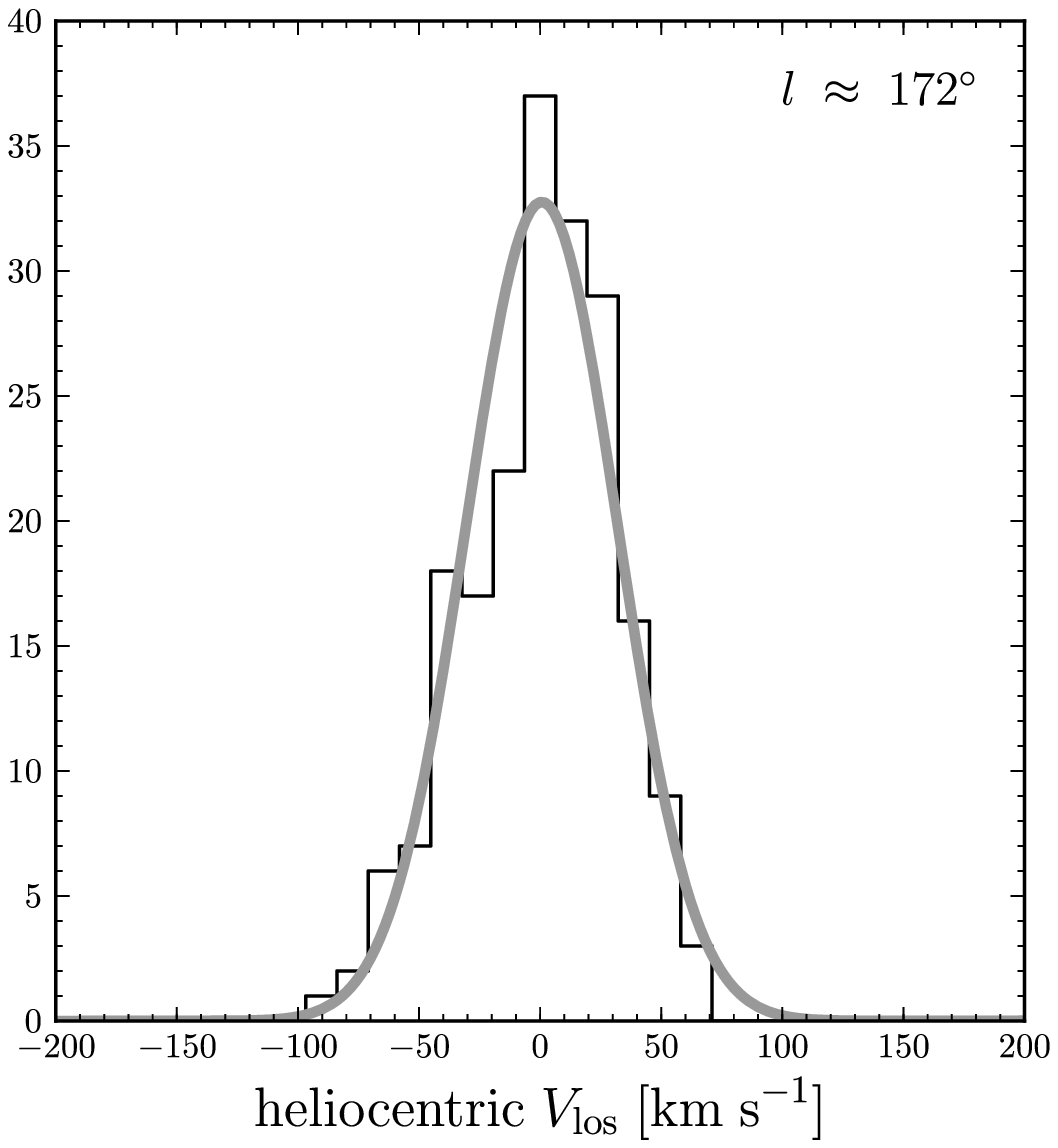}
\includegraphics[width=0.24\textwidth,clip=]{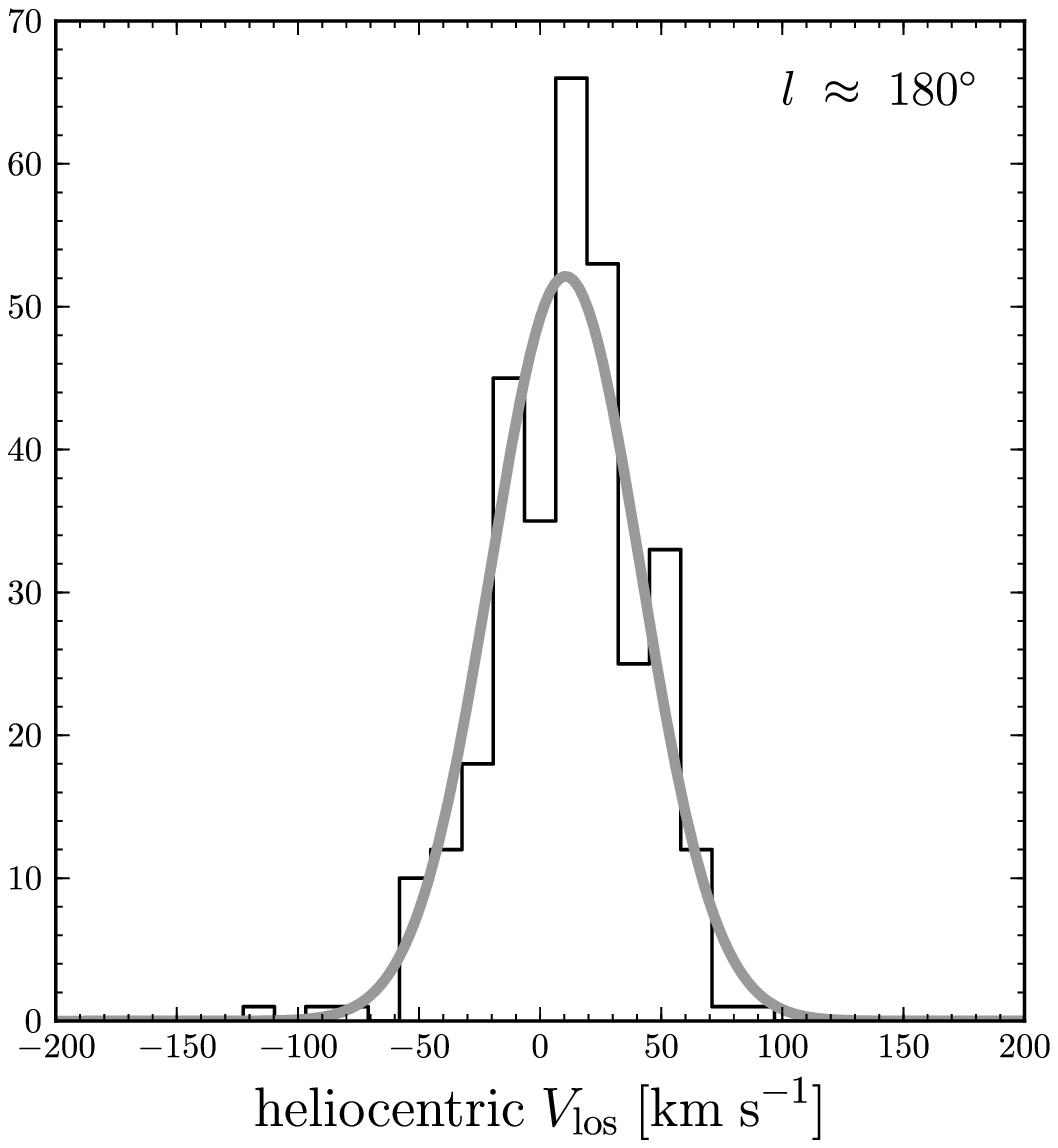}\\
\includegraphics[width=0.24\textwidth,clip=]{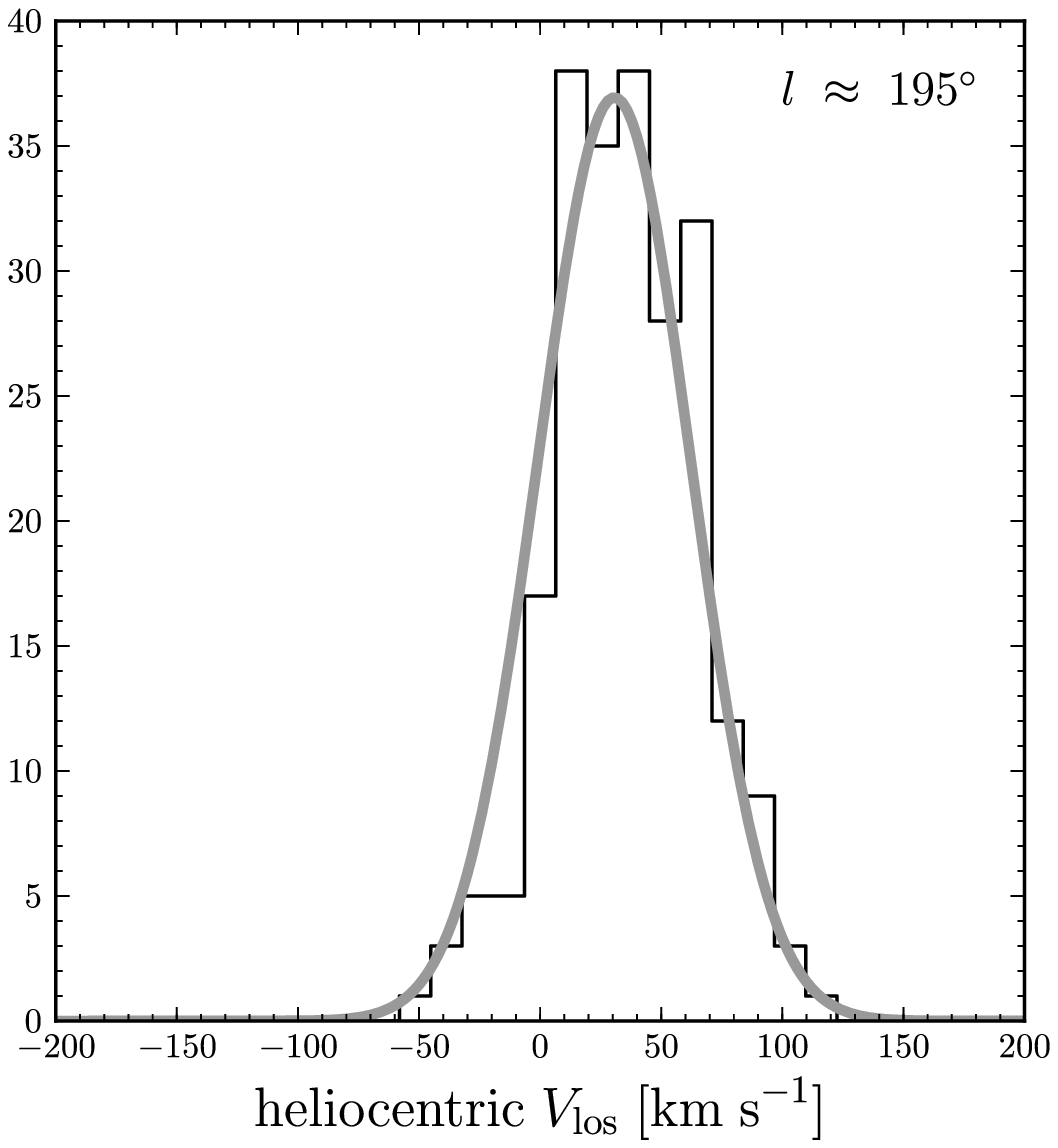}
\includegraphics[width=0.24\textwidth,clip=]{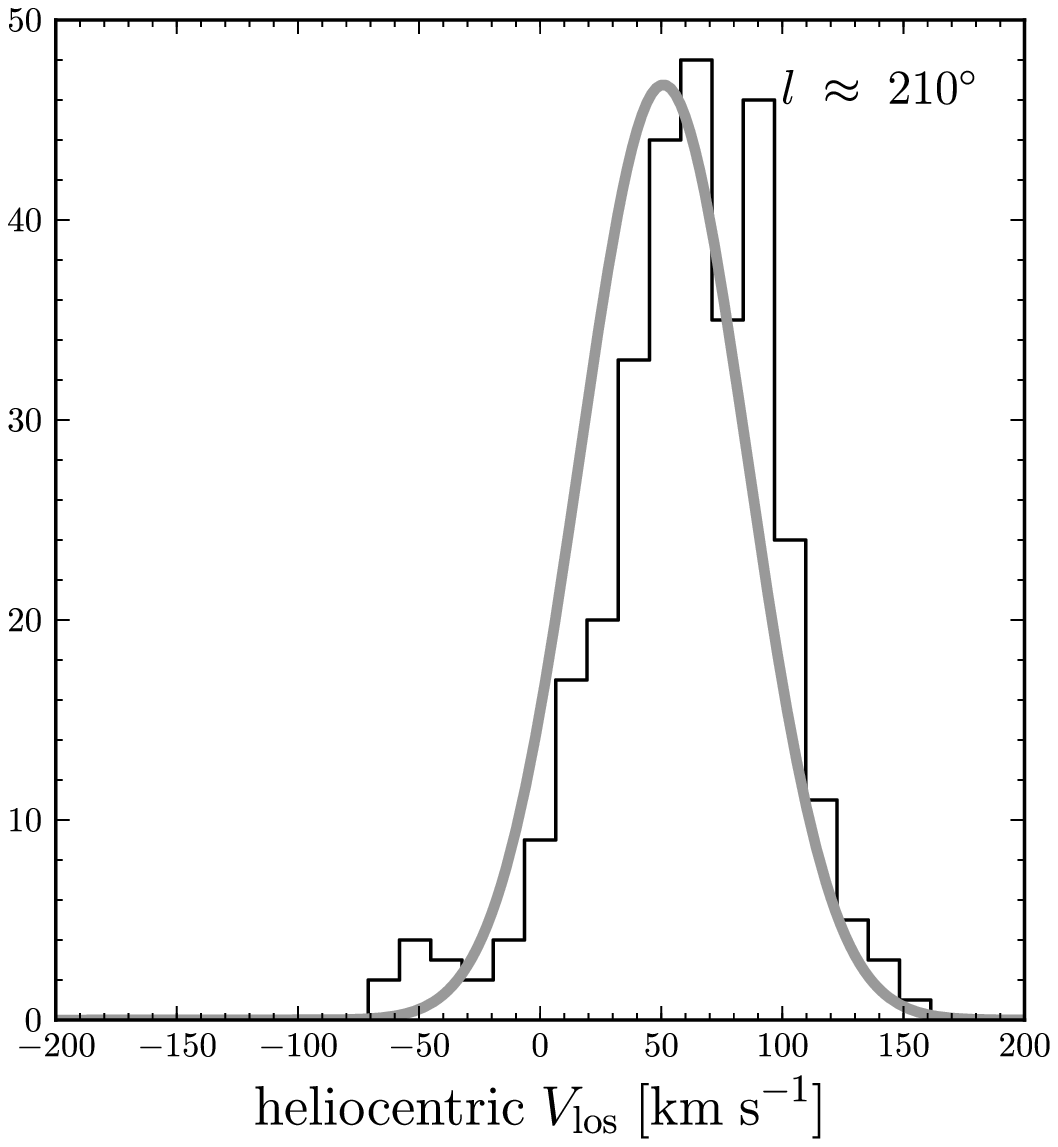}
\caption{Comparison between the full line-of-sight heliocentric
  velocity distributions of the data and the best-fit
  flat-rotation-curve model. Fields are ordered by increasing
  longitude. The agreement between the gray model curve and the data
  is good, except for some offsets in the mean of the distributions,
  as is also clear
  from \figurename~\ref{fig:bestfit}.}\label{fig:locations}
\end{figure*}

We compare the best-fit flat-rotation curve model with the data in
\figurenames~\ref{fig:bestfit} and
\ref{fig:locations}. \figurename~\ref{fig:bestfit} shows the raw
heliocentric velocities versus Galactic longitude for all of the stars
in the sample, as well as the mean for each field, and the prediction
for the mean from the best-fit model. This ``model mean'' is computed
by constructing the predicted line-of-sight velocity distribution
using
\eqnname~(\ref{eq:pvlosplate}) and then calculating its mean. The full
predicted line-of-sight velocity distribution for each field is shown
in \figurename~\ref{fig:locations} as the smooth curve, which is to be
compared with the histogrammed data points for each field. It is clear
from
\figurename~\ref{fig:locations} that the fits capture the overall
smooth features of the distribution of line-of-sight velocities in
each field, including any asymmetry in the distribution. However, it
is also obvious that some of the model predictions are shifted from
the observed histograms. This effect also appears in the comparison
between the data mean and model mean for each field in the lower panel
of
\figurename~\ref{fig:bestfit}. There appears to be a distinct pattern
in these residuals. Thus, \emph{the best-fit axisymmetric model fails
  to account for $\approx 10\,\kms$ smooth deviations in the velocity
  field}. These deviations are likely due to non-axisymmetric
  streaming motions, which are expected at this level of accuracy; We
  discuss this further in \sectionname~\ref{sec:discuss-solar}. With
  the exception of the $l=60^\circ$ field, the data seem to follow the
  smooth velocity distribution quite well, albeit somewhat shifted on
  average.

\subsection{Systematics}\label{sec:systematics}

The basic models presented in the previous section are potentially
subject to systematic uncertainties, which we discuss in this
section. These uncertainties are mainly related to the distances of
the stars in our sample, as these are obtained from somewhat uncertain
models of the color--magnitude distribution of giant stars in the
near-infrared $J,H,$ and $\ks$ bands. We discuss here the influence of
our assumed value for the scale length of the stars, of potential
systematic offsets in the preliminary \apogee\ metallicities used to
inform the photometric distances, of systematic distance
uncertainties, and of the impact of mis-estimated interstellar
extinction and absorption. We also discuss the effect of binary
contamination, of using a different set of stellar isochrones for
distance estimation, and of modeling the data using
many underlying populations of stars, with velocity dispersions
smoothly varying with age.

In our fits, we have simply assumed a value of $3\,\kpc$ for the
radial scale length, $h_R$, of the sample of stars used in the
analysis. This assumption is based on the fact that, locally, the
dominant solar-metallicity disk population has this scale length
\citep{Bovy12a}. The parameter $h_R$ enters our analysis in two ways,
the first being in the asymmetric-drift correction
(\eqnname~(\ref{eq:va})), and the second being in the prior on the
distance distribution (\eqnname~(\ref{eq:distprior})). The former of
these is the most important for determining $V_c(R)$. Changing $h_R$
to $2\,\kpc$---which is highly unlikely, given that such a short scale
length is not observed for any of the metal-rich, ``thin disk''
populations in the solar neighborhood \citep{Bovy12a}---increases the
best-fit value to $V_c = 223\,\kms$ for a model with a flat rotation
curve, and to $V_c = 220\,\kms$ for a power-law rotation
curve. Increasing $h_R$ to $4\,\kpc$ decreases the best-fit value to
$V_c(R_0) = 217\,\kms$. Similar changes happen for the mock data sets
described in
\appendixname~\ref{sec:appmock}. Therefore, we conclude that the
assumed radial scale length has a negligible influence on the inferred
$V_c$.

To determine the influence of systematic offsets in the metallicities
used in the photometric distances, we have performed fits allowing for
a single metallicity offset, $\Delta \feh$, for all of the stars in the
sample, for a model with a flat rotation curve. We find that the
best-fit $\Delta \feh = -0.15\pm 0.02\,\dex$, but the best-fit value
of $V_c$ is unchanged, as is its uncertainty (see
\figurename~\ref{fig:pdfs} for the $\Delta \feh$--$V_c$ PDF). Although
an offset of $-0.15\,\dex$ might seem large, this offset should not be
thought of as a measurement of $\feh$ for the stars in our sample, because
the photometric-distance PDFs, such as that in
\figurename~\ref{fig:imf_h_jk}, are relatively insensitive to large
changes in \feh, especially at the high-metallicity end. We have also
performed a fit with a different $\Delta \feh$ for inner ($l \leq
90^\circ$) and outer-disk fields. Such a fit finds $\Delta \feh =
-0.12\pm0.03\,\dex$ for the outer-disk fields, and $\Delta \feh =
0.70\pm0.06\,\dex$ for the inner-disk fields, with a best-fit $V_c =
220\,\kms$; see \figurename~\ref{fig:pdfs} for the full PDF of the
$\Delta \feh$ and $V_c$. As discussed previously, the
photometric-distance PDFs are largely insensitive to changes in \feh\
at high
\feh; the Padova isochrones also have an upper limit of $\feh_{\mathrm{max}} = 0.45\,\dex$, such that increasing a star's $\feh$ beyond this has no effect, because we then use $\feh_{\mathrm{max}}$. Therefore, even a change of $\Delta \feh = 0.7\,\dex$ has a
negligible influence on the inner, high-metallicity, disk fields. In
none of these fits is our inferred range for $R_0$ affected.

Similarly, we have performed fits allowing for a single
distance-modulus offset, $\Delta \mu$, to be applied to all of the
stars in our sample, or for a separate offset for inner- and
outer-disk fields. In the former case, we find $\Delta \mu =
-0.27\pm0.06$, corresponding to a $12\pm3$\% distance offset, without
any influence on the inferred value of $V_c$. In the inner/outer
offset fit, we find 23\% smaller distances at $l \leq 90^\circ$, and
20\% larger distances in the outer disk, again with a negligible
influence on the best-fit $V_c$ ($V_c = 221\,\kms$), but with a second
minimum around $V_c=190\,\kms$, and a wider $V_c$ PDF that is,
however, at $V_c < 235\,\kms$ at 99\% confidence. A more physical
effect is to allow for an offset in the extinction, $\Delta A_H$,
again using a single offset for the whole sample, or splitting the
sample into inner- and outer-disk fields. A single extinction offset
is well-constrained to be small: $\Delta A_H =
-0.01\pm0.01$. Similarly, the best-fit extinction offset in the outer
disk is $\Delta A_H = 0\pm0.02$, while in the inner disk we find some
evidence for mis-estimated extinction, with $\Delta A_H =
-0.2\pm0.03$. In both of these cases, the best-fit $V_c$ increases by
$1\,\kms$ or less, and the uncertainties remain approximately the same
(see \figurename~\ref{fig:pdfs}).

As discussed in \sectionname~\ref{sec:data}, we do not limit the
sample to stars for which multiple velocity epochs are available that
indicate that they are not part of a multiple system. Such a cut
significantly reduces the longitudinal coverage of our sample. When we
do apply this cut, we find that the best fits for $V_c$ and all other
parameters are unchanged from the results for our basic
flat-rotation-curve model in \sectionname~\ref{sec:basic}, albeit with
larger uncertainties. We find in particular that $V_c =
217\pm7\,\kms$, but the uncertainties in the position and velocity of
the Sun are much increased.

In \appendixname~\ref{sec:appdist}, we describe how we employ Padova
isochrones to estimate photometric distances to the stars in our
sample. When instead we use isochrones from the BaSTI library
\citep{Pietrinferni04a}, using the filter transformation from
\citet{Carpenter01a} to transform \ks\ to the $K$-band
filter used by BaSTI, we find that the results of our basic models
(both with a flat and a power-law rotation curve) are essentially
unchanged (with changes in $V_c$ of about $1\,\kms$).

To determine whether our results change when we fit the data as a mix
of multiple stellar populations, with velocity dispersions varying
from that of a young population to that of the oldest disk population,
we have performed a fit where the likelihood for $V_{\mathrm{los}}$ in
\eqnname~(\ref{eq:pvlos}) is generalized to
\begin{equation}\begin{split}
    p(& \vlos|l,b, (J-\ks)_0,H_0,\feh,\vc(R),\Ro,V_{R,\odot},V_{\phi,\odot},\iso) \\
    & = \sum_iP(i) \, p(\vlos|l,b,
      (J-\ks)_0,H_0,\feh,\vc(R),\Ro,\\
& \qquad \ \ \ \ \qquad \qquad \ \ V_{R,\odot},V_{\phi,\odot},\df_i,\iso) ,
\end{split}\end{equation}
where $\df_i$ is the single-Gaussian population model used before,
with dispersion at age $\tau_i$
\begin{equation}\label{eq:multivdisp}
\sigma_{R,i} = \sigma_{R,0}\,\left(\frac{\tau_i + 0.1\,\mathrm{Gyr}}{10.1\,\mathrm{Gyr}}\right)^{0.38}\,
\end{equation}
and using an exponentially-declining star-formation history, such that
$P(i) \propto e^{\tau_i/(8\,\mathrm{Gyr})}$. This model follows the
fit of \citet{Aumer09a} in the solar neighborhood. We use a mix of 20
equally age-spaced populations between $1$ and $10\,\mathrm{Gyr}$, and
find $V_c = 216\pm7\,\kms$, showing that this alternative model does
not change our results; the full $V_c$--$R_0$ PDF is shown in
\figurename~\ref{fig:pdfs}. This model has $\sigma_{R,0} = 38\pm2\,\kms$,
as expected for the old, ``thin disk'' population.

\section{Discussion}\label{sec:discussion}

\subsection{Effect of Non-Axisymmetric Features}\label{sec:discuss-nonaxi}

\begin{figure*}[hbtp]
\begin{center}
\includegraphics[width=0.75\textwidth]{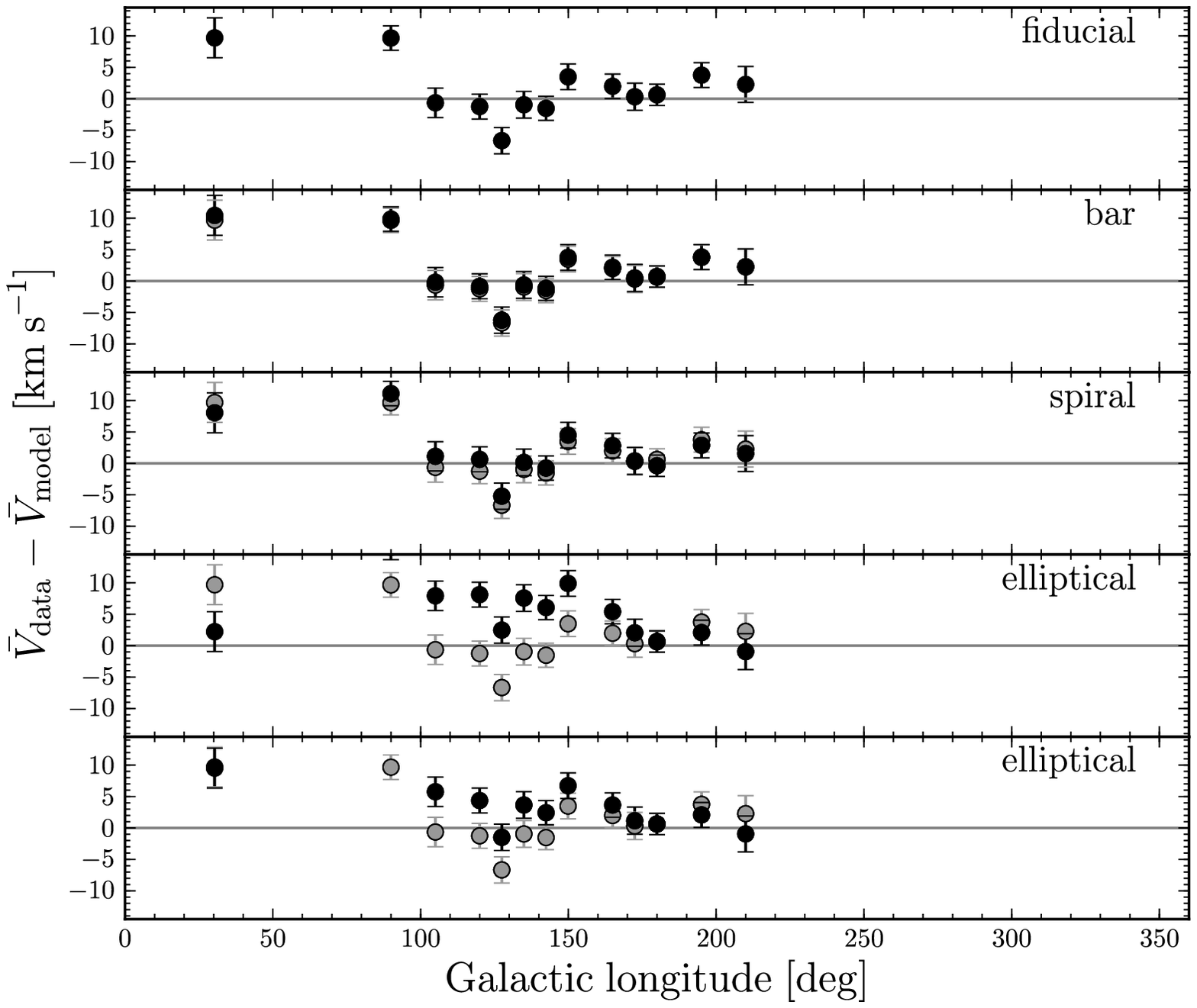}
\caption{Influence of non-axisymmetry on the predicted mean of the
  $V_{\mathrm{los}}$ distributions. The flat-rotation curve model
  of \tablename~\ref{table:results} and \figurename
  s~\ref{fig:bestfit} and \ref{fig:locations} is displayed in the
  uppermost panel. Non-axisymmetric models shown are: the model for
  the Milky Way bar of \citet{Dehnen00a}; a two-armed logarithmic
  spiral with a fractional amplitude of $1\%$ and a pitch angle of
  $-15^\circ$; elliptical disk models of \citet{Kuijken94a} with an
  amplitude of $5\%$ and position angles of $0^\circ$ (fourth panel)
  and $-45^\circ$ (bottom panel). All models are calculated for a warm
  disk population with $\sigma_{R}(R_0)=44\,\kms$ and are
  adiabatically grown in an initially equilibrium, axisymmetric
  disk. While the influence of bar and spiral structure perturbations
  is small, an elliptical disk model can reduce the size of the
  residuals. This figure only shows the influence of non-axisymmetry
  on the mean of the line-of-sight velocities in each field; much more
  information is of course contained in the full distribution of
  line-of-sight velocities.}\label{fig:nonaxi}
\end{center}
\end{figure*}

In the fits in \sectionname~\ref{sec:results}, we have assumed
axisymmetric models for the velocity field in the Galactic
disk. However, the Milky Way is known to have some non-axisymmetric
structures, such as the central bar \citep{Blitz91a,Binney91a}, spiral
arms \citep[\eg,][]{Drimmel01a}, and a potentially triaxial
halo \citep{Law09a}, that may influence the measurement performed in
this paper. Because we use a warm disk population as dynamical
tracers, we can expect the influence of any non-axisymmetry to be
smaller than for the colder gas tracers used in other
studies \citep[\eg,][]{Levine08a}, because warm tracers respond less
strongly to non-axisymmetric perturbations to the Milky Way potential
than cold tracers \citep{Lin69a}.

We have calculated the velocity field for a warm tracer population in
various non-axisymmetric models (see J.~Bovy, 2013, in preparation, for
full details). We follow the procedure of \citet{Dehnen00a} and
\citet{Bovy10a} to calculate the velocity distribution at a given
position for a given non-axisymmetric model for the disk, by
performing backward orbit integrations until before the onset of the
non-axisymmetric perturbation in an axisymmetric disk. At that time, the
distribution function of the pre-existing axisymmetric disk is
evaluated. By the conservation of phase-space density, this
probability is equal to the current probability of the phase-space
starting point of the orbit integration. We assume a flat rotation
curve for the axisymmetric part of the potential, and we model the
initial, axisymmetric distribution function as a Dehnen distribution
function (\eqnname~(\ref{eq:fdehnen})) with a velocity dispersion of
$0.2\,V_c$, a radial scale length of $R_0/3$ and a dispersion scale
length of $R_0$. These parameters are close to the best-fit parameters
for the \apogee\ tracer population.

We calculate the mean velocity field for the bar model of
\citet{Dehnen00a}, and use it to construct the mean, non-axisymmetric
$V_{\mathrm{los}}$ field. This field is then applied as part of the
model, and the result for the mean line-of-sight velocity in each
field is shown in \figurename~\ref{fig:nonaxi}---there is of course
much more discretionary power in the full distribution functions. In
the same manner, we calculate the mean $V_{\mathrm{los}}$ field for a
two-armed logarithmic spiral with a fractional potential-amplitude of
1\% of the background, axisymmetric potential, a pitch angle of
$-15^\circ$, an angular frequency of $0.65\,\Omega_0$ (placing the Sun
near the $4$:$1$ inner Lindblad resonance), and an angle between the
Sun--Galactic-center line and the line connecting the peak of the
spiral pattern at the solar radius of $20^\circ$. We do the same for
models with a flat elliptical (stationary, $m=2$) distortion to the
potential (see \citealt{Kuijken94a}), with a fractional amplitude of
$5\%$ and position angles of $0^\circ$ and $-45^\circ$ with respect
the the Sun--Galactic center line. All of these perturbations were
adiabatically grown. We see in
\figurename~\ref{fig:nonaxi} that the influence of the bar and spiral
structure is negligible in the mean $V_{\mathrm{los}}$ velocity
field. An elliptical, $m=2$, perturbation could distort the mean
velocity field in a way that would affect our analysis. Naively, the
model in the bottom panel of \figurename~\ref{fig:nonaxi}
significantly reduces the residuals between the best-fit
$\bar{V}_{\mathrm{los}}$ and the data $\bar{V}_{\mathrm{los}}$.

It is clear that the data used in this paper could be used to
constrain non-axisymmetric perturbations to the axisymmetric potential
assumed here. We defer a full treatment of this to a subsequent paper.

\subsection{Comparison with Other Determinations of the Local Circular
  Velocity}\label{sec:discuss-compare}

There have been many previous determinations of the local circular
velocity, and we discuss here how our new measurement compares to
these. Determinations of the local circular velocity can be roughly
divided into two groups: measurements of the Sun's velocity with
respect to an object or population assumed to be at rest with respect
to the Galactic center (\eg, Sgr A$^*$ or a population of halo
objects), or direct measurements of the local radial force (\eg, by
determining the Oort constants or the orbit of a stream of stars). The
former directly measure the Sun's Galactocentric velocity,
$V_{\phi,\odot}$, but must assume a value for the Sun's motion with
respect to the circular orbit at $R_0$ to arrive at $V_c(R_0)$. The
most direct measurement of this kind is the combination of the
precisely measured proper motion of Sgr A$^*$ \citep{Reid04a} with the
distance to the Galactic center determined from the Keplerian orbits
of S stars in the innermost parsec \citep{Ghez08a,Gillessen09a}. These
measurements give $R_0 \approx 8\,\kpc$, which combined with the
proper motion of Sgr A$^*$ of $30.24\,\kms\kpc^{-1}$, yields a total
solar velocity of $242\,\kms$. As noted before, this result is
entirely consistent with our inferred value of the angular motion of
the Galactic center, and with our $V_{\phi,\odot} =
242^{+10}_{-3}\,\kms$. The discrepancy between the value of
$V_c=229\,\kms$ in \citet{Ghez08a} (or higher measurements also based
on the proper motion of Sgr A$^*$) and our $V_c = 218\,\kms$ is
therefore \emph{entirely due to our different value for the solar
velocity with respect to $V_c$}, and intrinsically these measurements
are consistent. Similarly, the recent measurements of $V_{\phi,\odot}
= 276\pm23\,\kms$ and $V_{\phi,\odot} = 244\pm14\,\kms$ using the
kinematics of the Sgr stream \citep{Carlin12a} are consistent with our
measurement of the solar velocity.

Measurements based on the Oort constants
\citep[\eg,][]{Feast97a}, or the dynamics of a cold stellar stream
\citep{Koposov10a}, also indicate that $V_c\!\sim\!220\,\kms$. The
measurement of the Oort constants from \emph{Hipparcos} proper motions
by \citet{Feast97a} directly measure the angular frequency of the
local circular orbit, independent (in principle) from the solar
motion: $\Omega_0 = V_c/R_0 = 27.2\pm0.9\,\kms\kpc^{-1}$. This value
is consistent with our determination (see
\tablename~\ref{table:results}) $\Omega_0 =
27.0^{+0.3}_{-3.5}\,\kms\kpc^{-1}$. The determination of $V_c =
221\pm18\,\kms$ of \citet{Koposov10a} from fitting an orbit to the
cold GD-1 stellar stream at $R\!\sim\!10\,\kpc$ is clearly consistent
with our measurement, although note that they assumed the
$V_{\phi,\odot} - V_c = 5.25\,\kms$ value for the solar motion
from \citet{Dehnen98a}. 

The recent measurement of $V_c = 254\pm16\,\kms$, from the kinematics
of masers in the Galactic disk \citep{Reid09a}, was shown to have an
overly optimistic precision by
\citet{Bovy09a}, who concluded from a more general model for the
distribution function of the masers that these data only imply $V_c =
246\pm30\,\kms$. The measurement of $V_c$ from the maser kinematics
also assumes that $V_{\phi,\odot}$ is given by $V_c$ plus the
locally-determined solar motion of $\sim\!10\,\kms$. Apart from
indicating a high value for $V_c$, albeit with a large uncertainty,
the masers were also found to be lagging with respect to circular
motion by about $15\,\kms$, a large offset for a young and relatively
cold tracer population. This offset cannot be explained by the
asymmetric drift and it is physically implausible since it requires
the masers to have more eccentric orbits than most of the young
stars \citep{Mcmillan10a}. Requiring the masers to be on circular
orbits assuming a flat rotation curve leads to a solar motion of
$V_{\phi,\odot} - V_c = 18.6\pm2.4\,\kms$ and $V_c =
232\pm24$ \citep{Mcmillan10a}, in good agreement with our
measurements. Our measurement, however, relies on the theoretically
well-motivated asymmetric-drift correction that is a direct
consequence of the collisionless Boltzmann equation.

Therefore, we conclude that our measurement of $V_c$ is consistent
with previously reported values. Compared to other determinations, our
measurement of $V_c$ is one of the most highly precise, and unlike
many other measurements, it does not require assumptions about the
relation between the solar motion and $V_c$.

\begin{figure}[htbp]
\includegraphics[width=0.5\textwidth]{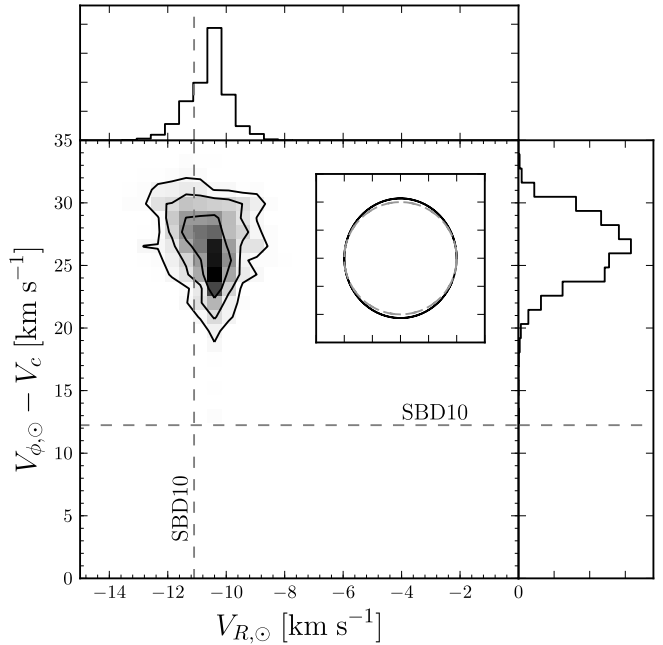}
\caption{Sun's peculiar velocity with respect to circular motion. This
  figure shows the posterior probability distribution function of the
  Sun's velocity with respect to the circular orbit at the solar
  radius. The measurement of the Sun's motion with respect to the
  orbit of a zero velocity dispersion population
  from \citeauthor{Schoenrich10a} (\citeyear{Schoenrich10a}; SBD10) is
  indicated by dashed lines. The fact that our measurement of the
  Sun's rotational peculiar velocity does not agree with the SBD10
  value may indicate that the closed orbit at the solar radius is not
  circular, due to non-axisymmetry at the level of 10 km s$^{-1}$. The
  inset shows this non-circular closed orbit (spatially to scale) if
  the non-axisymmetry is due to ellipticity of the
  disk \citep[\eg,][]{Kuijken94a}; the dashed gray curve is the
  circular orbit. Alternatively, the locally-determined solar motion
  may be off by $\sim\!10\,\kms$.}\label{fig:vpec}
\end{figure}

Our data \emph{strongly} rule out that the Milky Way's circular
velocity at the solar radius is $> 235\,\kms$. Marginalizing over all
of the systematics discussed in \sectionname~\ref{sec:systematics},
$V_c < 235\,\kms$ at $> 99\%$ confidence. Fixing $V_c$ to $250\,\kms$
with $R_0 = 8\,\kpc$ or $8.4\,\kpc$ leads to a best-fit with much
larger $\chi^2$: $\Delta \chi^2 = 47$ ($\sim\!5\sigma$) and $\Delta
\chi^2 = 34$ ($\sim\!3.7\sigma$), respectively.

Our measurement that the Milky Way's rotation curve is essentially
flat over $4\,\kpc < R < 14\,\kpc$ agrees with measurements based on
the kinematics of HI emission \citep[\eg,][]{GKT79,Merrifield92a} and
with the local measurement of the Oort constants \citep{Feast97a}.

\subsection{Implications for the Motion of the LSR}\label{sec:discuss-solar}

We found in \sectionname~\ref{sec:results}, in both the flat- and
power-law-rotation-curve models, that the Sun's velocity with respect
to the center of the Galaxy---a distinct parameter from $V_c$ in our
fit---is larger than $V_c(R_0)$ by $\sim\!24\,\kms$. Defining the
Rotational Standard of Rest (RSR; \citealt{Shuter82a}) as the circular
velocity in the axisymmetric approximation to the full potential, this
solar motion with respect to the RSR is much larger than that measured
by applying Str\"{o}mberg's asymmetric-drift relation to local samples
of stars. The asymmetric-drift relation is used to estimate the
velocity of the zero-dispersion orbit by extrapolating the relation
given in \eqnname~(\ref{eq:va}) from warmer samples of stars to
$\sigma_R = 0$; it is this zero-dispersion orbit that is typically
what is meant by the LSR \citep{Fich91a}. Such analyses of the
\emph{Hipparcos} data yield a solar motion with respect to the LSR
that is somewhere between $5$ and $13\,\kms$
\citep{Dehnen98a,Hogg05a,Schoenrich10a}; $13\,\kms$ is also the offset
of the average rotational velocity of nearby stars
\citep[\eg,][]{AllendePrieto04a}. The discrepancy between our
determination of the solar velocity with respect to the RSR, and that
with respect to the LSR, is clearly shown in
\figurename~\ref{fig:vpec}, where the PDF for
$V_{\phi,\odot}-V_c,V_{R,\odot}$ is shown for the flat-rotation curve
model, together with the locally-measured motion with respect to the
LSR of \citet{Schoenrich10a}. The analogous figure for the power-law
rotation-curve fit has an essentially identical appearance. It is
clear that the locally-measured solar motion $V_{\phi,\odot}$ is not
equal to our globally-measured solar motion to very high
significance. Or, equivalently, the motion of the RSR is not the same
as that of the LSR: the LSR seems to rotate $\sim\!12\,\kms$ faster
than the RSR.

This discrepancy may result from a breakdown in the assumptions of the
locally-measured solar motion using Str\"{o}mberg's asymmetric-drift
relation. As convincingly shown by \citet{Schoenrich10a}, the neglect
of the radial metallicity gradient in the analysis of
\citet{Dehnen98a} leads to a correction of $\sim\!7\,\kms$, but
further improvements in the chemical-evolution model of
\citet{Schoenrich10a} might lead to further corrections. The local
velocity distribution has also long been known to contain various
streams or ``moving groups''
\citep[\eg,][]{Kapteyn05a,Schwarzschild07a,Eggen86a,Dehnen98b,Bovy09b}
that are likely of a dynamical origin
\citep{Bensby07a,Famaey08a,Bovy10b,Sellwood10a} and contain a
significant fraction of the nearby stars \citep{Bovy09b}. These moving
groups make an accurate determination of the solar motion challenging.
As such, the locally-determined correction for the Sun's motion with
respect to the LSR may well be incorrect by $\sim\!5\,\kms$ or more. A
solar motion of $\sim\!24\,\kms$ rather than $\sim\!12\,\kms$ is not
much more unlikely given the distribution function for a
$\sim\!5\,\mathrm{Gyr}$ population: using a Dehnen distribution
function (\equationname~(\ref{eq:fdehnen})) with a radial dispersion
of $30\,\kms$ (\equationname~(\ref{eq:multivdisp})) and assuming our
best-fit model gives $P(V_{\phi,\odot}-V_c > 12\,\kms) \approx 0.2$
and $P(V_{\phi,\odot}-V_c > 24\,\kms) \approx 0.1$.

However, if Str\"{o}mberg's relation leads to a solar motion with
respect to the LSR that is different from the globally-determined
solar motion with respect to the RSR, the most straight-forward
interpretation is that LSR's orbit is not circular, but deviates from
circularity by about $10\,\kms$ because of large-scale,
non-axisymmetric streaming motions. The deviation cannot be much
larger than this value, as the
\emph{radial} component of the solar motion as determined from local
samples agrees with our global measurement. As our measurement of the
circular velocity is mainly at azimuths $\phi\lesssim45^\circ$, this
$10\,\kms$ deviation must happen over this angular scale.

There is mounting evidence for the existence of streaming motions in
the Galactic disk of this magnitude. There has been a long-standing
discrepancy of $\sim\!7\,\kms$ between the rotation curves determined
from first ($l < 90^\circ$) and fourth ($l > 270^\circ$) quadrant
tangent-point measurements \citep[\eg,][]{Kerr62a,GKT79,Levine08a},
which could indicate streaming motions. Similarly, the analysis of the
extreme line-of-sight velocity of HI emission toward
$l\!\sim\!90^\circ$ leads to differences in $V_c$ of approximately
$30\,\kms$, albeit with large uncertainties
\citep{KTG79,Jackson81a,Jackson85a}. The fact that the maximum
line-of-sight velocity in both CO and 21-cm data reaches zero at
$l\!\sim\!70^\circ$, rather than at $l = 90^\circ$, has been used to
argue that the LSR moves ahead of $V_c$ with a speed of
$\sim\!7.5\,\kms$ \citep{Shuter82a,Clemens85a}, similar to the offset
we find. More recently, an analysis of line-of-sight velocities from
the \emph{RAVE} has found a gradient, $\dd \bar{V}_R / \dd R$, in the
mean radial velocity, $\bar{V}_R$, of $\sim\!3\,\kms\,\kpc^{-1}$, and
large streaming motions in both $\bar{V}_R$ and the mean tangential
velocity, $\bar{V}_\phi$, within a few $\kpc$ from the
Sun \citep{Siebert11a}.

As a simple exploration of this possibility, we have computed the
closed orbit at $R_0$ in a model for the Milky Way disk where this
streaming motion is due to ellipticity of the disk, following
\citet{Kuijken94a}. We use a flat rotation curve, with a
$\cos(2\,\phi)$ perturbation of constant ellipticity having an
amplitude of $14\,\kms$. Such a perturbation could arise if, for
example, the halo is triaxial \citep[\eg,][]{Law09a}. The closed orbit
is shown to scale in the inset in
\figurename~\ref{fig:vpec}; it has an eccentricity of 0.06.

Regardless of the origin of the discrepancy between our
globally-measured solar motion and the locally-measured value, our
larger preferred value of $V_{\phi,\odot} - V_c$ means that the Sun is
likely closer to the pericenter of its orbit around the Galactic
center than previously believed. Revising $V_{\phi,\odot} - V_c$
upward by $12\,\kms$ increases the eccentricity of the Sun's orbit to
$\sim\!0.1$ from $\sim\!0.06$ and increases its mean Galactocentric
radius to $\sim\!8.9\,\kpc$ from $\sim\!8.5\,\kpc$. Such a large mean
radius increases the tension between the Sun's high metallicity
compared to local stars \citep{Wielen96a}, although the peak of the
local metallicity distribution may be closer to solar metallicity than
previously believed \citep[\eg,][]{Casagrande11a}.

Finally, we address what our results imply for dark-matter
direct-detection experiments, and for correcting the motion of
Galactic and extra-galactic objects for the motion of the Sun. Both of
these applications essentially require knowledge of the \emph{total}
Galactocentric solar velocity, $V_{\phi,\odot}$, to transform
velocities into the Galactocentric (dark-matter-halo) rest
frame\footnote{The one exception is that in the Standard Halo Model,
the velocity dispersion of the isothermal halo is given by
$V_c/\sqrt{2}$. However, the velocity dispersion of the halo is better
determined directly from observations than by using this
assumption.}. Although this motion is often decomposed as $V_c$ plus
the Sun's motion with respect to $V_c$, our results in this paper show
that this is unnecessarily dangerous, as it requires the \emph{strong
assumption} that the LSR is on a circular orbit, such that
$V_{\phi,\odot} = V_c + V_{\phi,\odot,\mathrm{LSR}}$, where the final
term is the solar velocity with respect to the LSR. Our results in
\sectionname~\ref{sec:results} strongly rule out that this assumption
holds for the currently accepted solar motion of
$\sim\!12\,\kms$. However, this approach is entirely unnecessary for
correcting velocities to the Galactocentric rest frame, as the proper
motion of Sgr A$^*$ combined with the ``clean'' estimate of $R_0$ from
Galactic-center dynamics shows that $V_{\phi,\odot} \approx
242\,\kms$, which agrees with our measured value of
$V_{\phi,\odot}$. Adopting a standard $V_{\phi,\odot} = 242\,\kms$ for
correcting velocities for the Sun's motion, and decoupling this
correction from the question of the true value of $V_c$, would
therefore be helpful.

\subsection{Implications for the Mass of the Milky Way}\label{sec:discuss-mass}

We can use our measurement of the inner rotation curve of the Milky
Way to estimate the Milky Way's total dark-halo mass. We combine the
measurements in this paper with those of the Milky Way's outer
rotation curve ($R \gtrsim 20\,\kpc$) of \citet{Xue08a} (specifically,
taking the measurements using simulation II in Table 3 of that
paper). Based on \figurename~\ref{fig:rotcurve} and
\tablename~\ref{table:results}, we add our measurement as a flat
rotation curve with $V_c = 218\,\kms$, evaluated at three radii with
uncertainties of $3\,\kms$, $6\,\kms$, and $10\,\kms$ at $R=4\,\kpc$,
$R=8\,\kpc$, and $R=12\,\kpc$, respectively (removing the measurements
at $R=7.5\,\kpc$ and $R=12.5\,\kpc$ from \citealt{Xue08a}). We also
use our recent measurements of the local density of dark matter,
$\rho_{\mathrm{DM}} = 0.008\pm0.003\,M_\odot\,\mathrm{pc}^{-3}$
\citep{Bovy12d}, and of the disk scale length, $h_R = 3.25\pm0.25\,\kpc$
\citep{Bovy12c}, revising the measurement from the latter paper
downward to correct for the influence of the dark halo on that
measurement.

We model the Milky Way's potential as the combination of a bulge
component with a Hernquist profile with a scale length of
$600\,\mathrm{pc}$, a disk component with a scale height of
$300\,\mathrm{pc}$ \citep{Bovy12b} and a scale length that is a free
parameter, and a Navarro-Frenk-White halo (NFW; \citealt{Navarro97a}),
with a scale radius that is a free parameter (\ie, without constraint
on the concentration). The relative contributions of these three
components are free parameters. We then fit the data given in the
previous paragraph, and we find that $M_{\mathrm{halo}} =
8^{+8}_{-2}\times10^{11}\,M_\odot$. The total mass of the Milky Way is
$\sim\!8.5\times10^{11}\,M_\odot$.

This ``low'' estimate for the mass of the Milky Way likely makes it
less massive than M31, which has an estimated dark-halo mass of
$14\times10^{11}\,M_\odot$ \citep{Watkins10a}. This measurement leads
to a combined mass for the Local Group of
$\sim\!2.4\times10^{12}\,M_\odot$, which is consistent with the
Local-Group-timing-argument within the large cosmic scatter
\citep{vdMarel12a}. Our measurement of $V_c = 218\pm6\,\kms$ combined with its estimated $I$-band magnitude of $-22.3$ mag makes the Milky Way underluminous  with respect to the Tully-Fisher relation of external spiral galaxies by about $1\sigma$ \citep[see][]{Flynn06a}.

\section{Conclusion}\label{sec:conclusion}

In this paper, we have measured the Milky Way's rotation curve over
the range $4\,\kpc < R < 14\,\kpc$ from the new \apogee\ data set of
kinematically-warm stellar tracers at large distances from the
Sun. Our measurement is not ``clean'', in the sense of being a
geometric measurement such as that provided by tangent-point
observations of HI emission or by a measurement of the Sun's motion
relative to an object assumed to be at rest with respect to the
Galactic center. Because we use a warm stellar population, we must
correct for the offset between the average rotational velocity of this
population and the circular velocity---the asymmetric drift---using a
dynamical Jeans model. However, the fact that our measurement uses a
dynamical effect, rather than being purely geometric, has the
advantage that we unambiguously measure $V_c(R)$ in the sense of the
radial force at $R$, and that we can measure the solid-body-rotation
contribution to the rotation curve, in contrast to HI measurements.

Our main results are discussed in \sectionname~\ref{sec:results}. We
find that the Milky Way's rotation curve is approximately flat over
$4\,\kpc < R < 14\,\kpc$, with $V_c(R_0) = 218 \pm
6\,\kms$. \tablename~\ref{table:results} summarizes our results, and
provides some alternative representations, such as the Oort constants,
the local rotational frequency, and the contribution of the
non-flatness of the rotation curve to a cylindrical-Poisson-equation
determination of the local surface-mass and dark-matter densities. We
simultaneously measure the Sun's velocity in the Galactocentric rest
frame---these are independent free parameters in our model---and find
that $V_{R,\odot} = -10 \pm 1\,\kms$ and the angular motion of the
Galactic center $\mu = 6.3^{+0.1}_{-0.7}\,\mathrm{mas\
  yr}^{-1}$. These values are consistent with previous studies. Our
measurement of $V_{\phi,\odot} = 242^{+10}_{-3}\,\kms$ then leads to a
solar offset from $V_c$ that is larger than the locally-measured value
\citep[$\!\sim\!12\,\kms$;][]{Schoenrich10a} by $14 \pm 3\,\kms$. This result
may indicate that the challenging local measurement of the solar
motion is incorrect or that the solar velocity is influenced by
non-axisymmetric streaming motions that put the LSR on a non-circular
orbit.

Looking forward, we expect to accurately measure the atmospheric
parameters and abundances other than \feh\ from the high-resolution
\apogee\ spectra in the near future. This will improve the
measurements described in this paper by (a) providing better
dwarf/giant separation, (b) allowing for more direct and more precise
distances to be derived, and (c) letting the analysis be conducted on
different chemically-defined populations of stars. In particular, this
will allow for a more detailed investigation into the signatures of
non-axisymmetric dynamics in our data.

The data upon which the measurements presented in this paper are based
will be released as part of \sdssiii's Data Release 10 in the summer
of 2013.

\acknowledgements It is a pleasure to thank Scott Tremaine for many
illuminating discussions about this work. We also thank Jennifer
Johnson, Mark Reid, Hans-Walter Rix, and Greg Stinson for helpful
comments. J.B. was supported by NASA through Hubble Fellowship grant
HST-HF-51285.01 from the Space Telescope Science Institute, which is
operated by the Association of Universities for Research in Astronomy,
Incorporated, under NASA contract NAS5-26555.  J.B. was partially
supported by SFB 881 funded by the German Research Foundation DFG and
is grateful to the Max-Planck Institut f\"ur Astronomie for its
hospitality during part of the period during which this research was
performed. T.C.B. acknowledges partial support by grants PHY 02-16783
and PHY 08-22648: Physics Frontiers Center/Joint Institute for Nuclear
Astrophysics (JINA), awarded by the U.S. National Science Foundation.
This research made use of the \emph{emcee} MCMC sampler \citep{FM12a}.

This publication makes use of data products from the Two Micron All
Sky Survey, which is a joint project of the University of
Massachusetts and the Infrared Processing and Analysis
Center/California Institute of Technology, funded by the National
Aeronautics and Space Administration and the National Science
Foundation.

This publication makes use of data products from the Wide-field
Infrared Survey Explorer, which is a joint project of the University
of California, Los Angeles, and the Jet Propulsion
Laboratory/California Institute of Technology, funded by the National
Aeronautics and Space Administration.

Funding for SDSS-III has been provided by the Alfred P. Sloan
Foundation, the Participating Institutions, the National Science
Foundation, and the U.S. Department of Energy Office of Science.
The SDSS-III web site is http://www.sdss3.org/.

SDSS-III is managed by the Astrophysical Research Consortium for the
Participating Institutions of the SDSS-III Collaboration including the
University of Arizona,
the Brazilian Participation Group,
Brookhaven National Laboratory,
University of Cambridge,
Carnegie Mellon University,
University of Florida,
the French Participation Group,
the German Participation Group,
Harvard University,
the Instituto de Astrofisica de Canarias,
the Michigan State/Notre Dame/JINA Participation Group,
Johns Hopkins University,
Lawrence Berkeley National Laboratory,
Max Planck Institute for Astrophysics,
Max Planck Institute for Extraterrestrial Physics,
New Mexico State University,
New York University,
Ohio State University,
Pennsylvania State University,
University of Portsmouth,
Princeton University,
the Spanish Participation Group,
University of Tokyo,
University of Utah,
Vanderbilt University,
University of Virginia,
University of Washington,
and Yale University.

\appendix

\section{Photometric Distance PDFs for Giant Stars}\label{sec:appdist}

In this appendix, we describe the calculation of the photometric-distance
distributions $p(d|l,b,(J-\ks)_0,H_0,\feh,\df,\iso)$ for the stars in our
sample. The photometric-distance PDF for each star is the combination
of the probability of its observed photometry, given a model for the
color--absolute-magnitude distribution of giant stars, and a prior on
the distance. Thus, we write
\begin{equation}
\begin{split}
p(d|l,b,(J-\ks)_0,H_0,\feh,\df,\iso) & \propto p((J-\ks)_0,H_0 | d,\feh,\iso)\,p(d,l,b|\df)\\
& = \iso_{\feh}(H_0-\dm(d),(J-\ks)_0)\,\dens(R,z|\df)\,d^2\cos b\,,
\end{split}\label{eq:distprior}
\end{equation}
where \dm\ is the distance modulus, and $d^2\cos b$ is the Jacobian
for the coordinate transformation $(R,\phi,z) \rightarrow (d,l,b)$,
because we have anticipated that the distance prior is written in
Galactocentric cylindrical coordinates $(R,z,\phi)$. This prior is an
exponential distribution in $R$, $\dens(R,z|\df) \propto
\exp\left(-R/\hR \right)$; our sample is sufficiently close to the
plane that vertical density gradients are unimportant (see
\figurename~\ref{fig:data_lb}). We have implicitly assumed that the
prior does not depend on \feh. Our fiducial model has $\hr = 3$ kpc,
as is appropriate for the metal-rich disk stars that make up our
sample \citep{Bovy12a}.

\begin{figure}[htbp]
\includegraphics[width=0.5\textwidth,clip=]{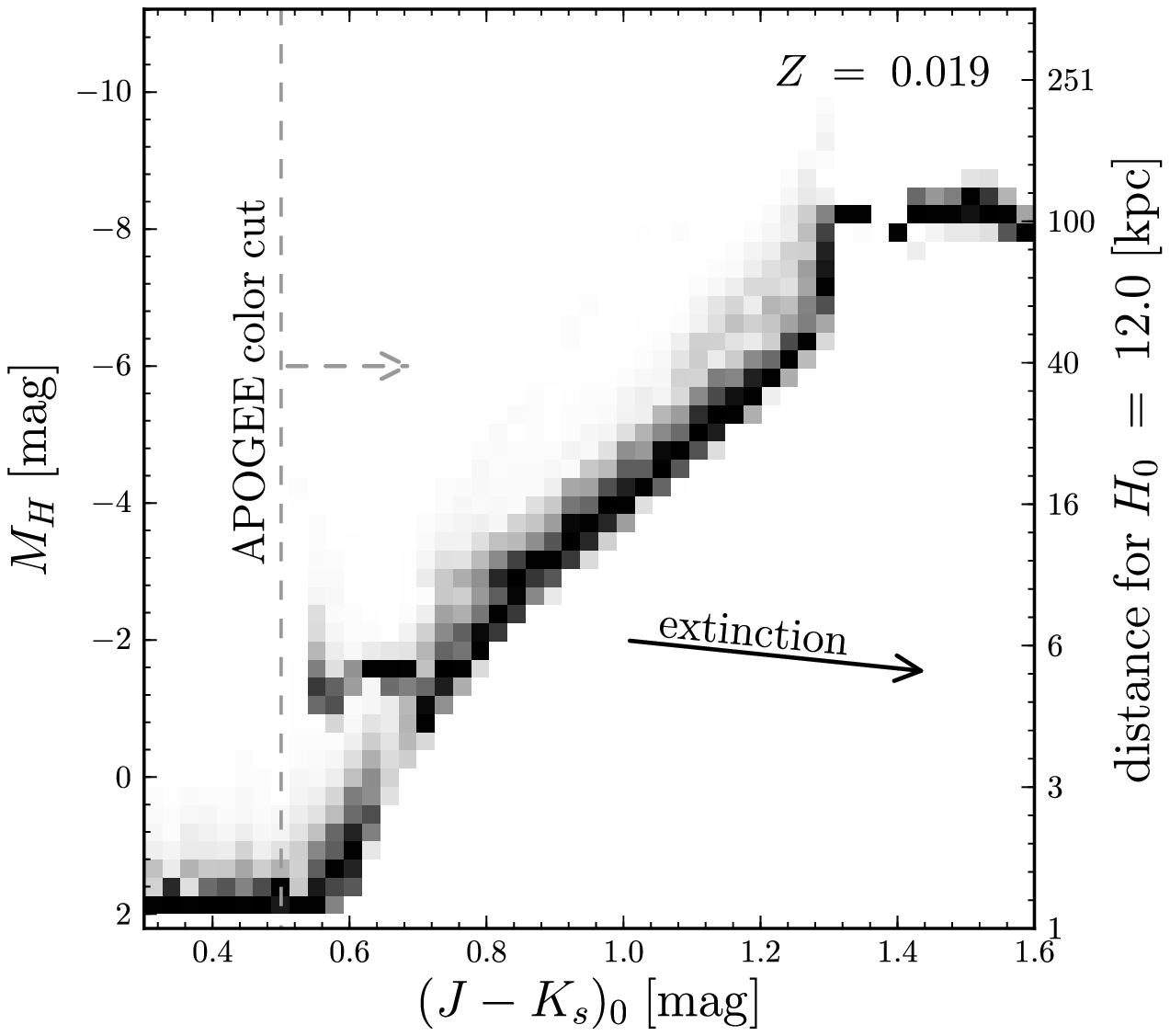}
\caption{Density in the color--magnitude plane, for the giant branch
  of stars with solar metallicity. Each color bin is normalized
  separately to show the absolute-magnitude distribution at each
  color.  This density is calculated using Padova
  isochrones \citep{Bertelli94a,Bonatto04a,Marigo08a,Girardi10a},
  assuming a lognormal \citet{Chabrier01a} IMF and a constant
  star-formation rate. This figure is an example of the density that
  we use as the photometric-distance distribution for each star. The
  length of the extinction arrow is the median extinction for the
  sample (0.45 in $J-\ks$).}\label{fig:imf_h_jk}
\end{figure}

In \eqnname~(\ref{eq:distprior}), we have also anticipated that we
obtain the probability of the observed photometry of the star, given
its distance from a model for the isochrones of giants, combined with
a model for the distribution along the isochrone from an
initial-mass-function model; this probability is labeled as `iso'. We
only use the color $(J-\ks)_0$ and the magnitude $H_0$, as little is
gained by adding $J_0$ or $\kso$ separately, and we compute the
density in the color--absolute-magnitude plane for a given age and
metallicity (which constitutes a single isochrone) by counting the
number of stars generated from an initial mass function in small boxes
in $(J-\ks)_0$ -- $H_0$. For stars in our sample, we find the nearest
$Z/Z_\odot$ (assuming $Z_\odot = 0.019$) on a grid with a
$0.0005\,\dex$ spacing, and we average Padova isochrones for
metallicities $Z/Z_\odot, Z/Z_\odot-0.0005$, and $Z/Z_\odot, +0.0005$,
except at the edges of the grid at $Z/Z_\odot = 0.0005$ and $Z/Z_\odot
= 0.03$. We average over age by assuming a constant star-formation
rate up to 12 Gyr (except for the fit
in \sectionname~\ref{sec:systematics} with multiple populations, where
we use an exponentially-declining star-formation rate). We use a
lognormal \citet{Chabrier01a} model for the IMF (but the resulting
distance PDFS are essentially the same when using a
\citealt{Kroupa03a} IMF). The resulting density for a star with solar
metallicity is shown in \figurename~\ref{fig:imf_h_jk}. For a given
distance and \feh, the probability of a star's magnitude given its
color is evaluated by computing its absolute magnitude $M_H$ and
evaluating a density like the one shown in
\figurename~\ref{fig:imf_h_jk}.

The distribution $\iso_{\feh}(H_0-\dm(d),(J-\ks)_0)$ is relatively
insensitive to changes in \feh\ at high metallicity ($\feh \gtrsim
-0.4$), where $90\%$ of our sample lies. Thus, for the majority of the
stars in our sample, the photometric distance PDF does not strongly
depend on the measured \feh\ and
resembles \figurename~\ref{fig:imf_h_jk}.

Also shown in \figurename~\ref{fig:imf_h_jk} is the direction that
extinction moves objects in color and magnitude. In contrast to the
case for the main sequence, this direction is at a large angle with
the locus of the giant branch, such that under- or over-estimated
extinction values can significantly change the photometric-distance
PDF. The length of the extinction arrow in this figure is the median
extinction in $(J-\ks)$ for the sample (0.45 mag).

\section{Analysis Tests on Mock Data Samples}\label{sec:appmock}

\begin{deluxetable*}{lr@{}lr@{}lr@{}lr@{}lr@{}lr@{}l}
\tablecaption{}
\tablecolumns{13}
\tablewidth{0pt}
\tabletypesize{\footnotesize}
\tablecaption{Galactic Parameters for a Flat-Rotation-Curve Fit to Mock Data Sets}
\tablehead{\colhead{Parameter} & \multicolumn{2}{c}{Data / Mock input} & \multicolumn{2}{c}{Mock 1} & \multicolumn{2}{c}{Mock 2} & \multicolumn{2}{c}{Mock 3} & \multicolumn{2}{c}{Mock 4} & \multicolumn{2}{c}{Mock 5}}
\startdata

$V_c(R_0)\ [\mathrm{km\ s}^{-1}]$ & $218$&$\pm6$ & $225$&$^{+1}_{-10}$ & $222$&$\pm$5 & $221$&$^{+1}_{-12}$ & $225$&$^{+1}_{-11}$ & $217$&$^{+1}_{-12}$\\
$R_0\ [\mathrm{kpc}]$ & $8.1$&$^{+1.2}_{-0.1}$ & $8.0$&$^{+1.2}_{-0.1}$ & $8.0$&$^{+1.1}_{-0.1}$ & $7.9$&$^{+0.9}_{-0.1}$ & $8.0$&$^{+0.6}_{-0.1}$ & $7.8$&$^{+0.9}_{-0.1}$\\
$V_{R,\odot}\ [\mathrm{km\ s}^{-1}]$ & $-10.5$&$^{+0.5}_{-0.8}$ & $-10.3$&$^{+0.8}_{-0.4}$ & $-9.3$&$^{+0.6}_{-0.4}$ & $-9.8$&$^{+1.0}_{-0.4}$ & $-10.8$&$^{+0.9}_{-0.4}$ & $-9.4$&$^{+0.9}_{-0.4}$\\
$\Omega_{\odot}\ [\mathrm{km\ s}^{-1}\ \mathrm{kpc}^{-1}]$ & $30.0$&$^{+0.3}_{-3.3}$ & $30.7$&$^{+0.4}_{-4.2}$ & $30.6$&$^{+0.3}_{-3.2}$ & $30.3$&$^{+0.4}_{-3.7}$ & $30.7$&$^{+0.3}_{-2.5}$ & $30.3$&$^{+0.4}_{-3.8}$\\
$\sigma_R(R_0)\ [\mathrm{km\ s}^{-1}]$ & $31.4$&$^{+0.1}_{-3.2}$ & $32.0$&$^{+0.2}_{-2.2}$ & $28.8$&$^{+1.0}_{-0.6}$ & $31.5$&$^{+1.0}_{-1.1}$ & $34.3$&$^{+0.1}_{-3.1}$ & $29.3$&$^{+1.8}_{-0.4}$\\
$R_0/h_\sigma$ & $0.03$&$^{+0.1}_{-0.27}$ & $0.10$&$^{+0.01}_{-0.17}$ & $-0.07$&$\pm$0.06 & $0.07$&$^{+0.04}_{-0.09}$ & $0.25$&$^{+0.01}_{-0.19}$ & $-0.04$&$^{+0.08}_{-0.04}$\\
$X^2 \equiv \sigma_\phi^2 / \sigma_R^2$ & $0.70$&$^{+0.30}_{-0.01}$ & $0.38$&$^{+0.21}_{-0.02}$ & $0.60$&$^{+0.12}_{-0.03}$ & $0.39$&$^{+0.13}_{-0.01}$ & $0.35$&$^{+0.19}_{-0.02}$ & $0.50$&$^{+0.09}_{-0.03}$\\\\
$\Delta\chi^2/\mathrm{dof}$ &  \ldots &  & $-0.19$ &  & $-0.18$ &  & $-0.20$ &  & $-0.17$ &  & $-0.22$ & 
\enddata

\tablecomments{\protect{T}he final line of this table has the
  difference in $\chi^2$ per degree-of-freedom between the best-fit to
  the mock data set and the best-fit to the real data. Parameters are
  as in \tablename~\ref{table:results}.\label{table:mocks}}
\end{deluxetable*}

\begin{deluxetable*}{llr@{}lr@{}lr@{}lr@{}l}
\tablecaption{}
\tablecolumns{10}
\tablewidth{0pt}
\tabletypesize{\footnotesize}
\tablecaption{Galactic Parameters for non-Flat-Rotation-Curve Fits to Mock Data Sets}
\tablehead{\colhead{Rotation curve} & \colhead{Parameter} & \multicolumn{2}{c}{Data / Mock input} & \multicolumn{2}{c}{Mock 1} & \multicolumn{2}{c}{Mock 2} & \multicolumn{2}{c}{Mock 3}}
\startdata

power-law & $V_c(R_0)\ [\mathrm{km\ s}^{-1}]$ & $218$&$\pm6$ & $224$&$^{+28}_{-2}$ & $222$&$^{+8}_{-21}$ & $224$&$^{+4}_{-27}$\\
  & $\beta$ & \ldots &  & $0.07$&$^{+0.06}_{-0.04}$ & $0.02$&$^{+0.02}_{-0.11}$ & $0.08$&$^{+0.01}_{-0.14}$\\
  & $R_0\ [\mathrm{kpc}]$ & $8.1$&$^{+1.2}_{-0.1}$ & $8.0$&$^{+1.2}_{-0.1}$ & $8.0$&$^{+0.7}_{-0.1}$ & $8.0$&$^{+0.7}_{-0.1}$\\
  & $V_{R,\odot}\ [\mathrm{km\ s}^{-1}]$ & $-10.5$&$^{+0.5}_{-0.8}$ & $-9.7$&$^{+0.7}_{-0.4}$ & $-9.3$&$^{+0.5}_{-0.7}$ & $-9.1$&$^{+0.7}_{-0.6}$\\
  & $\Omega_{\odot}\ [\mathrm{km\ s}^{-1}\ \mathrm{kpc}^{-1}]$ & $30.0$&$^{+0.3}_{-3.3}$ & $30.8$&$^{+0.5}_{-2.2}$ & $30.4$&$^{+0.2}_{-3.9}$ & $30.5$&$^{+0.5}_{-4.6}$\\
  & $\sigma_R(R_0)\ [\mathrm{km\ s}^{-1}]$ & $31.4$&$^{+0.1}_{-3.2}$ & $31.8$&$^{+0.2}_{-2.2}$ & $29.4$&$\pm$1.0 & $33.3$&$^{+0.4}_{-1.6}$\\
  & $R_0/h_\sigma$ & $0.03$&$^{+0.01}_{-0.27}$ & $0.07$&$^{+0.01}_{-0.14}$ & $-0.05$&$^{+0.05}_{-0.08}$ & $0.17$&$^{+0.04}_{-0.11}$\\
  & $X^2 \equiv \sigma_\phi^2 / \sigma_R^2$ & $0.70$&$^{+0.30}_{-0.01}$ & $0.44$&$^{+0.16}_{-0.01}$ & $0.59$&$^{+0.10}_{-0.04}$ & $0.36$&$^{+0.09}_{-0.02}$\\\\
linear & $V_c(R_0)\ [\mathrm{km\ s}^{-1}]$ & $218$&$\pm6$ & $223$&$\pm$12 & $222$&$^{+20}_{-11}$ & $224$&$^{+11}_{-32}$\\
  & $\mathrm{d} V_c / \mathrm{d} R \left(R_0 \right)\ [\mathrm{km\ s}^{-1}\ \mathrm{kpc}^{-1}]$ & \ldots &  & $2.0$&$^{+0.8}_{-2.7}$ & $1$&$^{+1}_{-3}$ & $2.8$&$^{+0.8}_{-5.2}$\\
  & $R_0\ [\mathrm{kpc}]$ & $8.1$&$^{+1.2}_{-0.1}$ & $8.0$&$^{+1.0}_{-0.1}$ & $8.0$&$^{+0.8}_{-0.1}$ & $8.0$&$^{+0.8}_{-0.1}$\\
  & $V_{R,\odot}\ [\mathrm{km\ s}^{-1}]$ & $-10.5$&$^{+0.5}_{-0.8}$ & $-9.7$&$\pm0.7$ & $-9.1$&$^{+0.9}_{-0.2}$ & $-8.9$&$\pm0.6$\\
  & $\Omega_{\odot}\ [\mathrm{km\ s}^{-1}\ \mathrm{kpc}^{-1}]$ & $30.0$&$^{+0.3}_{-3.3}$ & $30.8$&$^{+0.3}_{-4.3}$ & $31$&$^{+1}_{-3}$ & $30.6$&$^{+0.3}_{-5.8}$\\
  & $\sigma_R(R_0)\ [\mathrm{km\ s}^{-1}]$ & $31.4$&$^{+0.1}_{-3.2}$ & $31.3$&$^{+0.5}_{-1.6}$ & $30.0$&$^{+0.2}_{-2.1}$ & $32.8$&$^{+0.6}_{-1.7}$\\
  & $R_0/h_\sigma$ & $0.03$&$^{+0.01}_{-0.27}$ & $0.04$&$^{+0.03}_{-0.09}$ & $-0.03$&$^{+0.01}_{-0.13}$ & $0.12$&$^{+0.02}_{-0.10}$\\
  & $X^2 \equiv \sigma_\phi^2 / \sigma_R^2$ & $0.70$&$^{+0.30}_{-0.01}$ & $0.46$&$^{+0.13}_{-0.03}$ & $0.59$&$^{+0.14}_{-0.01}$ & $0.40$&$^{+0.13}_{-0.02}$\\\\
cubic & $V_c(R_0)\ [\mathrm{km\ s}^{-1}]$ & $218$&$\pm6$ & $223$&$^{+21}_{-3}$ & $223$&$^{+11}_{-5}$ & $223$&$^{+16}_{-6}$\\
  & $\mathrm{d} V_c / \mathrm{d} R \left(R_0 \right)\ [\mathrm{km\ s}^{-1}\ \mathrm{kpc}^{-1}]$ & \ldots &  & $2.6$&$^{+0.6}_{-1.9}$ & $-1$&$\pm$1 & $1$&$\pm$1\\
  & $\mathrm{d}^2 V_c / \mathrm{d} R^2 \left(R_0 \right)\ [\mathrm{km\ s}^{-1}\ \mathrm{kpc}^{-2}]$ & \ldots &  & $-0.2$&$^{+0.1}_{-0.2}$ & $0.1$&$^{+0.1}_{-0.3}$ & $-0.3$&$^{+0.3}_{-0.1}$\\
  & $\mathrm{d}^3 V_c / \mathrm{d} R^3 \left(R_0 \right)\ [\mathrm{km\ s}^{-1}\ \mathrm{kpc}^{-3}]$ & \ldots &  & $0.002$&$^{+0.040}_{-0.007}$ & $0.05$&$^{+0.04}_{-0.01}$ & $0.08$&$\pm$0.02\\
  & $R_0\ [\mathrm{kpc}]$ & $8.1$&$^{+1.2}_{-0.1}$ & $8.0$&$^{+1.2}_{-0.1}$ & $8.1$&$^{+0.9}_{-0.1}$ & $8.0$&$^{+0.8}_{-0.1}$\\
  & $V_{R,\odot}\ [\mathrm{km\ s}^{-1}]$ & $-10.5$&$^{+0.5}_{-0.8}$ & $-9.3$&$^{+0.2}_{-0.7}$ & $-9.6$&$^{+0.8}_{-0.5}$ & $-9.2$&$^{+0.3}_{-0.7}$\\
  & $\Omega_{\odot}\ [\mathrm{km\ s}^{-1}\ \mathrm{kpc}^{-1}]$ & $30.0$&$^{+0.3}_{-3.3}$ & $30.6$&$^{+0.3}_{-2.7}$ & $30.7$&$^{+0.3}_{-2.9}$ & $30.3$&$^{+0.5}_{-2.4}$\\
  & $\sigma_R(R_0)\ [\mathrm{km\ s}^{-1}]$ & $31.4$&$^{+0.1}_{-3.2}$ & $32.6$&$^{+0.2}_{-2.0}$ & $28.6$&$^{+0.8}_{-0.7}$ & $33.0$&$^{+0.3}_{-1.3}$\\
  & $R_0/h_\sigma$ & $0.03$&$^{+0.01}_{-0.27}$ & $0.11$&$^{+0.03}_{-0.10}$ & $-0.12$&$^{+0.05}_{-0.08}$ & $0.13$&$^{+0.06}_{-0.07}$\\
  & $X^2 \equiv \sigma_\phi^2 / \sigma_R^2$ & $0.70$&$^{+0.30}_{-0.01}$ & $0.42$&$^{+0.12}_{-0.02}$ & $0.62$&$\pm$0.05 & $0.35$&$^{+0.07}_{-0.02}$
\enddata
\tablecomments{\protect{P}arameters in this table are as in
  \tablename~\ref{table:results}.\label{table:mocks-nonflat}}
\end{deluxetable*}

In this section, we perform extensive mock data analyses to test the
methodology and approximations described
in \sectionname~\ref{sec:method}, and to determine at what level the
data are sensitive to changes in the Galactic parameters. We create
mock data sets by re-sampling the line-of-sight velocity for each data
point from its PDF
$p(\vlos|l,b,(J-\ks)_0,H_0,\feh,\vc(R),\Ro,V_{R,\odot},V_{\phi,\odot},\df,\iso)$
(\eqnname~(\ref{eq:pvlos})) in the best-fit model with a flat rotation
curve for the real data (\tablename~\ref{table:results}). However,
rather than using the simple Gaussian-with-asymmetric-drift-offset
model, we sample from a Dehnen distribution function
(\eqnname~(\ref{eq:fdehnen})), with a radial scale length of $h_R =
3\,\kpc$, and a radial-velocity-dispersion scale length taken from the
best-fit model. Because the Dehnen distribution function has $X^2$
built-in, we do not use the best-fit $X^2$. For a Dehnen DF, the
marginalization of the planar velocity distribution over the component
of the velocity that is tangential to the line-of-sight cannot be done
analytically, and we numerically integrate over this component.  We do
not add any observational error to the mock line-of-sight velocities,
as the observational uncertainties are vanishingly small
(see \sectionname~\ref{sec:data}). In this manner we produce five mock
data sets that are exactly like the real data, except for the
line-of-sight velocities.

We apply the exact same fitting procedure to the mock data as is used
for the real data. We fit the five mock data sets with a flat-rotation
curve model, and obtain the best-fits and uncertainties given in
\tablename~\ref{table:mocks} (for comparison, this table also includes
the best-fit model for the real data that was used to generate the
mock data). We see that the true values for the parameters of interest
are recovered well by the methodology
of \sectionname~\ref{sec:method}. We also see that the uncertainty
ranges on the Galactic parameters are similar, both in size and
asymmetry, to those derived for the real data. Therefore, the best-fit
parameters do not appear to be biased by the approximate methodology
used in this paper, and the uncertainties are as should be expected
from these mock data tests.

\begin{figure}[htbp]
\includegraphics[width=0.5\textwidth]{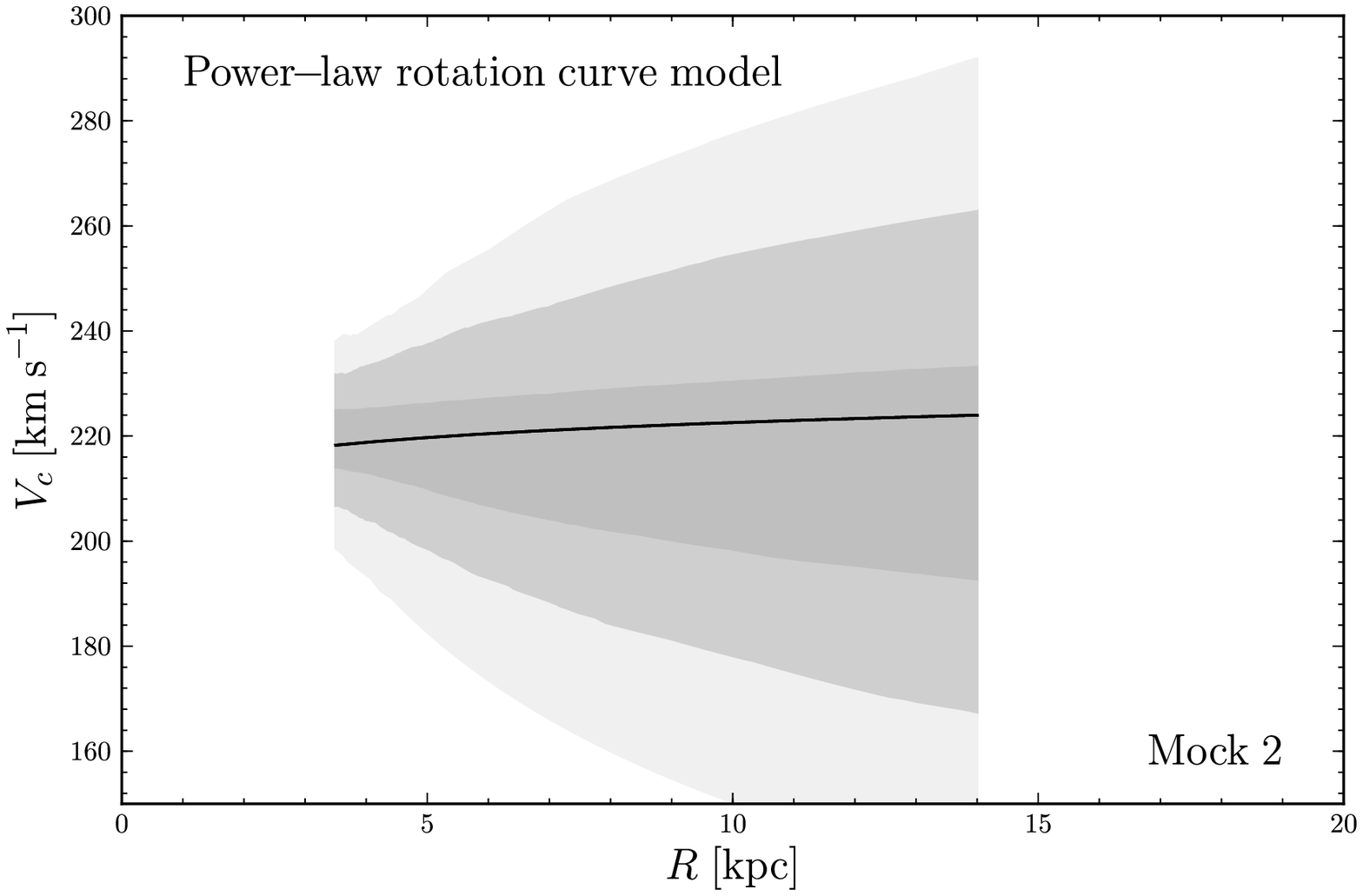}
\includegraphics[width=0.5\textwidth]{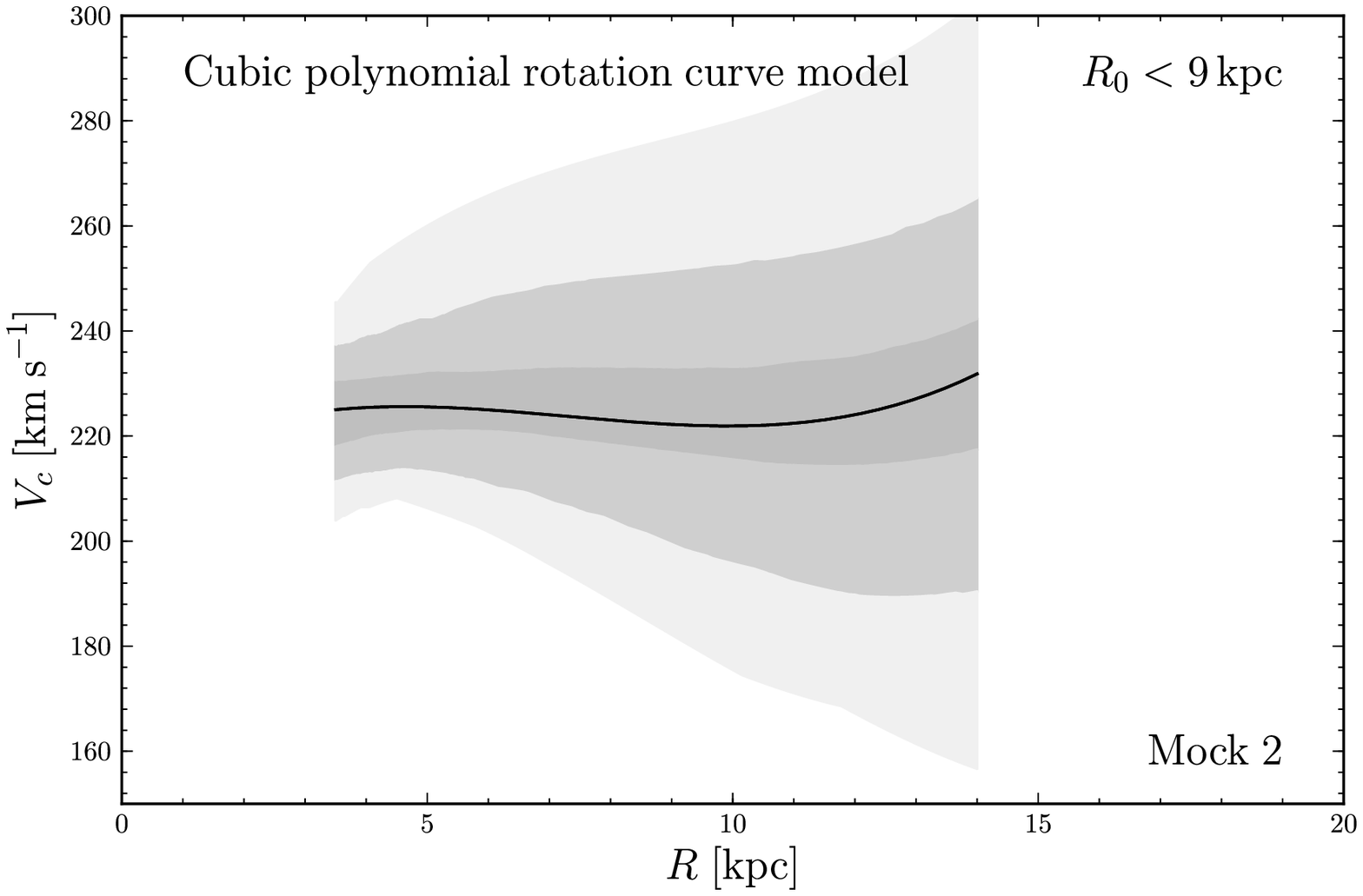}
\caption{Same as \figurename~\ref{fig:rotcurve}, but for mock data set
  2. To completely mimic the procedure applied to the real data we
  have imposed a $R_0 < 9\,\kpc$ prior for the cubic-polynomial fit to
  the rotation curve, even though this only excludes tens of the
  10,000 points that were sampled from the PDF (see
  \tablename~\ref{table:mocks-nonflat}).}\label{fig:rotcurve-mocks}
\end{figure}

We also fit models with a non-flat rotation curve to the mock data
sets. We only do this for three of the mock data sets; the results for
fitting a power-law rotation curve, or a linear or cubic polynomial
curve, are given in \tablename~\ref{table:mocks-nonflat}. We see that,
in these cases, the Galactic parameters are recovered within the
uncertainties, with perhaps a slight bias of a few $\kms$ in the
best-fit parameters, although the (asymmetric) PDFs for the best-fit
parameters do contain the ``true'' values within the 68\% confidence
intervals given. Again, the size of the uncertainties on the best-fit
parameters for the mock data are similar to those found for the real
data in \tablename~\ref{table:results}.

In \figurename~\ref{fig:rotcurve-mocks}, we show for one of the mock
data sets the same representation of the constraints on the shape of
the rotation curve as is given for the real data in
\figurename~\ref{fig:rotcurve}. Comparing these two figures, we see
that the constraints are similar, with the real data giving a slightly
narrower range around the best-fit, which is approximately flat. It is
clear that around $R = 14\,\kpc$, the constraints on the rotation
curve from the current \apogee\ data become weak, so we limit our
discussion to $4\,\kpc < R < 14\,\kpc$ in this paper.

\end{document}